\documentclass[aps, prb, preprint, groupedaddress]{revtex4-2}

\usepackage{graphicx}
\usepackage{amsmath}
\usepackage{amssymb}
\usepackage{bm}
\usepackage{color}
\usepackage{colortbl}
\usepackage{xcolor}

\definecolor{tud0d}{RGB}{83,83,83}
\definecolor{tud0c}{RGB}{137,137,137}
\definecolor{tud0b}{RGB}{181,181,181}
\definecolor{tud0a}{RGB}{220,220,220}

\definecolor{tud1a}{RGB}{93,133,195}
\definecolor{tud2a}{RGB}{0,156,218}
\definecolor{tud3a}{RGB}{80,182,149}
\definecolor{tud4a}{RGB}{175,204,80}
\definecolor{tud5a}{RGB}{221,223,72}
\definecolor{tud6a}{RGB}{255,224,92}
\definecolor{tud7a}{RGB}{248,186,60}
\definecolor{tud8a}{RGB}{238,122,52}
\definecolor{tud9a}{RGB}{233,80,62}
\definecolor{tud10a}{RGB}{201,48,142}
\definecolor{tud11a}{RGB}{128,69,151}

\definecolor{tud1b}{RGB}{0,90,169}
\definecolor{tud2b}{RGB}{0,131,204}
\definecolor{tud3b}{RGB}{0,157,129}
\definecolor{tud4b}{RGB}{153,192,0}
\definecolor{tud5b}{RGB}{201,212,0}
\definecolor{tud6b}{RGB}{253,202,0}
\definecolor{tud7b}{RGB}{245,163,0}
\definecolor{tud8b}{RGB}{236,101,0}
\definecolor{tud9b}{RGB}{230,0,26}
\definecolor{tud10b}{RGB}{166,0,132}
\definecolor{tud11b}{RGB}{114,16,133}

\definecolor{tud1c}{RGB}{0,78,138}
\definecolor{tud2c}{RGB}{0,104,157}
\definecolor{tud3c}{RGB}{0,136,119}
\definecolor{tud4c}{RGB}{127,171,22}
\definecolor{tud5c}{RGB}{177,189,0}
\definecolor{tud6c}{RGB}{215,172,0}
\definecolor{tud7c}{RGB}{210,135,0}
\definecolor{tud8c}{RGB}{204,76,3}
\definecolor{tud9c}{RGB}{185,15,34}
\definecolor{tud10c}{RGB}{149,17,105}
\definecolor{tud11c}{RGB}{97,28,115}

\definecolor{tud1d}{RGB}{36,53,114}
\definecolor{tud2d}{RGB}{0,78,115}
\definecolor{tud3d}{RGB}{0,113,94}
\definecolor{tud4d}{RGB}{106,139,55}
\definecolor{tud5d}{RGB}{153,166,4}
\definecolor{tud6d}{RGB}{174,142,0}
\definecolor{tud7d}{RGB}{190,111,0}
\definecolor{tud8d}{RGB}{169,73,19}
\definecolor{tud9d}{RGB}{156,28,38}
\definecolor{tud10d}{RGB}{115,32,84}
\definecolor{tud11d}{RGB}{76,34,106}

\usepackage{framed}
\usepackage{float}
\usepackage{enumitem}
\usepackage{soul}
\usepackage{siunitx}
\DeclareSIUnit\angstrom{\text{\AA}}
\usepackage{miller}
\usepackage{chemformula} 
\usepackage{verbatim}
\usepackage{booktabs}
\usepackage{comment}

\usepackage{tikz}
\usetikzlibrary{arrows}
\usetikzlibrary{automata}
\usetikzlibrary{calc}
\usetikzlibrary{intersections}
\usetikzlibrary{plotmarks}
\usetikzlibrary{positioning}
\usetikzlibrary{shapes}
\usetikzlibrary{spy}

\usetikzlibrary{external}
\usepackage{pgfplots}
\usepgfplotslibrary{fillbetween}
\usepgfplotslibrary{groupplots}
\usepgfplotslibrary{colormaps}
\pgfkeys{/pgf/number format/.cd,
    1000 sep={\,},
    min exponent for 1000 sep=4}
\pgfplotsset{unit code/.code 2 args={\si{#1#2}}}
\usepackage{siunitx}
    \DeclareSIUnit\fu{f.u.}
    \DeclareSIUnit\uc{u.c.}
\pgfplotsset{compat=newest}
\usepackage{pgfkeys}
    \def\addlegendimage{\csname pgfplots@addlegendimage\endcsname}

\usepackage{varioref}
\usepackage[colorlinks]{hyperref}
\usepackage[nameinlink]{cleveref}

\begin{document}

\title{Thermal stability of nano-scale ferroelectric domains by molecular dynamics modeling}
\author{Arne J. Klomp}
\email[]{klomp@mm.tu-darmstadt.de}
\affiliation{Department of Materials Science, Technical University of Darmstadt, 64287 Darmstadt, Germany}
\author{Ruben Khachaturyan}
\email[]{ruben.khachaturyan@rub.de}
\author{Theophilus Wallis}
\author{Anna Gr{\"u}nebohm}
\affiliation{Interdisciplinary Centre for Advanced Materials Simulation (ICAMS) and Center for Interface-Dominated High Performance Materials (ZGH), Ruhr-University Bochum, Universit\"atsstr. 150, 44801 Bochum, Germany}
\author{Karsten Albe}
\affiliation{Department of Materials Science, Technical University of Darmstadt, 64287 Darmstadt, Germany}

\date{August 31, 2022}

\newcommand{\AK}[1]{{\color{blue}{#1}}}
\newcommand{\KA}[1]{{\color{red}{#1}}}
\newcommand{\RK}[1]{{\color{green}{#1}}}
\newcommand{\AG}[1]{{\color{purple}{#1}}}
\newcommand{\todo}[1]{{\color{red}{#1}}}

\newcommand{\com}[1]{{\color{blue}{#1}}}

\newcommand{\heff}[0]{$H^{\text{eff}}$}
\newcommand{\RomanNumeralCaps}[1]{\MakeUppercase{\romannumeral #1}}

\begin{abstract}
    Ultra-dense domain walls are increasingly important for many devices but their 
    microscopic properties are so far not fully understood.
    Here we use molecular dynamic simulations to study the domain wall stability in the prototypical ferroelectric \ch{BaTiO3} combining core-shell pair potentials and a coarse-grained effective Hamiltonian.
    We transfer the discussion of the field-driven nucleation and motion of domain walls to thermally induced modifications of the wall without an external driving force.
    Our simulations show that domain wall dynamics and stability depend crucially on microscopic thermal fluctuations.
    Enhanced fluctuations at domain walls may result in the formation of critical nuclei for the permanent shift of the domain wall.
    If two domain walls are close -- put in other words, when domains are small -- thermal fluctuations can be sufficient to bring domain walls into contact and lead to the annihilation of small domains.
    This is even true well below the Curie temperature and when domain walls are initially as far apart as 6 unit cells.
    Such small domains are, thus, not stable and limit the maximum achievable domain wall density in nanoelectronic devices.
\end{abstract}

\maketitle

\newpage
\clearpage

\section{Introduction}
\label{sec:introduction}
    
        In the past two decades ferroelectric domains and especially the domain walls (DWs) separating these domains have come to be viewed in a different light.
        This was fueled by the ever-increasing demand for micro- and nanoscopic devices and electronic components.
        While domains carry the permanent polarization in a ferroelectric, it has been recognized that DWs in themselves offer a variety of intriguing functionalities.
        On the one hand, DWs can enhance existing functional properties such as macroscopic piezoelectric response, dielectric coefficients and conductivity \cite{Liu2017, Kampfe2014, Zuo2014}, and influence phase transitions~\cite{Grunebohm2020, Grunebohm2022}. 
        On the other hand, novel functionality opens the route to future devices~\cite{Catalan2012, Said2017,Bednyakov2018,Sharma2019}, such as data storage devices~\cite{Garcia2014}, diodes~\cite{Whyte2015}, memristors~\cite{Bai2018,McConville2020}, and ferroelectric transistors \cite{Chai2020}.
        Consequently, the research field of "domain wall nanoelectronics" has emerged using DWs as their functional component \cite{Catalan2012}.
        Naturally, in order to maximize the impact of DWs on a material a high density of DWs is desired \cite{Wada2005, Hlinka2009, Liu2017,  Grunebohm2022}.
        
        Yet, surprisingly few microscopic studies tackle the problem of the stability of nano-domains in ferroelectrics at finite temperatures~\cite{Prosandeev2021}.
        Thus, there is it the need for a comprehensive study of the dynamics of closely spaced ferroelectric DWs.
        In this work we address the important questions,
        (a) how the dynamics of ultra-dense DWs can be investigated efficiently, and
        (b) what the highest possible DW density and its temperature-dependence are.
        Is there a lower limit to the distance between individual \qty{180}{\degree} DWs in the prototypical ferroelectric tetragonal phase of \ch{BaTiO3}?
        We show that DW spacings on the order of few unit cells are not stable against thermal fluctuations even well below the Curie temperature.
        The time it takes for a small domain to collapse depends on its thickness and the magnitude of thermal fluctuations.

        We recognize that modern imaging techniques give impressive spatial resolution of ferroelectric structures \cite{Grunebohm2022}.
        However, only computer simulations allow to study ferroelectric systems with quasi-arbitrary temporal resolution and without superimposed impact of surfaces and possible defects.
        We can, therefore, give details and explanations on the microscopic processes related to the stability of nano-domains.
        Several computational models on different scales exist to study ferroelectric effects.
        On the micro- and mesoscopic scale, Landau-type or phase field models are often employed \cite{Levanyuk2020}.
        However, thermal fluctuations cannot be treated directly by these methods and are only rarely mimicked using random fields \cite{Yang2020}.
        Thus, we decided to use classical molecular dynamics (MD) simulations for two different types of potentials: First, we use  pair potentials of core-shell type that capture all the relevant fluctuations and dynamics on an atomic scale \cite{Mitchell1993} as have been pioneered by \textcite{Tinte1999}. Recent applications of core-shell atomistic models include Refs.~\cite{Shin2007, Boddu2017}.
        Second, this approach is supplemented by the computationally more efficient effective Hamiltonian (\heff) approach that is built around the material response on a unit cell level which have been pioneered by Rabe \textit{et al.} \cite{Rabe1987, Zhong1994}.
        Recent studies involving this method involve Refs.~\cite{Khachaturyan2022,Nahas2020}.
        Both approaches are based on ab initio calculations.
        So far, a reliable cross-validation for the two models on a relevant case such as a ferroelectric DW is missing in literature.

        Using the electric field as a driving force, \textcite{Shin2007} found that the propagation of DWs in lead titanate occurs by nucleation and nucleus growth on the DW.
        The study found that the energy barrier for growth of a 2D nucleus on the DW is much smaller than for the formation of an initial nucleus inside a domain far away from any DW.
        Moreover, the growth of nuclei occurs preferentially in the DW plane.
        As the nucleus formation at the DW is easy, we deem it possible that nucleation also occurs randomly through activation by thermal fluctuations in the absence of electric fields.
        Such thermal fluctuations occur naturally and it has been shown recently by \textcite{Caballero2020} that the resulting roughening of ferroelectric DWs is an essential and intrinsic feature for ferroelectrics.
        Currently, we lack information about how the process of roughening of DWs and the motion of DWs behaves as temperatures approach the Curie temperature and when DWs are close to each other, i.e., at high DW density.\\
        
        \ch{BaTiO3} is a prototype material for ferroelectrics \cite{Acosta2017} and many of the relations investigated with this material could be transferred to other materials, too.
        Because thermal fluctuations and DW roughening are ubiquitous in ferroelectrics we expect good transferability to related scenarios.
        
        As the ground to study nanoscopic domains and validate the agreement of atomistic core-shell model and \heff -model we choose charge neutral and stress-free DWs in the tetragonal phase of \ch{BaTiO3}.
        Therefore, we construct \qty{180}{\degree} DWs located on the \hkl(010) plane separating regions of anti-parallel polarization.
        While \qty{90}{\degree} DWs are also charge neutral because of the head-to-tail orientation of the polarization vectors at the DW, they carry a strain which significantly complicates our analysis.
        The \qty{180}{\degree} DW in tetragonal \ch{BaTiO3} is referred to as T180 DW for brevity.

\newpage
\clearpage

\section{Modeling}
\label{sec:modeling}

\begin{figure*}[tb] 
            \centering
            \begin{tikzpicture}
                \node (CSunitcell) [] at (0cm,0cm) {\includegraphics[width=4cm]{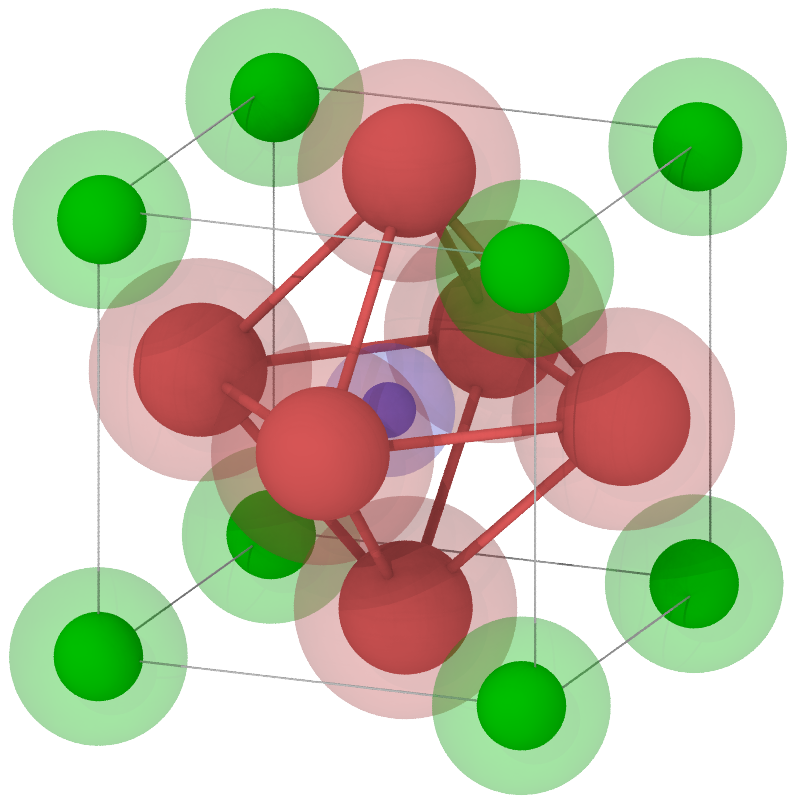}};
                \node (heff) [anchor= west] at (3.5cm, 0cm) {\includegraphics[width=4.5cm]{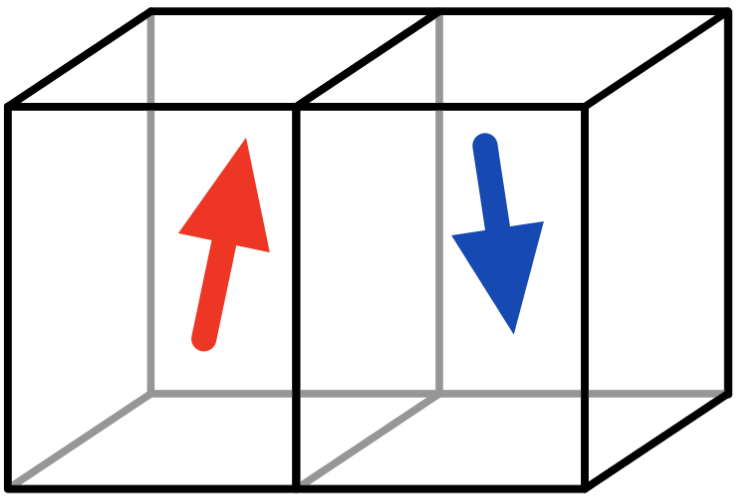}};
                \node (3d) [anchor=west] at (9.5cm, 0cm) {\includegraphics[width=3.8cm] {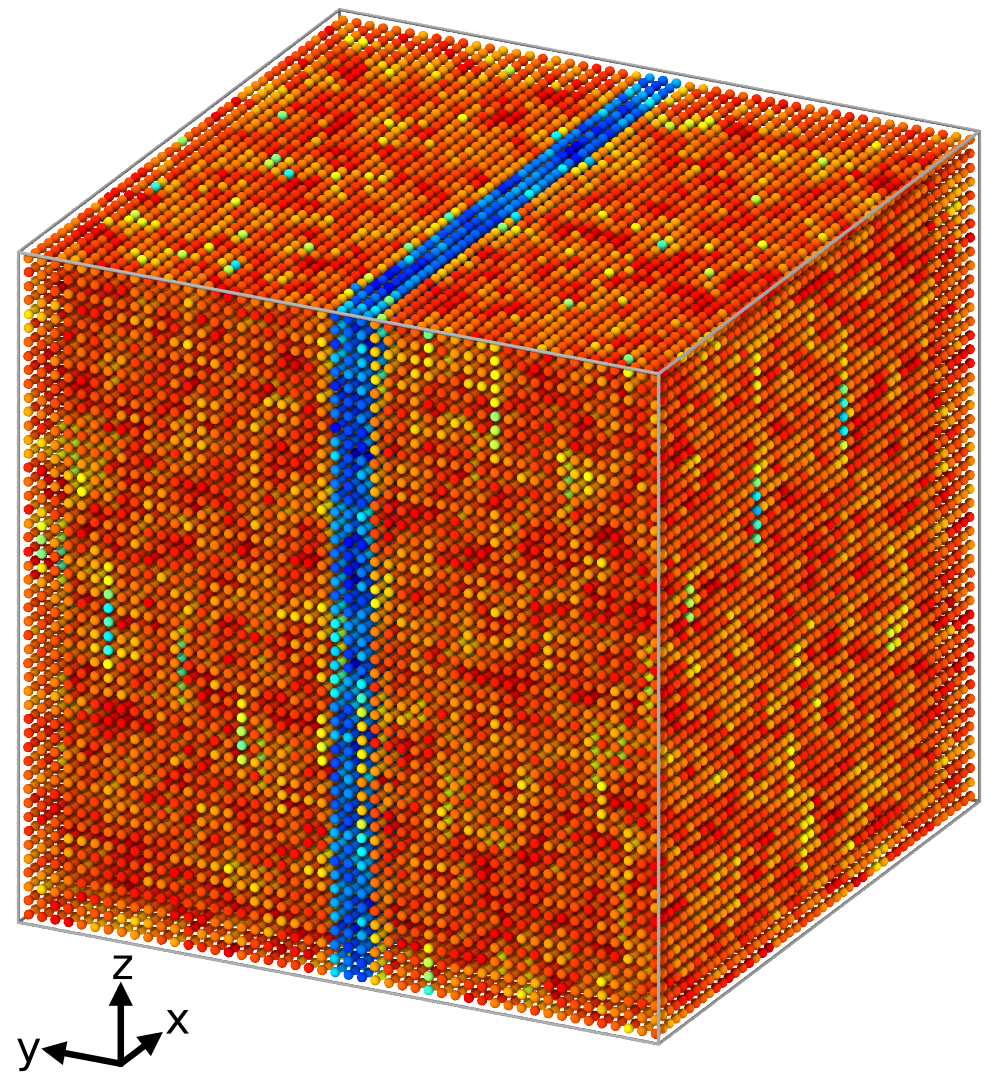}};
                \draw [->, very thick] (CSunitcell.east) -- (heff.west) node [above, pos=0.5] {coarse-} node [below, pos=0.5] {graining};
                \draw [->, very thick] (heff.east) -- (3d.west) node [above, pos=0.5] {mapping};
                \node[inner sep=3pt, rounded corners=0.1cm, fill={green!20}, opacity=0.9, text opacity=1] (Balabel) at ($(CSunitcell)+(-1.5,1.5)$) {Ba};
                \node[inner sep=3pt, rounded corners=0.1cm, fill={blue!20}, opacity=0.9, text opacity=1] (Tilabel) at ($(CSunitcell)+(0.3,-0.2)$) {Ti};
                \node[inner sep=3pt, rounded corners=0.1cm, fill={red!20}, opacity=0.9, text opacity=1] (Olabel) at ($(CSunitcell)+(-1.7,0.0)$) {O};
                \node [inner sep=3pt, rounded corners=0.1cm, fill={red!20}, opacity=0.9, text opacity=1] (ui) at ($(heff)-(1.5cm,0)$) {\textcolor{red}{$\bm{u_i}$}};
                \node [inner sep=3pt, rounded corners=0.1cm, fill={blue!20}, opacity=0.9, text opacity=1] (uj) at ($(heff)+(0.1cm,0)$) {\textcolor{blue}{$\bm{u_j}$}};
                \node (wi) at ($(heff)+(-0.45cm,1.2cm)$) {{$\bm{w_i}$}};
                \node (wj) at ($(heff)+(1.3cm,1.2cm)$) {{$\bm{w_j}$}};
                \node (labela) at ($(CSunitcell.south)-(0cm, 0.2cm)$) {a)};
                \node (labelb) at ($(labela)+(6.25cm,0cm)$) {b)};
                \node (labelc) at ($(labela)+(11.4cm,0cm)$) {c)};
            \end{tikzpicture}
            \caption{
            \label{fig:model}
            Illustration of the modeled degrees of freedoms.
            (a) Atomistic core-shell model for \ch{BaTiO3}:
            Each ion is modeled by one positively charged core and one negatively charged shell particle with small mass which are connected by a spring. 
            (b) Coarse-grained effective Hamiltonian:
            the atomic degrees of freedom per formula unit are mapped on the local dipole moment $\bm{u}_i$, i.e., the local polarization $\bm{P}_i$, and the local strain $\bm{w}_{i}$ which is internally optimized during each time step.
            (c) Exemplary simulation cell with \qtyproduct{48 x 48 x 48}{\uc} and a small reversal domain of \qty{3}{\uc} thickness.
            For simplicity, each unit cell is shown as one dot and only the z-component of each $\bm{P}_i$ is color coded (red = positive; blue = negative). 
        	}
        \end{figure*}
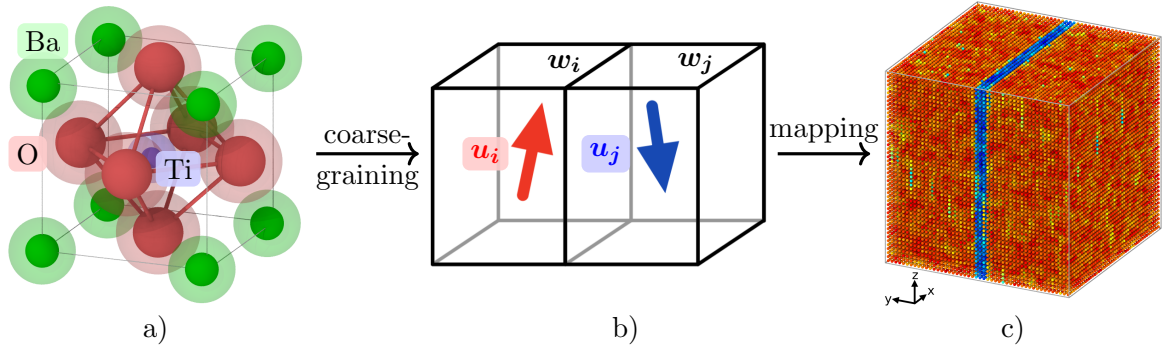

    We simulate single crystalline \ch{BaTiO3} (BTO) in an orthogonal simulation cell with typical side lengths of \qtyproduct{48 x 48 x 48}{\uc}, i.e., a total of \qty{110592}{\uc} (corresponds to approx. 5.5m\,particles in the atomistic model), see \Cref{fig:model}.
    Note that this system size guarantees convergence of energies and polarization profiles.
    We focus on the tetragonal phase with main polarization direction along $+z$ and introduce smaller reversed domains (polarization along $-z$) with thicknesses between \qty{1}{\uc} and \qty{24}{\uc}.
    DWs are in the \hkl(010) plane, i.e., their normal is along the $y$-direction.
    We initialize these domain structures by poling with local electric fields.
    By varying the regions of the applied electric field, domains of different size separated by T180\hkl(010) domain walls are created.
    After equilibrating, we remove the field and study the polarization and DW dynamics in the absence of electric fields.
    
    We use two different modeling approaches -- atomistic core-shell potentials and the coarse-grained effective Hamiltonian (\heff) approach -- to speed up our simulations and verify our results using a second independent technique.
    The poling procedures are slightly different for the different models, as explained below.
    We confirmed that both approaches yield the same microscopic properties (polarization and velocity distributions) and identical trends for the properties discussed.
        
    Regarding temperature, we focus on the range from \qty{70}{\kelvin} below to \qty{10}{\kelvin} above the tetragonal to cubic phase transitions temperature $T_{T\rightarrow C}$ upon heating.

    \subsection{Atomistic Simulations with Core-shell potential}

        For the atomistic simulations we employ the adiabatic core-shell model from \textcite{Vielma2013}.
        All MD simulations with the core-shell potential use the RESPA integrator \cite{Tuckerman1992}, evaluating the pair potential every \qty{0.4}{\femto\second} and the Fourier-space contribution of the electrostatic energy only every \qty{1.2}{\femto\second}. This significantly speeds up the calculation compared to a single time step without sacrificing accuracy.
          
        The interatomic interactions are integrated explicitly up to \qty{10}{\angstrom}, beyond which a particle-particle particle-mesh solver takes care of the long ranged electrostatic forces \cite{Hockney1988} (as implemented in \texttt{LAMMPS} \cite{Plimpton1995}). The thermostat time constant is set to \qty{0.2}{\pico\second} and the barostat time constant to \qty{2.0}{\pico\second}.
        The thermostat uses the center of mass of each core-shell pair for a Nos\'{e}-Hoover velocity scaling.

        After equilibration in the cubic phase (\qty{100}{\pico\second} at \qty{400}{\kelvin}) an electric field $E_z = \qty{+-5e7}{\volt\per\meter}$ is applied while cooling the system to the target temperature for \qty{36}{\pico\second}. Then the field is removed and the system -- now containing two T180\hkl(010) DWs -- is allowed to evolve freely at constant temperature \cite{Boddu2017}.
        Due to the large computational expense we remove the electric field instantaneously.
        We checked that instantaneous and gradual removal of electric field lead to same qualitative results using the \heff -model.

        The evolution of the local polarization is tracked by the local polarization vector per unit cell \(\bm{P}_i\) which we define as \cite{Sepliarsky2011}:
        \begin{align}
            \bm{P}_i = \frac{1}{\Omega} \sum_j{\frac{q^j}{n^j} (\bm{r}_i^j - \bm{r}_i^{\text{Ti}})} \text{ .}
        \end{align}
        Starting from the central Ti atom the atoms $j$ belonging to the same unit cell are identified and their distance to the central atom $(\bm{r}_i^j - \bm{r}_i^{\text{Ti}})$ is calculated.
        It is then multiplied with its ionic charge $q_j$ and divided by its multiplicity $n_j$ by which it appears in the unit cell (e.g. $n^{\text{Ba}} = 6$, because 6 Ba ions surround each Ti).
        This gives the local dipole moment and can be divided by the unit cell volume \(\Omega\) to obtain polarization.
        To reduce noise we average the ionic positions over \(100\)\,time steps, i.e., \qty{120}{\femto\second}.

        We also study high symmetry DWs in static calculations.
        To this end, we take the tetragonal unit cell, and use the conjugate gradient method to relax ionic positions along the tetragonal axis only.
        In this case the simulation cell with size \qtyproduct{1 x 1 x 20}{\uc} has one positive and one negative domain of equal size and dipoles are initially set to zero in the \ch{BaO} or \ch{TiO2} planes in the center of the wall.
        During relaxation the system stays in the high symmetric configurations creating DWs localized in the \ch{BaO} or \ch{TiO2} planes.
        The results are discussed in \Cref{fig:dw_profile_location}.

    \subsection{Coarse-grained Effective Hamiltonian Simulations}

        As gathering statistics on the DW evolution while treating all 5.5m\,atoms, with 6 degrees of freedom each, explicitly is unfeasible, we use the common coarse-graining to the effective Hamiltonian approach \cite{Zhong1995,Zhong1994} and only evolve the three components of the local soft mode vector $\bm{u}$ corresponding to the local dipole moment in time, see \Cref{fig:model}.
        This method successfully describes ferroelectric phase diagrams~\cite{Marathe2017,Kornev2006}, domain structures~\cite{Grunebohm2020,Paul2007,Lai2007}, and functional properties~\cite{Marathe2016,Ponomareva2012,Gui2011} of ferroelectric materials.
        
        The Hamiltonian in \Cref{eq:EffectiveHamiltonian} is based on the local soft mode vectors \(\bm{u}\) which are evolved dynamically in Molecular Dynamics (MD) simulations.
        During each MD step the local strain tensor \(\bm{w}\) and the six components of the global strain tensor \(\eta_1,\dots,\eta_6\) in Voigt notation are optimized.
        In the effective Hamiltonian, 
        \begin{equation}
        \begin{aligned}\label{eq:EffectiveHamiltonian}
            H^{\rm eff}
             & = \frac{M^*_{\rm dipole}}{2} \sum_{i,\alpha}\dot{u}_{\alpha,i}^2\\
             & + V^{\rm self}(\{\bm{u}\})+V^{\rm dpl}(\{\bm{u}\})+V^{\rm short}(\{\bm{u}\})\\
             & + V^{\rm elas,\,homo}(\eta_1,\dots\!,\eta_6)+V^{\rm elas,\,inho}(\{\bm{w}\})\\
             & + V^{\rm coup,\,homo}(\{\bm{u}\}, \eta_1,\cdots\!,\eta_6)+V^{\rm coup,\,inho}(\{\bm{u}\}, \{\bm{w}\})\\
             & -\sum_i Z^*\bm{E}_i.\bm{u}_i  \text{,}
        \end{aligned}
        \end{equation}
        the first term represents the kinetic energy of the dipoles with effective mass $M^*_{\rm dipole}$, and $V^{\rm self}(\{\bm{u}\})$, $V^{\rm dpl}(\{\bm{u}\})$, and  $V^{\rm short}(\{\bm{u}\})$  denote the self-energy and the interactions (long-range and short-range) between local modes in different unit cells \(i\), respectively. 
        The consecutive terms include the elastic energy by the homogeneous strain $V^{\rm elas,\,homo}(\eta_1,\dots,\eta_6)$ and inhomogeneous strain $V^{\rm elas,\,inho}(\{\bm{w}\})$, their corresponding coupling terms, and the coupling to external electrical fields ${\bm{E}}_i$, where $Z^*$ is the Born effective charge of the local soft mode. 
        The different terms have been fitted using  density functional theory calculations by \textcite{Nishimatsu2010}. 

        The effective Hamiltonian is used for molecular dynamics simulations via the \texttt{feram} code~\footnote{\url{http://loto.sourceforge.net/feram/}, accessed 2021} developed by \textcite{Nishimatsu2008} with use of the Nos\'e-Poincar\'e thermostat~\cite{Bond1999}.
        The simulation cells are, first, equilibrated at the desired temperature and, second, the multi-domain structure is created by poling with local electric fields of \qty{+-10e6}{\volt\per\meter}. The field is gradually removed in four steps: \qty{+-7e6}{\volt\per\meter}, \qty{+-4e6}{\volt\per\meter}, \qty{+-1e6}{\volt\per\meter}, and with equilibration over 30~ps per each field value. 
        In the given temperature interval, however, the fast equilibration of dipoles even allows to use instantaneous field changes without qualitative change of the results. 
        
    \subsection{Width and Energy of a Domain Wall}

        Domain walls (DWs) disturb the long-range ordering of the polarization vector of the system.
        Geometrically a DW can be characterized by its domain wall width \(d_{\text{DW}}\) and thermodynamically we can define its energy \(\Delta E^*_{\text{DW}}\).

        The width of the DW is judged based on the z-component of local polarization vectors $P_z$ across the wall,
        \begin{align}
            P_z = P_z^0 \tanh{\left(\frac{x - x_{\text{DW}}}{d_{\text{DW}}}\right)} \text{,}
        \end{align}
        where $x_{\text{DW}}$ is the position of the DW center and $d_{\text{DW}}$ is the width of the DW.

        Additionally, we track the energy penalty induced by the domain wall  by:
        \begin{align}
            \Delta E^*_{\text{DW}} = E_{\text{w/ DW}} - E_{\text{w/o DW}} \label{eq:apparent_DW_energy}\text{ .}
        \end{align}
        Here, \(E_{\text{w/ DW}}\) and \(E_{\text{w/o DW}}\) are the energy of a simulation cell with and without a DW \cite{Grunebohm2020}.
        In order to compare different system sizes it is convenient to define the energy penalty density, i.e., \(\Delta E^*_{\text{DW}}\) per area of a sharp and flat DW.
        In our simulation of simulation cells with two T180 DWs the area is, therefore, twice the simulation cell cross section.
        
        We note that the energy penalty is often used as synonymous to domain wall energy in literature, strictly speaking the latter is however based on free energies while we neglect the entropy contribution.

\newpage
\clearpage

\section{Results}
\label{sec:results}

    Both models used here qualitatively reproduce the phase sequence of \ch{BaTiO3}, but they cannot reproduce the phase transition temperatures exactly \cite{Tinte1999, Sepliarsky2004, Vielma2013}.
    To ease comparison between the two models we give temperatures as differences \(\Delta T = T - T_{t\rightarrow c}\) to the temperature of the tetragonal to cubic phase transition \(T_{t\rightarrow c}\).

    \subsection{Model validation and domain wall properties}
    \label{subsec:model_validation}

        In the following we evaluate our multi-model approach for the tetragonal phase of \ch{BaTiO3} based on the properties of T180\hkl(010) walls and compare to results in literature.

    \subsubsection{Static DW properties}
    
        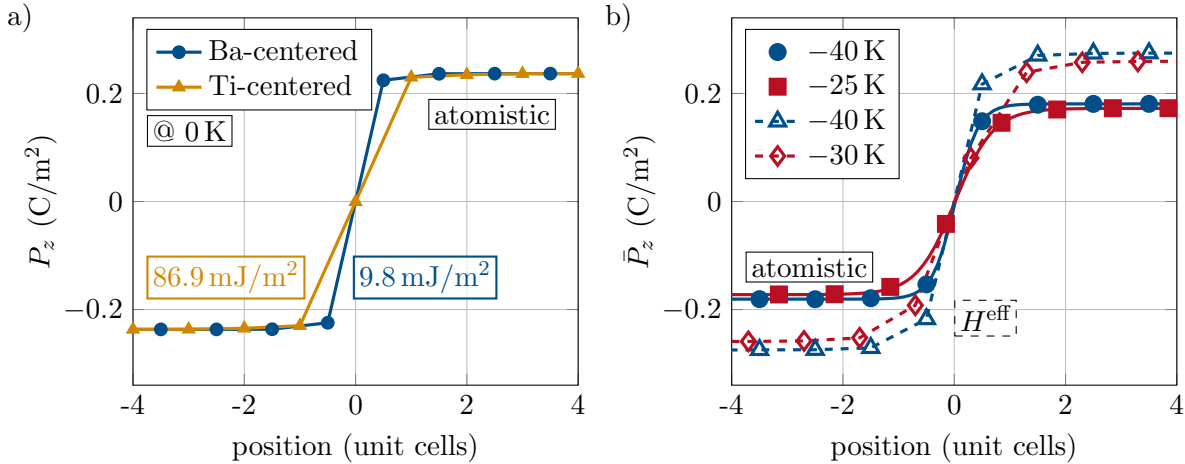
\begin{figure*}[htb] 
            \centering
            \begin{tikzpicture}[/pgfplots/tick scale binop=\times]
                \begin{axis} [
                    name=pzmap,
                    width=7.5cm,
                    anchor = west,
                    xmin=5,
                    xmax=13,
                    xlabel=position (unit cells),
                    ylabel=$P_z$ (\unit{\coulomb\per\square\meter}),
                    ylabel shift = -5pt,
                    xtick={5, 7, 9, 11, 13},
                    xticklabels={-4, -2, 0, 2, 4},
                    ymin=-0.34,
                    ymax=0.34,
                    legend pos=north west,
                    legend style={cells={anchor=west}},
                    grid=major,
                    ]
                    \addplot [tud1c, very thick, mark=*, solid] table [x expr=\thisrow{position}-0.5, y=setup1] {data/dw_Pz_static_v4.dat};
                    \addlegendentry{Ba-centered};
                    \addplot [tud7c, very thick, mark=triangle*, solid] table [x=position, y=setup2] {data/dw_Pz_static_v4.dat};
                    \addlegendentry{Ti-centered};
                    \node[black, inner sep=2pt, draw, anchor=north east, fill=white, fill opacity=0.7, text opacity=1] (atomisticlabel) at (12.6,0.19) {atomistic};
                    \node[tud1c, inner sep=2pt, draw, thick, anchor=north east, fill=white, fill opacity=0.7, text opacity=1] (gammaBa) at (11.55,-0.1) {\qty{9.8}{\milli\joule\per\square\meter}};
                    \node[tud7c, inner sep=2pt, draw, thick, anchor=north east, fill=white, fill opacity=0.7, text opacity=1] (gammaTi) at (8.1,-0.1) {\qty{86.9}{\milli\joule\per\square\meter}};
                    \node[black, inner sep=2pt, draw, anchor=north west, fill=white, fill opacity=0.7, text opacity=1] (temperature) at (5.25,0.16) {@ \qty{0}{\kelvin}};
                \end{axis}
                \begin{axis} [
                    name=atomisticfits,
                    at={($(pzmap)+(5cm,0cm)$)},
                    anchor = west,
                    width=7.5cm,
                    xmin=19.0,
                    xmax=27.0,
                    xlabel=position (unit cells),
                    ylabel=$\bar{P}_z$ (\unit{\coulomb\per\square\meter}),
                    ylabel shift = -5pt,
                    xtick={19,21,23,25,27},
                    xticklabels={-4,-2,0,2,4},
                    ymin=-0.34,
                    ymax=0.34,
                    grid=major,
                    legend pos=north west,
                    legend style={cells={anchor=west},},
                    ]
                    \addplot [tud1c, thick, only marks, mark size=3, mark=*] table [x expr=\thisrow{unitcell}-0.5, y=309K] {data/collected_profiles_v4.dat};
                    \addplot [tud1c, very thick, solid, no marks, forget plot] table [x expr=\thisrow{position}-0.5, y=fitPz] {data/fit_profile_309K_v4.dat};
                    \addlegendentry{\(T_C \qty{-40}{\kelvin}\)}; 
                    \addplot [tud9c, thick, only marks, mark size=3, mark=square*] table [x expr=\thisrow{unitcell}-0.15, y=324K] {data/collected_profiles_v4.dat};
                    \addplot [tud9c, very thick, solid, no marks, forget plot] table [x expr=\thisrow{position}-0.15, y=fitPz] {data/fit_profile_324K_v4.dat};
                    \addlegendentry{\(T_C \qty{-25}{\kelvin}\)}; 
                    \addplot [tud1c, very thick, dashed, mark=triangle, mark size=3.5, mark options={solid}, y filter/.code={\pgfmathparse{\pgfmathresult/100}}] table [x expr=\thisrow{plane}-0.5, y=260] {data/dw_profile_Heff.dat};
                    \addlegendentry{\(T_C \qty{-40}{\kelvin}\)};
                    \addplot [tud9c, very thick, dashed, mark=diamond, mark size=3.5, mark options={solid}, y filter/.code={\pgfmathparse{\pgfmathresult/100}}] table [x expr=\thisrow{plane}-0.7, y=270] {data/dw_profile_Heff.dat};
                    \addlegendentry{\(T_C \qty{-30}{\kelvin}\)};
                    \node[black, inner sep=2pt, draw, anchor=south west, fill=white, fill opacity=0.7, text opacity=1] (atomisticlabel) at (19.25,-0.15) {atomistic};
                    \node[black, inner sep=2pt, draw, dashed, anchor=south west, fill=white, fill opacity=0.7, text opacity=1] (atomisticlabel) at (23,-0.25) {\heff};
                \end{axis}
                \node (labela) at ($(pzmap.north west)-(1.5cm, 0.0cm)$) {a)};
                \node (labelb) at ($(atomisticfits.north west)-(1.5cm, 0.0cm)$) {b)};
            \end{tikzpicture}
            \caption{
            \label{fig:dw_profile_location}
            Polarization profiles across \qty{180}{\degree} DWs.
            (a) Comparison between Ba-centered and Ti-centered walls at \qty{0}{\kelvin} from the atomistic core-shell model.
            Insets give the corresponding domain wall energies. 
            (b) Comparison at different temperatures $\Delta T$ for both models.
            }
        \end{figure*}

        Using  static simulations (atomic relaxation at \qty{0}{\kelvin}) we investigate the energy landscape for the static shift of the wall. 
        As it has been shown by nudged elastic band simulations (NEB) \cite{Meyer2002,Beckman2009,Li2018} the energy difference between the high-symmetry \ch{Ba}-centered and \ch{Ti}-centered DWs corresponds to the energy barrier for a rigid shift of the DW.
        In agreement to literature \cite{Padilla1996}, we find that the \ch{Ti}-centered DW is a factor of 10 higher in energy as the dipole on the wall center is in an unfavorable state with zero local polarization, see \Cref{fig:dw_profile_location}~(a)
        \footnote{Quantitatively, the atomistic model overestimates the domain wall energies found in DFT simulations (using LDA at \qty{0}{\kelvin}) of \qty{6.2}{\milli\joule\per\square\meter}and \qty{62}{\milli\joule\per\square\meter} \cite{Padilla1996}.} 
        and, thus, the energy barrier for a rigid DW shift corresponds to \( \Delta \gamma = \qty{77}{\milli\joule\per\square\meter}\).
        As discussed by \textcite{Shin2007} this upper bound for the energy barrier is, however, never realized.
        Instead, the shift of the DW follows a nucleation and growth process with lower energy barriers.
        Therefore, the fact that the \heff -approach does not allow to model the Ti-centered wall explicitly does not hinder its application.
        
        Instead, the reliability of the methods depends crucially on their ability to reproduce the relevant thermal fluctuations and local relaxations.
        For the former point, we confirmed that the magnitude of fluctuations inside the bulk material as well as at the DW is comparable for the two models (see Supplemental Material 1 at \_) and in agreement with literature \cite{Shin2007, Liu2016b}.
        Regarding the latter point,  we recorded the pair-distribution function using the atomistic model and did not find any structural relaxations or distortions deviating from tetragonal \ch{BaTiO3} in the vicinity of the T180 DW further justifying the absence of these atomistic details in the \heff -model, see Supplemental Material 1 at \_ .
        
        Taking the different approximations and parametrizations of both models into account, DW energies and widths agree surprisingly well with each other and with values in literature.
        In terms of local dipole moments and strain the coarse-grained description and the interatomic potential reproduce the properties of Ba-centered T180 DWs at finite temperatures quite well.

    \subsubsection{DW properties at finite temperature}   
        
        \Cref{fig:dw_profile_location}~(b) compares the polarization profiles across the DW for different temperatures showing that DWs are only few unit cells wide, which is in agreement to accurate DFT calculations and Landau theory \cite{Grunebohm2012,Padilla1996,Marton2010}.
        For the \heff -model and the atomistic model we find bulk polarizations of \(P_z = \qty{0.28}{\coulomb\per\square\meter}\) and \(\qty{0.18}{\coulomb\per\square\meter}\), respectively, which is close to the experimental value of remnant polarization \(P_r = \qty{0.21}{\coulomb\per\square\meter} \) at room temperature \cite{Shieh2009}.

        With increasing temperature both methods yield a decreasing spontaneous polarization (red symbols in  \Cref{fig:dw_profile_location}~(b)).
        Additionally, the apparent energy penalty of the DW \(\Delta E^*_{\text{DW}}\), see \Cref{eq:apparent_DW_energy}, also increases with temperature, see Supplemental Material 3 at \_ .
        This result is in contrast to predictions by Landau theory that the domain wall energy decreases with increasing temperature (and decreasing magnitude of $P_z$) \cite{Marton2010}.
        How can we explain this apparent discrepancy?

    \subsubsection{Reliable estimation of DW energy from dynamic calculations}
       
        In the bulk as well as on the DW, thermal fluctuations result in the switching of local dipoles against their surrounding polarization.
        These fluctuations also appear as small clusters of needle-like shape in a single domain state, as illustrated in \Cref{fig:polarization_switching_series}~(b) on the right.
        Such clusters form and disappear spontaneously anywhere in the simulation cell at finite temperature.
        When these clusters appear on the DW the DW becomes rough and increases the effective wall area \cite{Caballero2020}.
        Consequently, we hypothesize that the energy per area does indeed decrease as suggested by Landau theory, but that the decrease is overcompensated by the increase in area due to roughening.
        
        In order to test this hypothesis we take a different approach to estimating the DW energy per area.
        With a look at \Cref{fig:polarization_switching_series}~(b) on the right,
        assume that needle-like clusters of height \(s\) (in unit cells) in a single-domain state carry an excess energy that only consists of contributions from its nearest neighboring unit cells.
        We deliberately consider the simulation setup without DW in order to avoid interactions of the clusters with the DW.
        In this case, one has to distinguish between the four interfaces parallel to the polarization (like T180 DWs) and the unfavorable head-to-head/tail-to-tail walls along the \(\pm z\)-direction.

        With the microscopic energy density per unit cell area for head-to-head/tail-to-tail contributions \(E_c\) and T180 contributions \(E_{\text{DW}}\), the excess energy of a single cluster inside the domain is:
        \begin{align}
            E_a = 2 E_c + 4 s E_{\text{DW}} \text{.}
        \end{align}
        
        \(E_c\) as well as \(E_{\text{DW}}\) are variable with temperature, mainly because polarization is a function of temperature.
        These clusters are appearing due to thermally activated fluctuations which commonly follow an Arrhenius behavior.
        Thus, the probability \(f\) that a needle-like cluster of size \(s\) exists at a given temperature \(T\) is given by:
        \begin{align}
            f = f_0 \exp{\left[ -\left( 2 E_c + 4 s E_{\text{DW}} \right) / \left( k_b T \right) \right]} \label{eq:frac_reversed_dipoles_energy}
        \end{align}
        If we assume \(E_c\) and \( E_{\text{DW}} \) to be single valued for a given temperature, we can obtain \( E_{\text{DW}} \) by counting the frequency with which a cluster of size \(s\) appears.
        
        As shown in Supplemental Material 4 at \_, we obtain DW energy densities \(E_{\text{DW}}\) that decrease from \qty{5.86}{\milli\joule\per\square\meter} at \( \Delta T = \qty{-90}{\kelvin} \) to \qty{4.73}{\milli\joule\per\square\meter} at \( \Delta T = \qty{-30}{\kelvin} \) from the atomistic model.
        Despite the global increase in energy penalty of the DW \(E^*_{\text{DW}}\) with increasing temperature, the actual energy per element of DW area decreases with increasing temperature due to the reduction of spontaneous polarization.
        The reason is that fluctuations lead to a roughening of the DW increasing its area.

        Our simulations, thus, nicely demonstrate that the temperature evolution of DW energy and width can only be fully understood if thermal fluctuations are taken into account.
        Consequently, it is important to combine accurate but costly DFT calculations, more coarse-grained atomistic models, and phenomenological models to bridge the scales and paint a complete picture of DW properties and behavior.

\newpage
\clearpage

    \subsection{\label{subsec:switching}Spontaneous Switching/Domain Collapse}
    
        \subsubsection{Phenomenological observations}

            \paragraph{Fluctuations \& driving force}

            DWs carry excess energy, thus, there is a driving force to remove DWs in a ferroelectric.
            However, with two parallel, flat, and infinitely thin DWs the system does not \textit{see} an energy gradient and has no immediate means of evolving towards the energy minimum.
            Can thermal fluctuations of local dipoles \cite{Kumar2010} enable two closely spaced DWs to activate this path and ultimately lead to the collapse of small domains?

            On the one hand, clusters of reversed polarization are ubiquitous and form and disappear on a sub-picosecond time scale.
            At the domain wall, the energy of reversed clusters is reduced by \(1 \times E_{\text{DW}}\) as one interface is anyway anti-parallel and as both \(E_{\text{DW}}\) and \(E_c\) are reduced due to lower polarization in the vicinity of the DW, see \Cref{fig:dw_profile_location}~(b).
            Thus the frequency \(f\) of flipped clusters is enhanced at the DW, see Supplemental Material 1 at \_ .
            For example, at $\Delta T = \qty{-49}{\kelvin}$ in the atomistic model,
            we find \qty{0.5}{\percent} and \qty{7.7}{\percent} of local dipoles flipped inside a single-domain state and at the DW (2 unit cells right and left of the DW), respectively . 
            The enhanced probability for switched dipoles on the wall can be understood as a flattening of the energy landscape and locally reduced polarization.
  
            On the other hand, DWs induce an energy penalty on the order of \qty{10}{\milli\joule\per\square\meter}, see calculations above.
            Thus, the system can gain energy if one of the domains vanishes with time.
            However, without thermal activation (\qty{0}{\kelvin}), there is an energy barrier of about \qty{77}{\milli\joule\per\square\meter} to rigidly shift a domain wall through the unfavorable TiO$_2$-centered position.
            Furthermore, without electric field, there is no force favoring one polarization direction over the other. 
            Without fluctuations, the multi-domain structure at \qty{0}{\kelvin} stays forever.
            
            \newcommand{\shiftright}{0.19\textwidth}
            \begin{figure*}[htbp] 
                \centering
                \begin{tikzpicture}
                    \node[anchor=north] (3dA) at (-5.5cm,-2cm) {\includegraphics[width=5cm] {img/polarization_evolution_3d_frame_31_full.png}};
                    \node[anchor=north] (3dB) at (0cm,-2cm) {\includegraphics[width=5cm]{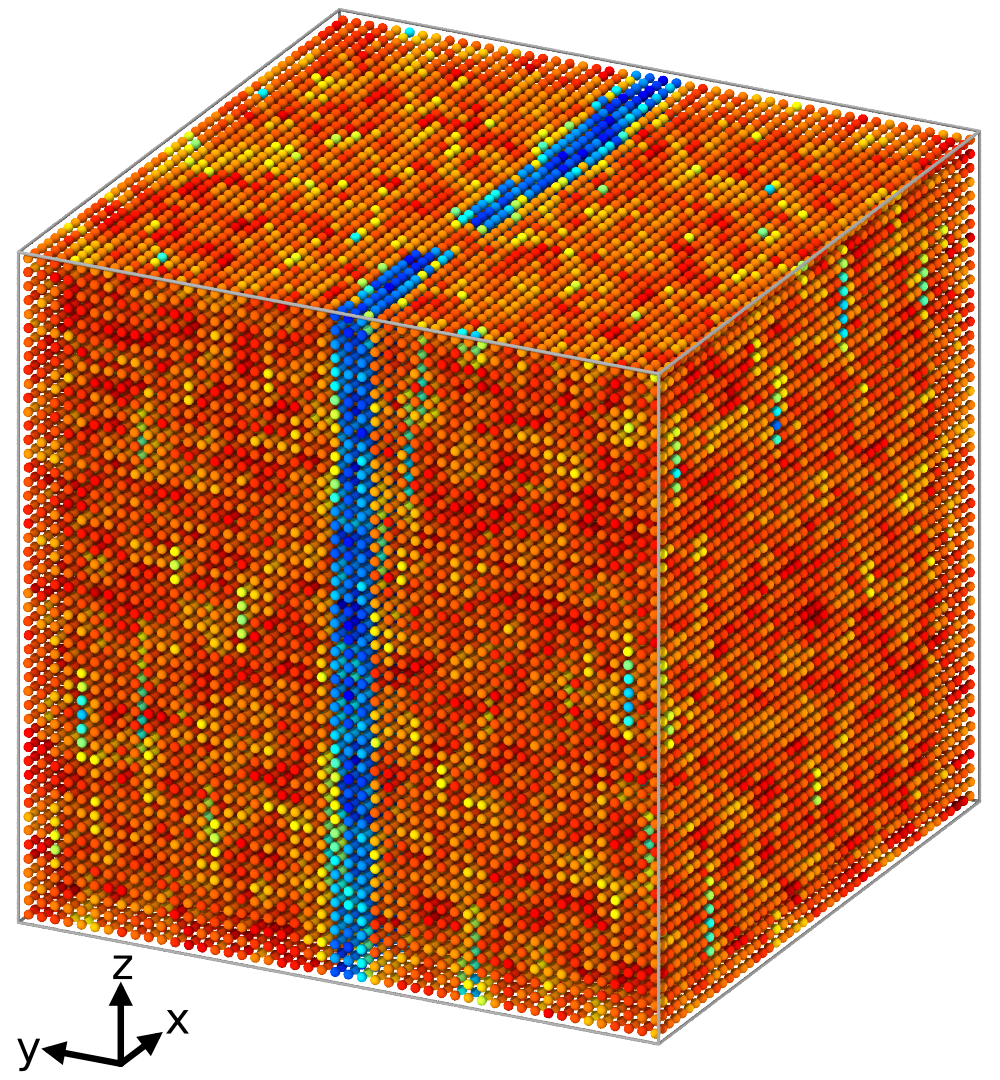}};
                    \node[anchor=north] (3dC) at (5.5cm,-2cm) {\includegraphics[width=5cm]{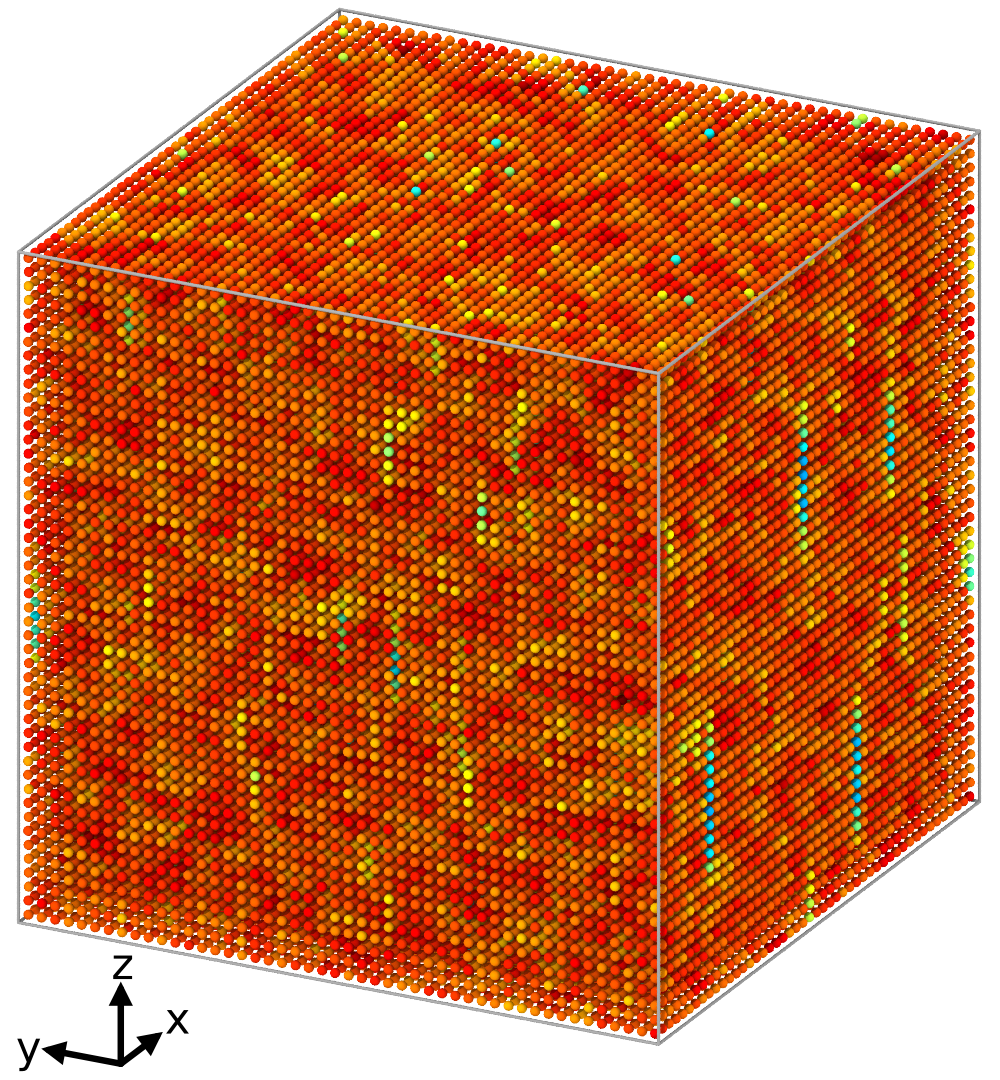}};
                    \node[inner sep=3pt, rounded corners=0.1cm, fill={black!20}, opacity=0.9, text opacity=1] (3Dalabel) at ($(3dA)+(0.0,-2.9)$) {$t = \qty{1.2}{\pico\second}$};
                    \node[inner sep=3pt, rounded corners=0.1cm, fill={black!20}, opacity=0.9, text opacity=1] (3Dblabel) at ($(3dB)+(0.0,-2.9)$) {$t = \qty{10.8}{\pico\second}$};
                    \node[inner sep=3pt, rounded corners=0.1cm, fill={black!20}, opacity=0.9, text opacity=1] (3Dclabel) at ($(3dC)+(0.0,-2.9)$) {$t = \qty{44.4}{\pico\second}$};
                    \node[anchor=north] (3dD) at ($(3dA)+(0.0,-3.2)$) {\includegraphics[width=5cm]{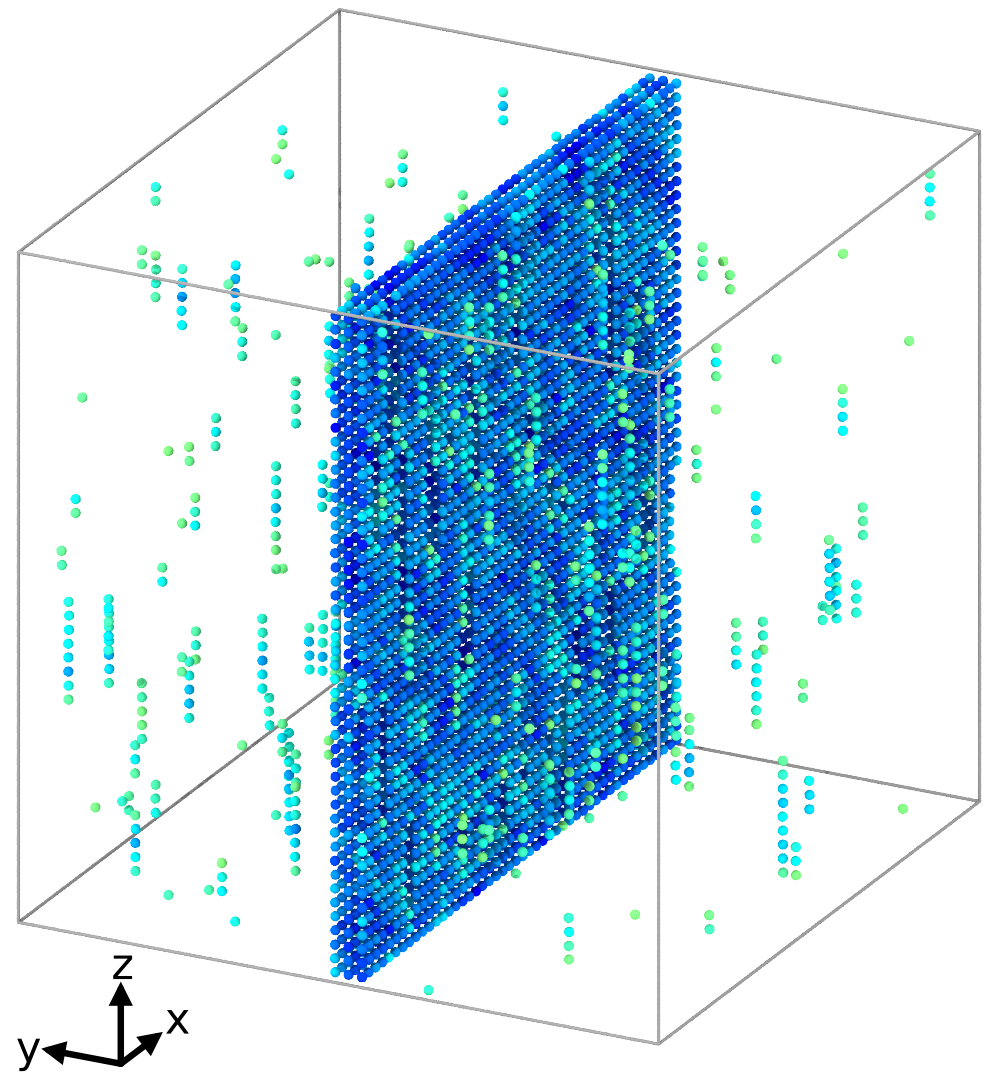}};
                    \node[anchor=north] (3dE) at ($(3dB)+(0.0,-3.2)$) {\includegraphics[width=5cm]{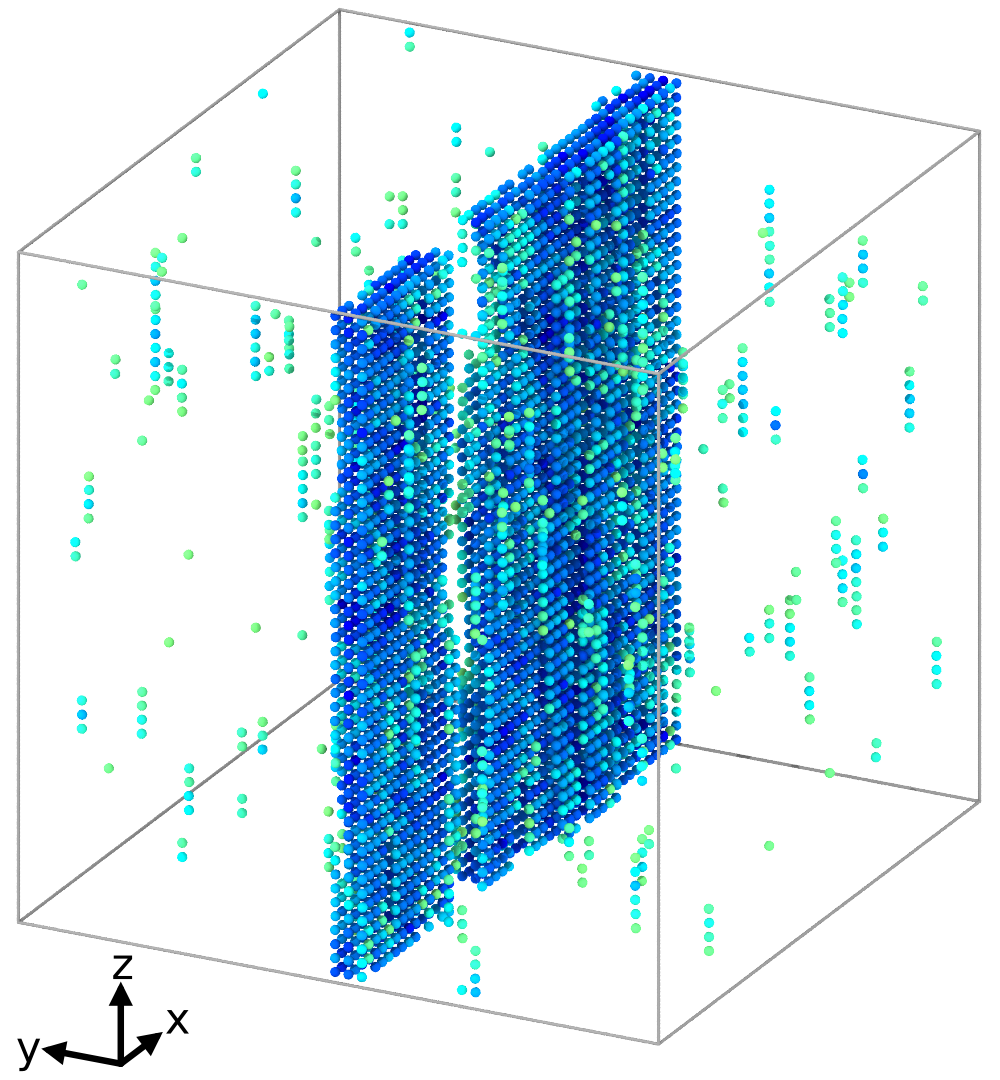}};
                    \node[anchor=north] (3dF) at ($(3dC)+(0.0,-3.2)$) {\includegraphics[width=5cm]{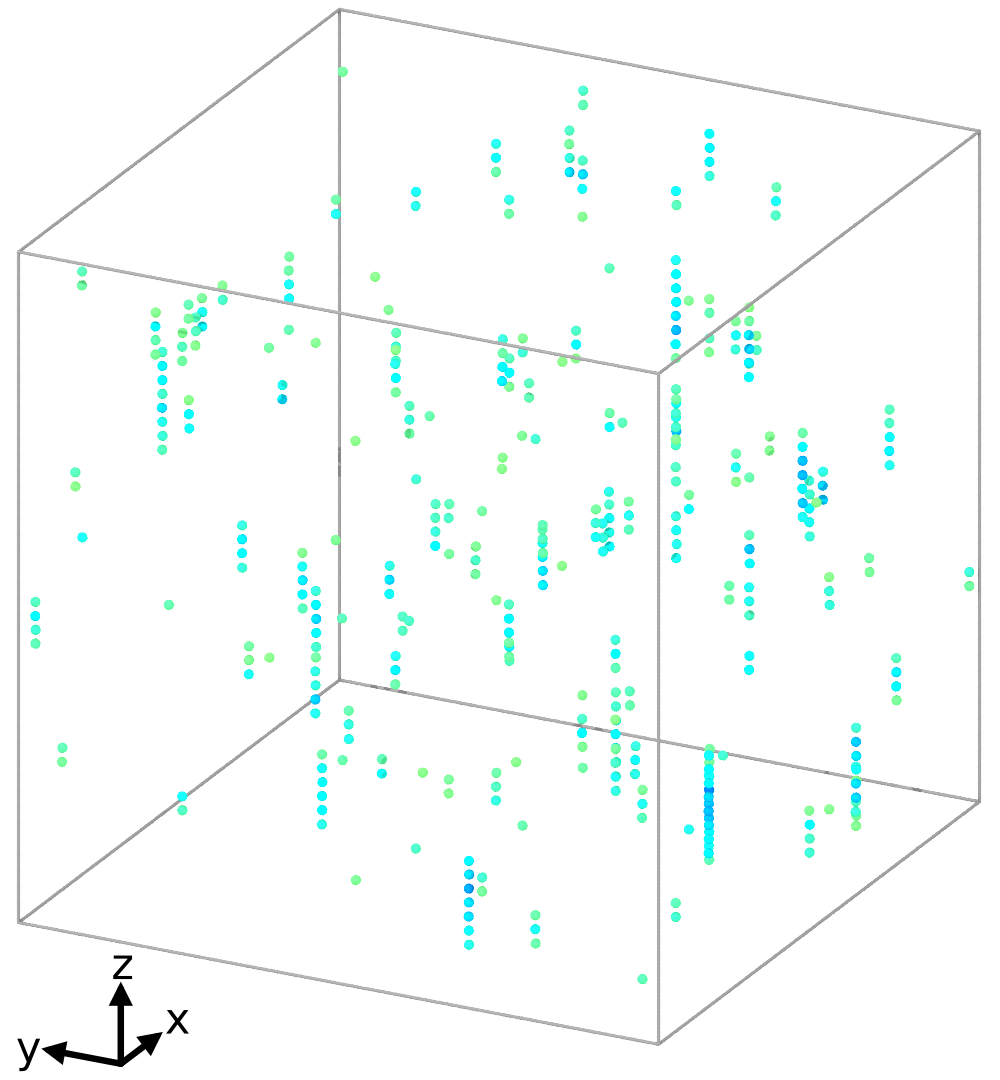}};
                    \node[anchor=north] (2dA) at ($(3dD)+(0.0,-2.9)$) {\includegraphics[width=5cm]{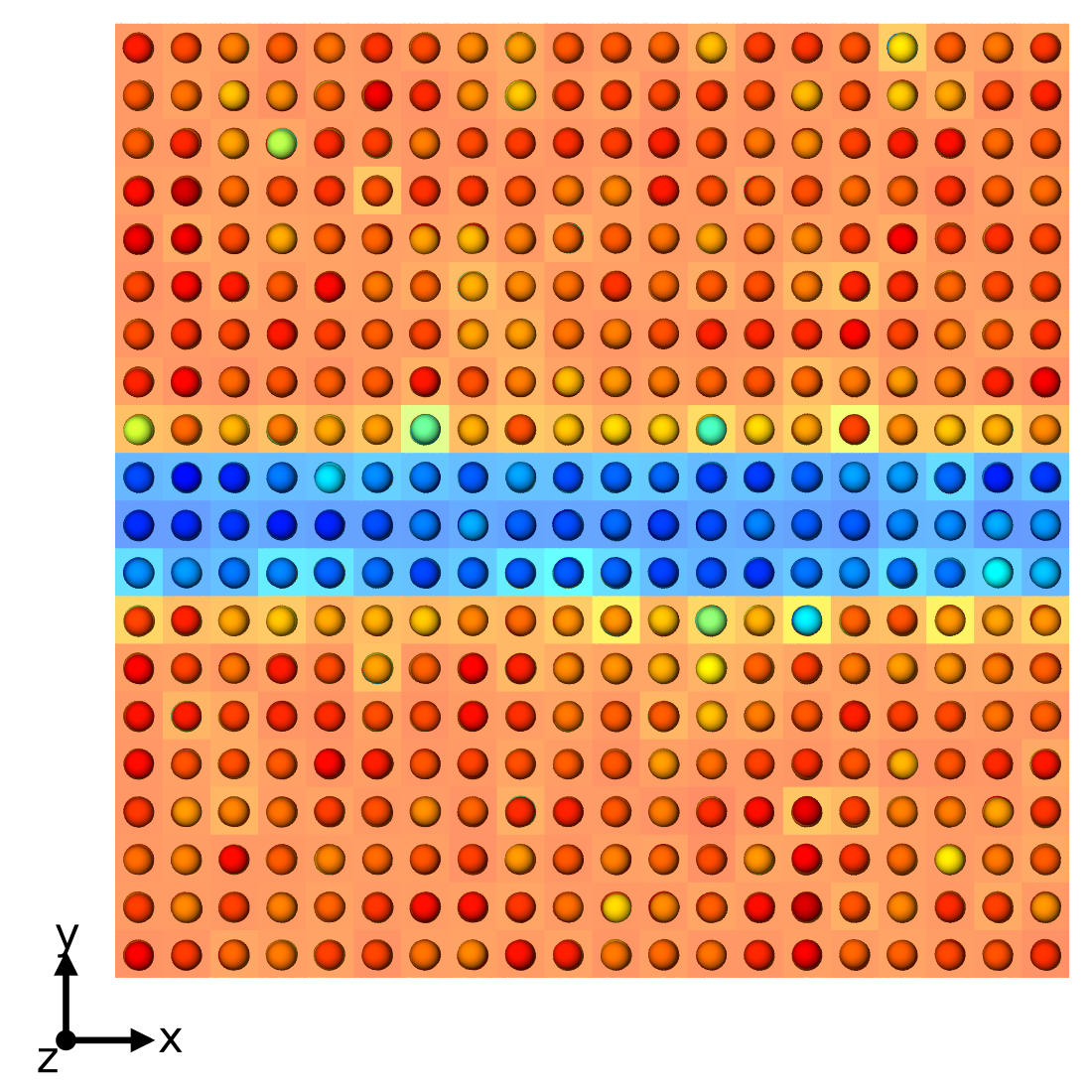}};
                    \node[anchor=north] (2dB) at ($(3dE)+(0.0,-2.9)$) {\includegraphics[width=5cm]{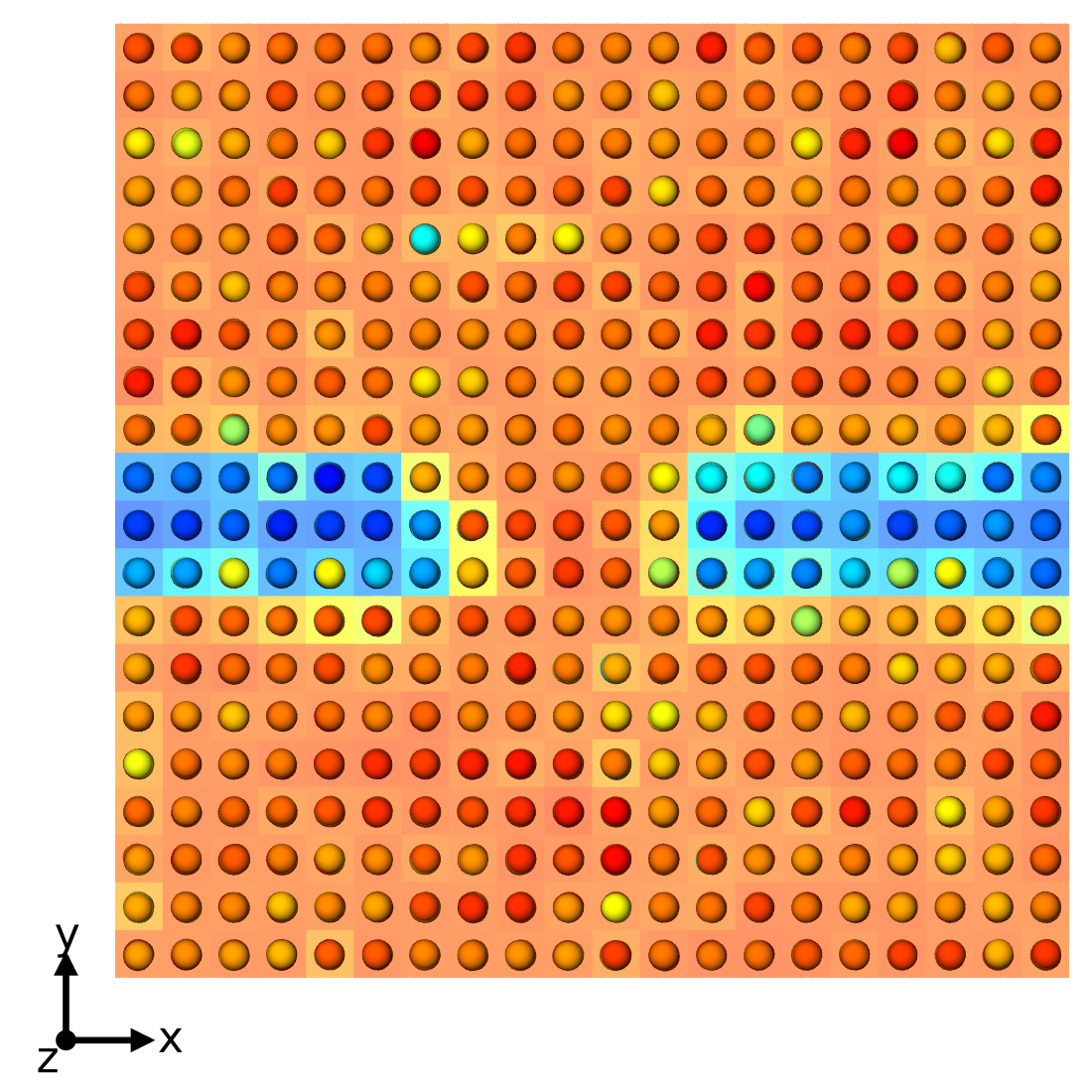}};
                    \node[anchor=north] (2cC) at ($(3dF)+(0.0,-2.9)$) {\includegraphics[width=5cm]{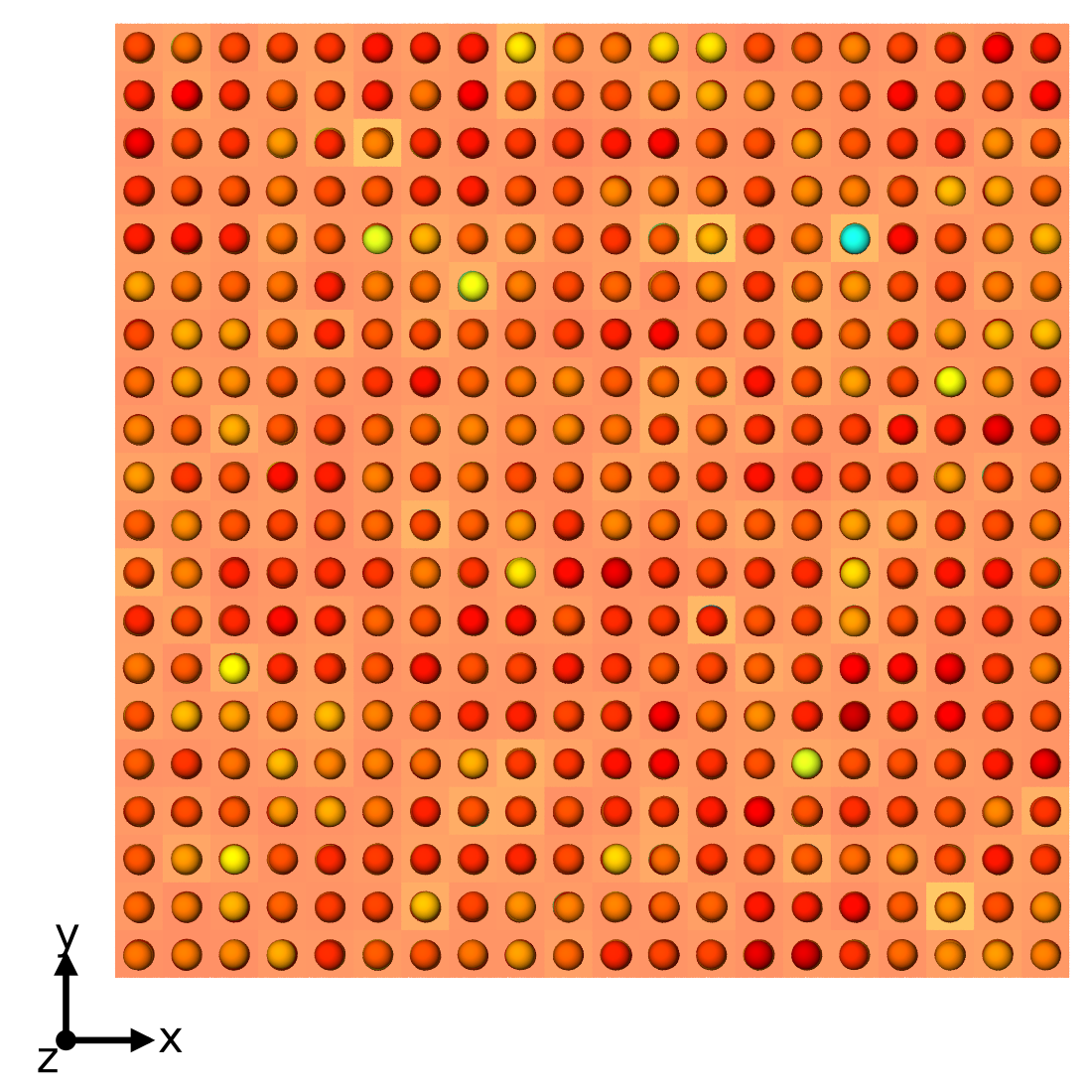}};
                    %
                    \node (subfiga) at ($(3dA.north west)+(0.2,-0.3)$) {a)};
                    \node (subfigb) at ($(3dD.north west)+(0.2,-0.3)$) {b)};
                    \node (subfigc) at ($(2dA.north west)+(0.2,-0.3)$) {c)};
                \end{tikzpicture}
                \caption{
                \label{fig:polarization_switching_series}
                Local polarization evolution and collapse of small domains from the atomistic simulation.
                (a) 3D representation of the full system;
                (b) only cells with negative polarization are shown;
                (c) top view (along \(z\)-direction) with column-wise averages of the polarization as the background color.
                In this scenario the initial thickness is \qty{3}{\uc} at \(\Delta T = \qty{-26}{\kelvin}\). 
                Each dot represents one unit cell color coded by its polarization $P_z$.
                Chains of switched dipoles form in both domains and disappear spontaneously on a sub-picosecond timescale.
                In addition to these fluctuations, the polarization pattern in the negative domain changes with time and vanishes after few tens of picoseconds.
                }
            \end{figure*}
            
            Due to these reasons -- the presence of thermal fluctuations and the excess energy of the DW -- we could indeed observe the collapse of small domains with atomistic as well as \heff -simulations.
            In \Cref{fig:polarization_switching_series} we monitor the collapse of domains of different thicknesses using different representations.
    
            Under the influence of the DWs, i.e., especially inside the small domain, fluctuations can remain over longer periods of time and do not vanish as quickly as inside the bulk.
            Over few tens of picoseconds the polarization pattern in the small negative domain changes and finally the small reversed domain collapses.
            We discuss the detailed sequence of collapse and varying scenarios in the following.
            At this point we should note that the time scales observed with both our models agree well with each other.
            However, we do not have experimental data that show how these time scales relate to the physically observed dynamics, yet.
            For the moment being, the focus is more on the discovered trends and mechanisms which are independent of e.g. the frequency of the fluctuations.
            
            \paragraph{Evolution of reversed domain and walls with time}

            \begin{figure*}[htbp] 
                \centering
                \begin{tikzpicture}
                    %
                    \begin{axis} [
                        name=totalupratio,
                        width=6cm,
                        anchor=west,
                        xlabel=time (\si{\pico\second}),
                        ylabel=fraction of positive \(P_z\),
                        xmin = 0,
                        xmax = 96,
                        ymin=0.89,
                        ymax=1.01,
                        ylabel shift = -4pt,
                        legend pos= south east,
                        legend style={cells={anchor=west}},
                        clip mode=individual,
                        ]
                        \addplot [black, mark=o, only marks] table [x=time, y=ratio_up] {data/pos_P_ratio_4uc_336K.dat};
                        \addlegendentry{cumulated};
                        \addplot [black, no marks, dashed, thick, forget plot] coordinates {(15,0) (15,2)}; 
                        \addplot [black, no marks, dashed, thick, forget plot] coordinates {(60,0) (60,2)}; 
                        \addplot [red, no marks, dotted, very thick, forget plot] coordinates {(7.2,0) (7.2,2)};
                        \draw [thick] (axis cs: 0.0, 1.02) -- (axis cs: 15, 1.02) node [pos = 0.4, above] {\small \( \Delta t_{\rm onset}\)};
                        \draw [thick] (axis cs: 15, 1.015) -- (axis cs: 60, 1.015) node [pos = 0.6, above] {\small \( \Delta t_{\rm switching}\)};
                    \end{axis}
                    %
                    \node [anchor=west] (feramlayer) at ($(totalupratio.east) + (0.25cm,0cm)$) {\includegraphics[width=4.3cm]{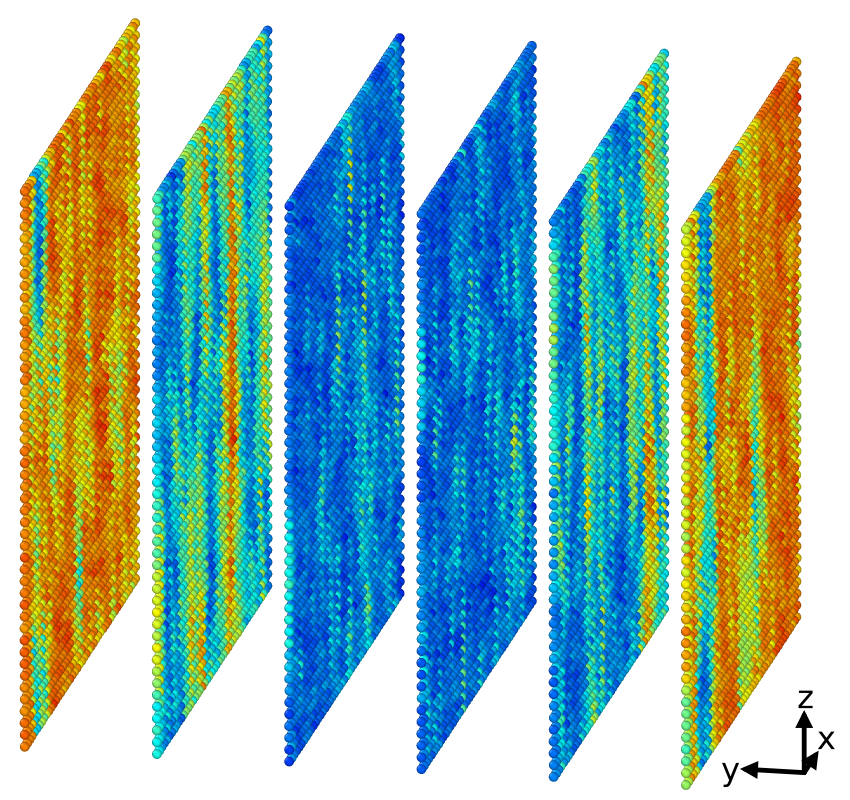}};
                    \node [anchor=north] (L0) at ($(feramlayer.south) + (-1.9,0.1)$) {\small \color{black!60}{L0}};
                    \node [anchor=north] (L1) at ($(feramlayer.south) + (-1.2,0.1)$) {\small \color{tud1c}{L1}};
                    \node [anchor=north] (L2) at ($(feramlayer.south) + (-0.5,0.1)$) {\small \color{tud4c}{L2}};
                    \node [anchor=north] (L3) at ($(feramlayer.south) + (0.2,0.1)$) {\small \color{tud7c}{L3}};
                    \node [anchor=north] (L4) at ($(feramlayer.south) + (0.9,0.1)$) {\small \color{tud9c}{L4}};
                    \node [anchor=north] (L5) at ($(feramlayer.south) + (1.6,0.1)$) {\small \color{black!60}{L5}};
                    \draw [dashed] ($(L0.north)!0.5!(L1.north)$) -- ($(L0.south)!0.5!(L1.south)$);
                    \draw [dashed] ($(L4.north)!0.5!(L5.north)$) -- ($(L4.south)!0.5!(L5.south)$);
                    %
                    \begin{axis}[
                        name=layers6b,
                        width=6cm,
                        at={($(feramlayer.east)+(0.25cm, 0cm)$)},
                        anchor = west,
                        set layers,
                        mark layer=axis background,
                        ylabel near ticks,
                        yticklabel pos=right,
                        xlabel={time (\si{\pico\second})},
                        ylabel={inter-layer correlation},
                        ylabel shift = -3pt,
                        xmin = 0,
                        xmax = 96,
                        ymin=-0.1,
                        ymax=1.1,
                        ytick={0,1},
                        legend pos=north east,
                        legend style={cells={anchor=west}},
                        ]
                        \addplot [tud4c, mark=square*, only marks, mark size=1.5] table [x=time_in_ps, y=corr_1_2] {data/correlation_atomistic_4L_336K.dat};
                        \addlegendentry{L1-L2};
                        \addplot [tud7c, mark=triangle*, only marks, mark size=2] table [x=time_in_ps, y=corr_1_4] {data/correlation_atomistic_4L_336K.dat};
                        \addlegendentry{L1-L4};
                        \addplot [black, no marks, dashed, thick, forget plot] coordinates {(15,-1) (15,2)};
                        \addplot [black, no marks, dashed, thick, forget plot] coordinates {(60,-1) (60,2)};
                        \addplot [black, no marks, forget plot, opacity=0.3] coordinates {(0,0) (96,0)};
                        \addplot [black, no marks, forget plot, opacity=0.3] coordinates {(0,1) (96,1)};
                    \end{axis}
                    \node [anchor=north] (lableA) at ($(totalupratio.north west) + (-1.2,1)$) {\small a)};
                    \node [] (lableB) at ($(lableA) + (6.25,0)$) {\small b)};
                    \node [] (lableC) at ($(lableA) + (10.7,0)$) {\small c)};
                    %
                    \draw[draw=black, fill=tud1c, opacity=0.5] (-1.2, -3) rectangle ++(0.7,-5.1);
                    \draw[draw=black, fill=tud1c, opacity=0.3] (-0.5, -3) rectangle ++(15.5,-5.1);
                    \node[inner sep=3pt, rounded corners=0.1cm, fill={black!20}, opacity=0.9, text opacity=1, rotate=90] (CSlabel) at (-0.85,-5.55) {atomistic model};
                    %
                    \begin{axis}[
                        name=scenario4L,
                        width=6.25cm,
                        anchor = north,
                        at={($(feramlayer.south)+(0.8cm, -1cm)$)},
                        axis background/.style={fill=white},
                        xlabel=time (\si{\pico\second}),
                        xtick={0,20,40,60},
                        yticklabels={},
                        xmin=0,
                        xmax=82,
                        ymin=-1.15,
                        ymax=1.15,
                        legend pos=south east,
                        legend style={cells={anchor=west}, font=\small},
                        label style={font=\small},
                        tick label style={font=\small}
                        ]
                        \addplot [tud1c, mark=*, only marks, mark size=1.5,] table [x=time_in_ps, y=avg_L1] {data/layerwise_scenarios_CS_4uc.dat};
                        \addlegendentry{L1};
                        \addplot [tud4c, mark=square*, only marks, mark size=1.5,] table [x=time_in_ps, y=avg_L2] {data/layerwise_scenarios_CS_4uc.dat};
                        \addlegendentry{L2};
                        \addplot [tud4c, mark=triangle, only marks, mark size=2.25,] table [x=time_in_ps, y=avg_L3] {data/layerwise_scenarios_CS_4uc.dat};
                        \addlegendentry{L3};
                        \addplot [tud1c, mark=diamond, only marks, mark size=2.25,] table [x=time_in_ps, y=avg_L4] {data/layerwise_scenarios_CS_4uc.dat};
                        \addlegendentry{L4};
                        \node () at (18, 0.95) {e) 4 layers};
                    \end{axis}
                    %
                    \begin{axis}[
                        name=scenario5L,
                        width=6.25cm,
                        anchor = east,
                        at={(scenario4L.west)},
                        axis background/.style={fill=white},
                        xmin=0,
                        xmax=126,
                        ymin=-1.15,
                        ymax=1.15,
                        xlabel=time (\si{\pico\second}),
                        ylabel=$P_z / P_s$,
                        ylabel shift = -7 pt,
                        legend pos=south east,
                        legend style={cells={anchor=west}, font=\small},
                        label style={font=\small},
                        tick label style={font=\small} 
                        ]
                        \addplot [tud1c, mark=*, only marks, mark size=1.5] table [x=time_in_ps, y=avg_L1] {data/layerwise_scenarios_CS_5uc.dat};
                        \addlegendentry{L1};
                        \addplot [tud4c, mark=square*, only marks, mark size=1.5] table [x=time_in_ps, y=avg_L2] {data/layerwise_scenarios_CS_5uc.dat};
                        \addlegendentry{L2};
                        \addplot [tud7c, mark=triangle*, only marks, mark size=2.25] table [x=time_in_ps, y=avg_L3] {data/layerwise_scenarios_CS_5uc.dat};
                        \addlegendentry{L3};
                        \addplot [tud4c, mark=diamond, only marks, mark size=2.25] table [x=time_in_ps, y=avg_L4] {data/layerwise_scenarios_CS_5uc.dat};
                        \addlegendentry{L4};
                        \addplot [tud1c, mark=pentagon, only marks, mark size=2.25] table [x=time_in_ps, y=avg_L5] {data/layerwise_scenarios_CS_5uc.dat};
                        \addlegendentry{L5};
                        \node () at (28, 0.95) {d) 5 layers};
                    \end{axis}
                    %
                    \begin{axis}[
                        name=scenario3L,
                        width=6.25cm,
                        anchor =  west,
                        at={(scenario4L.east)},
                        axis background/.style={fill=white},
                        xlabel=time (\si{\pico\second}),
                        yticklabels={},
                        xmin=0,
                        xmax=62,
                        ymin=-1.15,
                        ymax=1.15,
                        legend pos=south east,
                        legend style={cells={anchor=west}, font=\small},
                        label style={font=\small},
                        tick label style={font=\small} 
                        ]
                        \addplot [tud1c, mark=*, only marks, mark size=1.5, ] table [x=time_in_ps, y=avg_L1] {data/layerwise_scenarios_CS_3uc.dat};
                        \addlegendentry{L1};
                        \addplot [tud4c, mark=square*, only marks, mark size=1.5,] table [x=time_in_ps, y=avg_L2] {data/layerwise_scenarios_CS_3uc.dat};
                        \addlegendentry{L2};
                        \addplot [tud1c, mark=triangle, only marks, mark size=2.25,] table [x=time_in_ps, y=avg_L3] {data/layerwise_scenarios_CS_3uc.dat};
                        \addlegendentry{L3};
                        \node () at (14, 0.95) {f) 3 layers};
                    \end{axis}
                \end{tikzpicture}
                \caption{
                    \label{fig:explanation_layerwise_P}
                    Collapse of nano-sized domains with width \( d= \qty{4}{\uc}\) at \( \Delta T = \qty{-13}{\kelvin}\).
                    (a) Representative course of the fraction of positively polarized unit cells in the full simulation cell  over time.
                    Switching starts at \qty{15}{\pico\second} and is complete at \qty{60}{\pico\second}.
                    (b) Snapshots of the layers L1 - L4 inside the small domain and its adjacent layers L0 \& L5 at \qty{7.2}{\pico\second}.
                    (c) Correlation between layers inside the small domain of (b).
                    (d) - (f) Different scenarios of collapsing domains.
                }
            \end{figure*}

            Starting from perfectly flat DWs, over time, the large fluctuations on the wall may result in permanent changes of the DW structure, see \Cref{fig:polarization_switching_series}.
            After some time, a bridging segment of positive polarization penetrates the reversed domain and further grows until the system reaches the favorable single-domain state.  
            It is easier to see the microscopic details of the process if only the negative polarization vectors are shown, see \Cref{fig:polarization_switching_series} (b).
            Because the fluctuations appear in a needle-like configuration and polarization flipping occurs column-wise (discussed below) it is convenient to change to the top view, see \Cref{fig:polarization_switching_series} (c).
        
            All these representations show that the largest changes of the local dipoles occur in the smaller reversed domain and in the direct vicinity of the DWs.
            \Cref{fig:explanation_layerwise_P} (b) shows slices from a simulation snapshot that only displays the relevant layers.

            One might expect that there is a \textit{cross-talk} between DWs when they are very closely spaced.
            However, already for domain size between \qty{3}{\uc} to \qty{6}{\uc} we find that opposite DWs are uncorrelated.
            In \Cref{fig:explanation_layerwise_P} (c) the layer-wise correlation between adjacent layers (L1-L2) and layers at opposite DWs (L1-L4) shows that only immediately adjacent layers are correlated slightly at the beginning.
            As soon as a bridging segment forms, correlation of opposite DWs is evident.
            We, therefore, conclude that the onset of the domain collapse is a process dominated by random fluctuations that occur independently throughout the sample due to thermal excitation.
    
            The fact that the collapse of the domain as well as the modification of the walls with time are governed by fluctuations shows their stochastic nature.
            Therefore, the time-evolution for varying thickness of the small domain at different temperatures have been recorded multiple times within the coarse-grained framework.
            In order to keep the physical simulation time affordable, we only considered cases which collapse within \qty{200}{\pico\second}.
            
            \begin{figure*}[htbp] 
                \centering
                \begin{tikzpicture}
                    \draw[draw=black, fill=tud7c, opacity=0.5] (0.0, 0.0) rectangle ++(0.7,-9.5);
                    \draw[draw=black, fill=tud7c, opacity=0.3] (0.7, 0.0) rectangle ++(15.5,-9.5);
                    \node[inner sep=3pt, rounded corners=0.1cm, fill={black!20}, opacity=0.9, text opacity=1, rotate=90] (hefflabel) at (0.35,-4.75) {\heff -model};
                    \coordinate (posA) at (2.1, -0.2);
                    \coordinate (posB) at (2.1, -9.5);
                    \begin{axis}[
                        name=scenarioA,
                        width=6.2cm,
                        anchor = north west,
                        at={(posA)},
                        axis background/.style={fill=white},
                        ymin=-1.15,
                        ymax=1.15,
                        ylabel=$P_z / P_s$,
                        legend pos=north west,
                        legend style={cells={anchor=west}, font=\small},
                        label style={font=\small},
                        tick label style={font=\small} 
                        ]
                        \addplot [tud1c, mark=*, only marks, mark size=1.5] table [x=time_in_ps, y=layer1] {data/feram_layerwise_polarization_scenario_A.dat};
                        \addlegendentry{L1};
                        \addplot [tud4c, mark=square*, only marks, mark size=1.5] table [x=time_in_ps, y=layer2] {data/feram_layerwise_polarization_scenario_A.dat};
                        \addlegendentry{L2};
                        \addplot [tud4c, mark=triangle, only marks, mark size=2.25] table [x=time_in_ps, y=layer3] {data/feram_layerwise_polarization_scenario_A.dat};
                        \addlegendentry{L3};
                        \addplot [tud1c, mark=diamond, only marks, mark size=2.25] table [x=time_in_ps, y=layer4] {data/feram_layerwise_polarization_scenario_A.dat};
                        \addlegendentry{L4};
                        \node () at (160, 0.95) {a) 4 layers};
                    \end{axis}
                    \begin{axis}[
                        name=scenarioB,
                        width=6.2cm,
                        anchor = west,
                        at={(scenarioA.east)},
                        axis background/.style={fill=white},
                        yticklabels={},
                        ymin=-1.15,
                        ymax=1.15,
                        legend pos=north west,
                        legend style={cells={anchor=west}, font=\small},
                        label style={font=\small},
                        tick label style={font=\small} 
                        ]
                        \addplot [tud1c, mark=*, only marks, mark size=1.5, forget plot] table [x=time_in_ps, y=layer1] {data/feram_layerwise_polarization_scenario_B.dat};
                        \addplot [tud4c, mark=square*, only marks, mark size=1.5, forget plot] table [x=time_in_ps, y=layer2] {data/feram_layerwise_polarization_scenario_B.dat};
                        \addplot [tud4c, mark=triangle, only marks, mark size=2.25, forget plot] table [x=time_in_ps, y=layer3] {data/feram_layerwise_polarization_scenario_B.dat};
                        \addplot [tud1c, mark=diamond, only marks, mark size=2.25, forget plot] table [x=time_in_ps, y=layer4] {data/feram_layerwise_polarization_scenario_B.dat};
                        \node () at (40, 0.95) {b) 4 layers};
                    \end{axis}
                    \begin{axis}[
                        name=scenarioC,
                        width=6.2cm,
                        anchor =  west,
                        at={(scenarioB.east)},
                        axis background/.style={fill=white},
                        yticklabels={},
                        ymin=-1.15,
                        ymax=1.15,
                        legend pos=north west,
                        legend style={cells={anchor=west}, font=\small},
                        label style={font=\small},
                        tick label style={font=\small} 
                        ]
                        \addplot [tud1c, mark=*, only marks, mark size=1.5, forget plot] table [x=time_in_ps, y=layer1] {data/feram_layerwise_polarization_scenario_C.dat};
                        \addplot [tud4c, mark=square*, only marks, mark size=1.5, forget plot] table [x=time_in_ps, y=layer2] {data/feram_layerwise_polarization_scenario_C.dat};
                        \addplot [tud4c, mark=triangle, only marks, mark size=2.25, forget plot] table [x=time_in_ps, y=layer3] {data/feram_layerwise_polarization_scenario_C.dat};
                        \addplot [tud1c, mark=diamond, only marks, mark size=2.25, forget plot] table [x=time_in_ps, y=layer4] {data/feram_layerwise_polarization_scenario_C.dat};
                        \node () at (15, 0.95) {c) 4 layers};
                    \end{axis}
                    \begin{axis}[
                        name=scenarioD,
                        width=6.2cm,
                        anchor = north,
                        at={($(scenarioA.south) + (0,-0.6cm)$)},
                        axis background/.style={fill=white},
                        ymin=-1.15,
                        ymax=1.15,
                        xlabel=time in ps,
                        ylabel=$P_z / P_s$,
                        legend pos=south east,
                        legend style={cells={anchor=west}, font=\small},
                        label style={font=\small},
                        tick label style={font=\small} 
                        ]
                        \addplot [tud1c, mark=*, only marks, mark size=1.5, forget plot] table [x=time_in_ps, y=layer1] {data/feram_layerwise_polarization_scenario_F.dat};
                        \addplot [tud4c, mark=square*, only marks, mark size=1.5, forget plot] table [x=time_in_ps, y=layer2] {data/feram_layerwise_polarization_scenario_F.dat};
                        \addplot [tud7c, mark=triangle*, only marks, mark size=2.25, forget plot] table [x=time_in_ps, y=layer3] {data/feram_layerwise_polarization_scenario_F.dat};
                        \addplot [tud7c, mark=diamond, only marks, mark size=2.25, forget plot] table [x=time_in_ps, y=layer4] {data/feram_layerwise_polarization_scenario_F.dat};
                        \addplot [tud4c, mark=pentagon, only marks, mark size=2.25] table [x=time_in_ps, y=layer5] {data/feram_layerwise_polarization_scenario_F.dat};
                        \addplot [tud1c, mark=x, only marks, mark size=2.25, forget plot] table [x=time_in_ps, y=layer6] {data/feram_layerwise_polarization_scenario_F.dat};
                        \node () at (15, 0.95) {d) 6 layers};
                    \end{axis}
                    \begin{axis}[
                        name=scenarioE,
                        width=6.2cm,
                        anchor = west,
                        at={(scenarioD.east)},
                        axis background/.style={fill=white},
                        xlabel=time in ps,
                        yticklabels={},
                        ymin=-1.15,
                        ymax=1.15,
                        legend pos=north west,
                        legend style={cells={anchor=west}, font=\small},
                        label style={font=\small},
                        tick label style={font=\small}
                        ]
                        \addplot [tud1c, mark=*, only marks, mark size=1.5, forget plot] table [x=time_in_ps, y=layer1] {data/feram_layerwise_polarization_scenario_E.dat};
                        \addplot [tud4c, mark=square*, only marks, mark size=1.5, forget plot] table [x=time_in_ps, y=layer2] {data/feram_layerwise_polarization_scenario_E.dat};
                        \addplot [tud7c, mark=triangle*, only marks, mark size=2.25, forget plot] table [x=time_in_ps, y=layer3] {data/feram_layerwise_polarization_scenario_E.dat};
                        \addplot [tud7c, mark=diamond, only marks, mark size=2.25, forget plot] table [x=time_in_ps, y=layer4] {data/feram_layerwise_polarization_scenario_E.dat};
                        \addplot [tud4c, mark=pentagon, only marks, mark size=2.25, forget plot] table [x=time_in_ps, y=layer5] {data/feram_layerwise_polarization_scenario_E.dat};
                        \addplot [tud1c, mark=x, only marks, mark size=2.25, forget plot] table [x=time_in_ps, y=layer6] {data/feram_layerwise_polarization_scenario_E.dat};
                        \node () at (40, 0.95) {e) 6 layers};
                    \end{axis}
                    \begin{axis}[
                        name=scenarioF,
                        width=6.2cm,
                        anchor =  west,
                        at={(scenarioE.east)},
                        axis background/.style={fill=white},
                        xlabel=time in ps,
                        yticklabels={},
                        ymin=-1.15,
                        ymax=1.15,
                        legend pos=south east,
                        legend style={cells={anchor=west}, font=\small},
                        label style={font=\small},
                        tick label style={font=\small} 
                        ]
                        \addplot [tud1c, mark=*, only marks, mark size=1.5] table [x=time_in_ps, y=layer1] {data/feram_layerwise_polarization_scenario_D.dat};
                        \addlegendentry{L1};
                        \addplot [tud4c, mark=square*, only marks, mark size=1.5] table [x=time_in_ps, y=layer2] {data/feram_layerwise_polarization_scenario_D.dat};
                        \addlegendentry{L2};
                        \addplot [tud7c, mark=triangle*, only marks, mark size=2.25] table [x=time_in_ps, y=layer3] {data/feram_layerwise_polarization_scenario_D.dat};
                        \addlegendentry{L3};
                        \addplot [tud7c, mark=diamond, only marks, mark size=2.25] table [x=time_in_ps, y=layer4] {data/feram_layerwise_polarization_scenario_D.dat};
                        \addlegendentry{L4};
                        \addplot [tud4c, mark=pentagon, only marks, mark size=2.25] table [x=time_in_ps, y=layer5] {data/feram_layerwise_polarization_scenario_D.dat};
                        \addlegendentry{L5};
                        \addplot [tud1c, mark=x, only marks, mark size=2.25] table [x=time_in_ps, y=layer6] {data/feram_layerwise_polarization_scenario_D.dat};
                        \addlegendentry{L6};
                        \node () at (15, 0.95) {f) 6 layers};
                    \end{axis}
                \end{tikzpicture}
                \caption{
                Collection of representative scenarios of the time-evolution of polarization in the reversed domain.
                The  normalized polarization per layer \(P_z\)/ \(P_s\) is shown for the \heff -model at \( \Delta T = \qty{-26}{\kelvin} \).
                }
                \label{fig:Pz_over_time}
            \end{figure*}
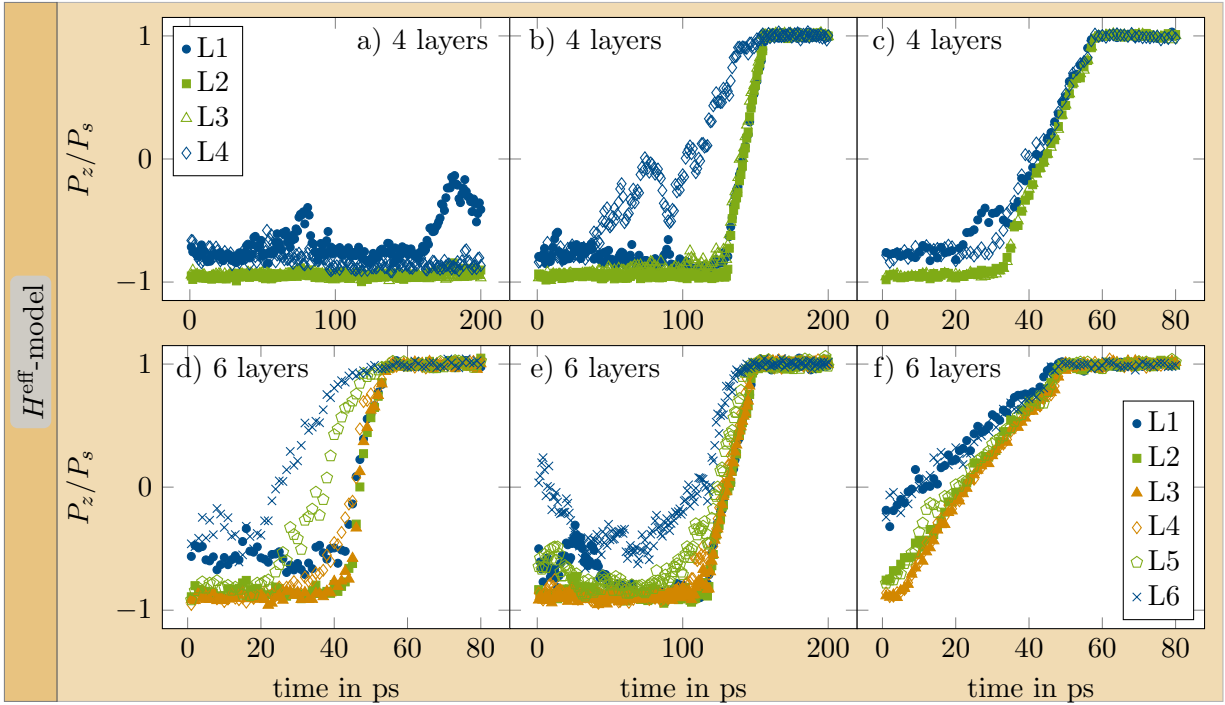

            Different scenarios can be distinguished in the time-evolution of the layer-wise cumulative polarization see \Cref{fig:explanation_layerwise_P}~(d)-(f) and \Cref{fig:Pz_over_time}:
            
            \begin{enumerate}
                \item At low temperature, fluctuations are small and for thick domains the positions and shapes of domain walls are mainly conserved (not shown). 
                \item With increasing temperature, the fluctuations on the DWs increase and there is an increasing probability for temporal and permanent modifications of the interface.
                \begin{enumerate}
                    \item The width of the DWs increases systematically with temperature.
                    \item Random thermal fluctuations result in back and forth bulging of walls as well as fluctuation of clusters on different scales on the wall, see e.g. \Cref{fig:explanation_layerwise_P}~(e) and \Cref{fig:Pz_over_time}~(e) in the first few picoseconds.
                    \item There is a small probability for the spontaneous shift of the domain wall by one unit cell by the 2D growth of the 2D nuclei, see \Cref{fig:Pz_over_time}~(d). 
                \end{enumerate}
                \item Bridging segments may form between both walls resulting in the collapse of the thinner domain. The probability for this collapse increases with temperature.
                However, also the bridging element is subject to thermal fluctuations and may be dissolved again, especially if it is only a few unit cells large.
                Bridging elements initially have a small extent along \(x\) but can quickly expand.
                    
                \item The probability for the collapse furthermore scales inversely with the domain width.
                \begin{enumerate}
                    \item For domain sizes of 3 - 6\,u.c. we find such a collapse in the tested temperature regime. Their statistical appearance and energetics are discussed in detail below.
                    \item For 1 or 2 reversed layers, both DWs are next to each other and, thus, the DWs are correlated from the very start.
                    Therefore, there is no nucleation process and we do not discuss them further.
                \end{enumerate}
            \end{enumerate}
            
            What we refer to as (cumulative) layer-wise polarization here is called "planar-averaged" polarization by \textcite{Kumar2010}.
            It was already recognized by \textcite{Miller1960} that the homogeneous nucleation of a new domain far from a DW is less likely than nucleation at the DW (or any other defect, for that matter).
            As a result the switching starts at the interfaces \cite{Shin2007}.

            Thus, the collapse of a small domain starts by the formation of a needle-like cluster at the DW.
            With time the switched clusters frequently grow along $x$, i.e., within the DW plane, resulting in an increased polarization within the layers at the interface.
            For example, refer to \Cref{fig:Pz_over_time}~(c) where the polarization is increased at the DW.
            For many of the samples we observe this enhanced polarization from the very beginning of the simulations.

            If the reversed cluster at the interface grows along $y$ (perpendicular to the DW) a bridging segment of positive polarization may penetrate the reversed domain.
            Once such bridging segment has formed any further expansion of it immediately reduces the DW area.
            Thus, the system now \textit{sees} the driving force leading to the energy minimum.
            Then the polarization starts to rapidly grow in all layers simultaneously, see e.g. \Cref{fig:Pz_over_time} (c) or (e).
            At this point the collapse of the small domain is inevitable.
            The collapse occurs by an expansion of the bridging segment to the sides (along \(\pm x\)) in a continuous manner, see \Cref{fig:polarization_switching_series}~(c) and (d).
            For more details see for more details see Supplemental Material 5 at \_ .
            
            During its collapse the polarization in the center of the reversed domain increases approximately linearly with time, see \Cref{fig:Pz_over_time}.
            Yet, the slope of \(P_z(t)\) may differ among the layers as polarization may already be enhanced before, see \Cref{fig:Pz_over_time}~(d) and (e).
            However, we do not find an obvious threshold or critical value of \(P_z\) beyond which the complete transition occurs.
            See for example \Cref{fig:Pz_over_time} (a), where almost half of a layer has changed polarization before the system returns to its initial state of a sharp DW.
            In contrast, \Cref{fig:Pz_over_time} (c)  shows only a small cluster in one layer before domain collapse commences.
            Consequently, we first embrace a statistical description and, second, view the problem from an ideal mechanistic viewpoint below.

\newpage
\clearpage

        \subsubsection{Trends in onset time}
        \label{subsec:statistical_description}

            For describing the probability of domain collapse, it is convenient to define two characteristic times, see \Cref{fig:explanation_layerwise_P}.
            First, the time until the onset of polarization switching in the central layers $\Delta t_{\text{onset}}$, i.e., the formation of a stable bridging segment connecting both DWs.
            And second, the switching time $\Delta t_{\text{switching}}$, i.e., the transient region with finite \(\frac{\partial P_z}{\partial t}\).

            To evaluate the stochastic process of domain collapse we simulate a larger number of samples and track their \(\Delta t_{\text{onset}}\) with temperature and varying initial domain thickness.
            However, for the reason of high computational demand, we run only 10 samples for \qty{3}{\uc} using the atomistic model and confirm the congruence of both models.
            More statics are obtained using the \heff -model for domain thicknesses of \(4 - \qty{6}{\uc}\) where we calculate hundreds of samples.
            Unfortunately, collecting enough statistics is a computationally demanding task and only few data points could be collected with sufficient accuracy.
            For each combination of temperature and thickness where more than \qty{50}{\percent} of the investigated samples collapsed we could calculate the median values displayed in \Cref{fig:characteristic_time_heff}.
            The fraction of samples where the small reversed domains did not collapse is also indicated.
            
            \newcommand{\barwidth}{0.22}
            \begin{figure}[htbp] 
                \centering
                \begin{tikzpicture}[/pgfplots/tick scale binop=\times]
                    \begin{axis} [
                        height=0.8*\axisdefaultheight,
                        width=\axisdefaultwidth,
                        name=fractions,
                        xlabel={domain size \(d\) (\si{\uc})},
                        ylabel={collapsed samples (\si{\percent})},
                        xmin=3.4,
                        xmax=6.6,
                        ymin=0,
                        ymax=100,
                        xticklabels={,,},
                        extra x ticks={4,5,6},
                        extra x tick labels={\qty{4}{\uc}, \qty{5}{\uc}, \qty{6}{\uc}},
                        xtick style={draw=none},
                        ]
                        \addplot [black, dashed] coordinates {
                            (3, 50)
                            (7, 50)
                        };
                        \filldraw [draw=black, fill=tud1c!50] (4-3/2*\barwidth,0) rectangle (4-1/2*\barwidth, 60); 
                        \node [rotate=90, anchor=west] at (4-\barwidth, 1) {\footnotesize \qty{-40}{\kelvin}};
                        \filldraw [draw=black, fill=tud1c!70] (4-1/2*\barwidth,0) rectangle (4+1/2*\barwidth, 90); 
                        \node [rotate=90, anchor=west] at (4, 1) {\footnotesize \qty{-35}{\kelvin}};
                        \filldraw [draw=black, fill=tud1c!90] (4+1/2*\barwidth,0) rectangle (4+3/2*\barwidth, 100); 
                        \node [rotate=90, anchor=west] at (4+\barwidth, 1) {\color{white} \footnotesize \qty{-30}{\kelvin}};
                        \filldraw [draw=black, fill=tud4c!30] (5-2*\barwidth,0) rectangle (5-\barwidth, 39); 
                        \node [rotate=90, anchor=west] at (5-3/2*\barwidth, 1) {\footnotesize \qty{-35}{\kelvin}};
                        \filldraw [draw=black, fill=tud4c!50] (5-\barwidth,0) rectangle (5, 57); 
                        \node [rotate=90, anchor=west] at (5-1/2*\barwidth, 1) {\footnotesize \qty{-30}{\kelvin}};
                        \filldraw [draw=black, fill=tud4c!70] (5,0) rectangle (5+\barwidth, 97); 
                        \node [rotate=90, anchor=west] at (5+1/2*\barwidth, 1) {\footnotesize \qty{-25}{\kelvin}};
                        \filldraw [draw=black, fill=tud4c!90] (5+\barwidth,0) rectangle (5+2*\barwidth, 100); 
                        \node [rotate=90, anchor=west] at (5+3/2*\barwidth, 1) {\color{white} \footnotesize \qty{-20}{\kelvin}};
                        \filldraw [draw=black, fill=tud7c!30] (6-2*\barwidth,0) rectangle (6-\barwidth, 54); 
                        \node [rotate=90, anchor=west] at (6-3/2*\barwidth, 1) {\footnotesize \qty{-30}{\kelvin}};
                        \filldraw [draw=black, fill=tud7c!50] (6-\barwidth,0) rectangle (6, 59); 
                        \node [rotate=90, anchor=west] at (6-1/2*\barwidth, 1) {\footnotesize \qty{-25}{\kelvin}};
                        \filldraw [draw=black, fill=tud7c!70] (6,0) rectangle (6+\barwidth, 100); 
                        \node [rotate=90, anchor=west] at (6+1/2*\barwidth, 1) {\footnotesize \qty{-20}{\kelvin}};
                        \filldraw [draw=black, fill=tud7c!90] (6+\barwidth,0) rectangle (6+2*\barwidth, 100); 
                        \node [rotate=90, anchor=west] at (6+3/2*\barwidth, 1) {\color{white} \footnotesize \qty{-15}{\kelvin}};
                    \end{axis}
                    \begin{axis} [
                        name=median,
                        set layers,
                        mark layer=axis background,
                        at={($(fractions.south)+(0cm,-1.4cm)$)},
                        anchor = north,
                        xlabel= {\(T - T_{T \rightarrow C}\) (\unit{\kelvin})}, 
                        ylabel= \( \Delta t_{\text{onset}}\) (\unit{\pico\second}),
                        xmin=-43,
                        xmax=-12,
                        ymin=0,
                        ymax=225,
                        legend pos=north east,
                        legend style={cells={anchor=west}},
                        ]
                        \addplot [tud1c, only marks, mark=o, mark options={scale = 1.0}, forget plot, opacity=0.3] table [x=temp, y=onset_time] {data/feram_gradual_removal_onset_time_4uc_all.dat};
                        \addplot [tud1c, mark=*, thick, mark options={solid, scale = 1.5}] table [x=temp, y=onset_time] {data/feram_gradual_removal_onset_time_4uc_median.dat};
                        \addlegendentry{4 uc};
                        5 uc
                        \addplot [tud4c, only marks, mark=square, mark options={scale = 1.0}, forget plot, opacity=0.3] table [x=temp, y=onset_time] {data/feram_gradual_removal_onset_time_5uc_all.dat};
                        \addplot [tud4c, mark=square*, mark options={solid, scale = 1.5}, thick] table [x=temp, y=onset_time] {data/feram_gradual_removal_onset_time_5uc_median.dat};
                        \addlegendentry{5 uc};
                        \addplot [tud7c, only marks, mark=triangle, mark options={scale = 1.0}, forget plot, opacity=0.3] table [x=temp, y=onset_time] {data/feram_gradual_removal_onset_time_6uc_all_upto_200ps.dat};
                        \addplot [tud7c, mark=triangle*, mark options={solid, scale = 2.0}, thick] table [x=temp, y=onset_time] {data/feram_gradual_removal_onset_time_6uc_median.dat};
                        \addlegendentry{6 uc};
                        \addplot [black, solid, no marks, forget plot] coordinates {
                            (-45, 200)
                            (-10, 200)
                        };
                        \addplot [tud4c, dashed, mark=square, mark options={solid, scale=1.5}, thick] coordinates {
                            (-30, 110.5833)
                            (-35, 200)
                            };
                        \draw [black, -stealth, thick] (-35, 208) -- (-35, 220);
                    \end{axis}
                    \node [] at ($(fractions.north west) + (-1.35,0.2)$) {a)};
                    \node [] at ($(median.north west) + (-1.35,0.2)$) {b)};
                \end{tikzpicture}
                \caption{
                    \label{fig:characteristic_time_heff}
                    Change of characteristic time \(\Delta t_{\text{onset}}\) with temperature for different initial width of the reversed domains.
                    Data is obtained with the \heff -model.
                    (a) Fraction of samples that actually showed domain collapse during the first \qty{200}{\pico\second}.
                    Median values for (b) can only be calculated when at least \qty{50}{\percent} of samples show a domain collapse.
                    (b) \(\Delta t_{\text{onset}}\) and its median for different \(d\) and \(T\).
                    All measurements are shown as small empty symbols, median values as full symbols.
                }
            \end{figure}
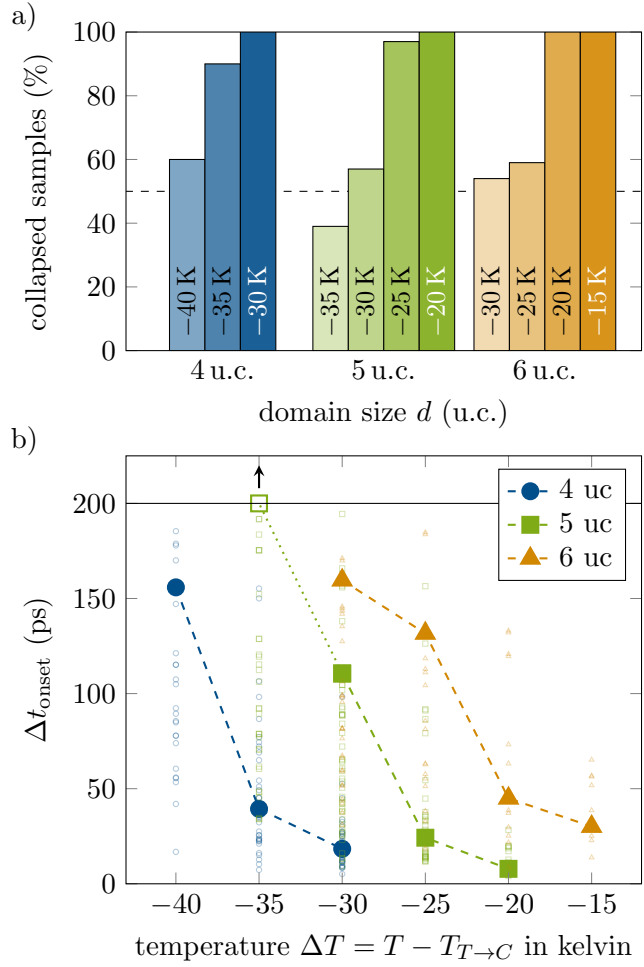

            Note that decreasing temperature generally leads to an increased driving force (due to higher DW energy density) but decreased DW mobility (less fluctuations).
            Still \(\Delta t_{\text{onset}}\) increases with decreasing temperature indicating that the collapse process is kinetically controlled.
            The trend of increasing \(\Delta t_{\text{onset}}\) with increasing domain thickness can be explained by the increased activation energy for the formation of larger bridging segments.
            
            With so few reliable data points fitting a model and extrapolating the data seems unfeasible.
            Originally, we had expected some kind of Arrhenius-type behavior because the formation of bridging elements does require activation energy.
            However, fitting a simple Arrhenius-type law does not describe the given data.
            In order to find the root to this behavior we next address the underlying energy landscape of the collapsing process.
            First, we devise a model picture based on first order interactions, and, second, we validate the approach with our simulations.

\newpage
\clearpage

        \subsubsection{Energy landscape of domain switching}
        
            \paragraph{Driving force}
            As domain collapse occurs spontaneously there must be a driving force for this process.
            Tracking the system's potential energy during the collapse of the small domain we indeed observe the reduction in energy of twice the energy of a single domain wall, see Supplemental Material 6 at \_ .
            However, the large thermal noise prevents the calculation of any activation barrier by simply tracking the system's potential energy.

            \paragraph{Steps of domain collapse}
            
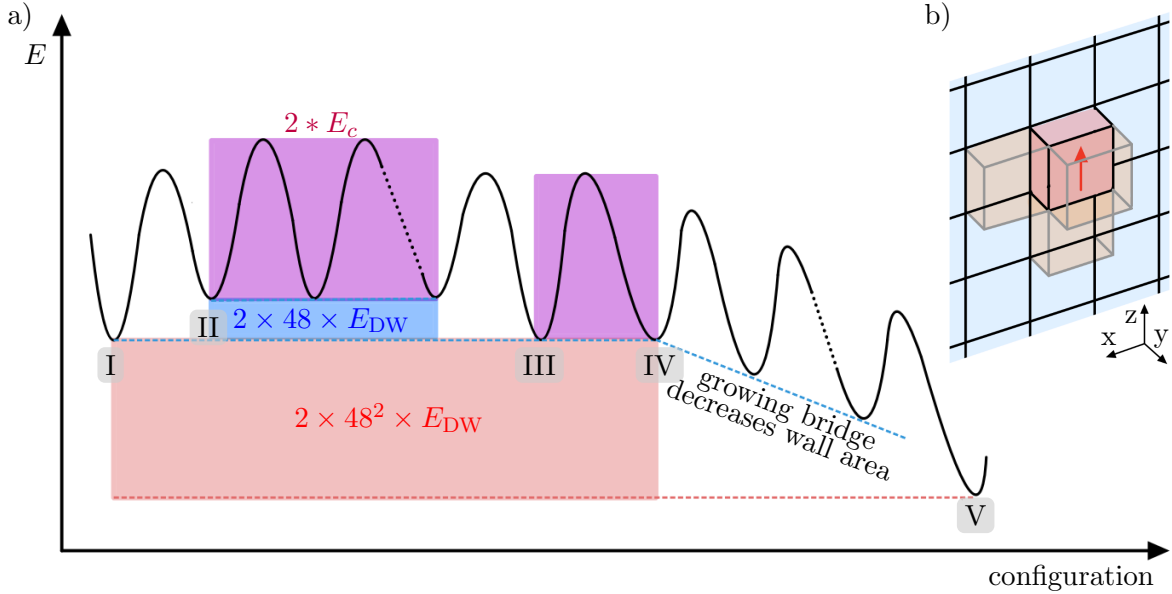
\begin{figure}[htb] 
                \centering
                \begin{tikzpicture}
                    \node [anchor=west] (schematic) at (0,0) {\includegraphics[width=15cm]{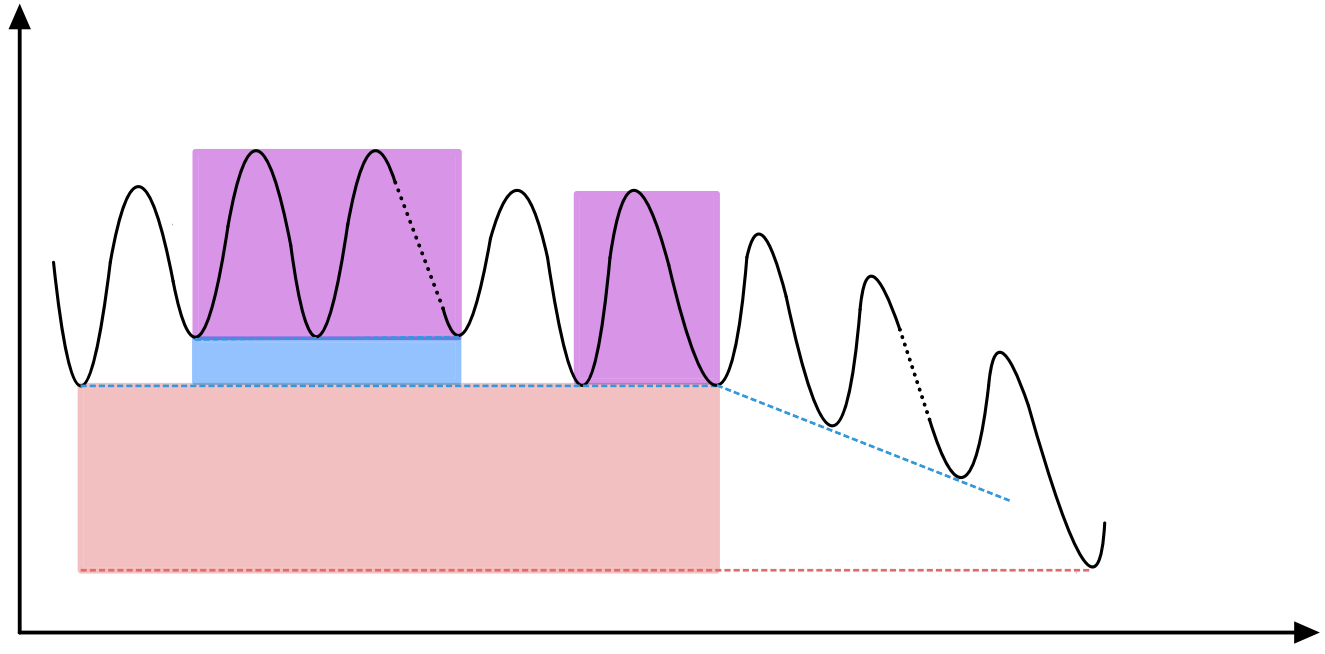}};
                    \node[] (E) at ($(schematic.north west)+(0cm, -0.7cm)$) {\(E\)};
                    \node[] (traj) at ($(schematic.south east)+(-1.5cm, 0cm)$) {configuration};
                    \node[] (Eflatdw) at ($(schematic.west)+(4.7cm, -1.7cm)$) { \color{red} \(2 \times 48^2 \times E_{\text{DW}}\)};
                    \node[] (Ecolumndw) at ($(schematic.west)+(3.8cm, -0.4cm)$) { \color{blue} \(2 \times 48 \times E_{\text{DW}}\)};
                    \node[] (Ec) at ($(schematic.west)+(3.8cm, 2.2cm)$) { \color{purple} $2*E_{c}$};
                    \node[] (labelc) at ($(schematic.north west)+(-0.2cm, -0.2cm)$) {a)};
                    \node[inner sep=3pt, rounded corners=0.1cm, fill={black!20}, opacity=0.6, text opacity=1,] (i) at (1,-1) {\RomanNumeralCaps{1}};
                    \node[inner sep=3pt, rounded corners=0.1cm, fill={black!20}, opacity=0.6, text opacity=1,] (ii) at (2.3,-0.5) {\RomanNumeralCaps{2}};
                    \node[inner sep=3pt, rounded corners=0.1cm, fill={black!20}, opacity=0.6, text opacity=1,] (iii) at (6.7,-1) {\RomanNumeralCaps{3}};
                    \node[inner sep=3pt, rounded corners=0.1cm, fill={black!20}, opacity=0.6, text opacity=1,] (iv) at (8.3,-1) {\RomanNumeralCaps{4}};
                    \node[inner sep=3pt, rounded corners=0.1cm, fill={black!20}, opacity=0.6, text opacity=1,] (v) at (12.5,-3) {\RomanNumeralCaps{5}};
                    \node[rotate=-21] (growing) at (10,-1.6) {\small growing bridge};
                    \node[rotate=-21] (growing) at (9.9,-1.85) {\small decreases wall area};
                    \node [anchor=north east] (directions) at (schematic.north east) {\includegraphics[width=3cm]{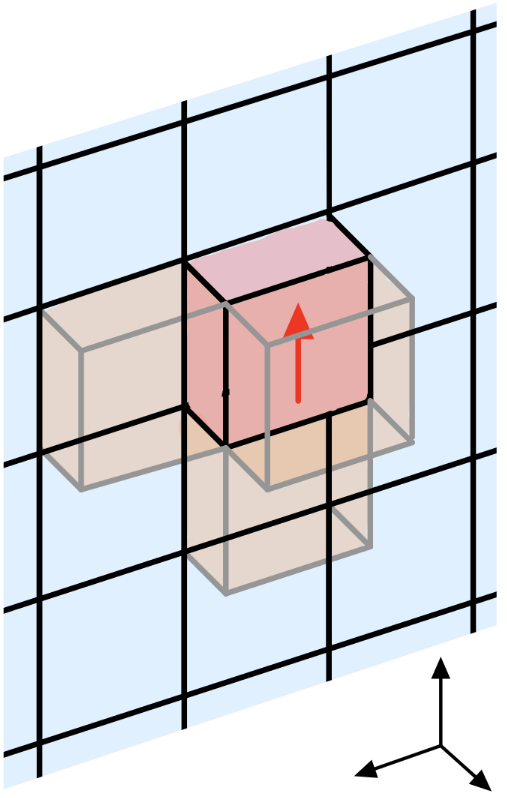}};
                    \node[] (xaxis) at ($(directions.south east) + (-1cm,0.5cm)$) {\small x};
                    \node[] (yaxis) at ($(directions.south east) + (-0.3cm,0.5cm)$) {\small y};
                    \node[] (zaxis) at ($(directions.south east) + (-0.7cm,0.8cm)$) {\small z};
                    \node[] (labeld) at ($(directions.north west)+(0cm, -0.2cm)$) {b)};
                \end{tikzpicture}
                \caption{
                    Simplified energy landscape of a possible trajectory for domain collapse only taking into account nearest neighbor interactions and fixed values of polarization per unit cell.
                    (a) 
                    Series of configurations during the collapse of a small domain with schematic representation of intermediate minima and maxima.
                    The energy penalty of the two flat DWs (state \RomanNumeralCaps{1}) shown in red amounts to \(2\times 48^2 \times E_{\text{DW}}\);
                    a complete row or column of switched dipoles at the wall (state \RomanNumeralCaps{2}) induces an energy penalty of \(2\times 48\times E_{\text{DW}}\) (blue);
                    charged interfaces induce the largest energy penalty of \( 2 \times E_c\) (purple).
                    State \RomanNumeralCaps{3} corresponds to the parallel shift of the complete DW.
                    Once a bridging segment between both walls has been formed (state \RomanNumeralCaps{4}), the DW area and the energy decrease with increasing size of the bridging segment until the single-domain state is obtained (state \RomanNumeralCaps{5}).
                    (b) This inset shows the definition of directions $x$, $y$, and $z$ and an initially reversed dipole on the wall (marked with red arrow). Possible directions for expansion of the nucleus are shaded in orange.
                }
                \label{fig:sketch_stepwise_switching}
            \end{figure}

            Since the collapse of the small domain happens during a finite time span, we try to single out the individual steps of switching and reveal the microscopic origin of the switching rates.
            Therefore, we first draw the schematic energy landscape of different configurations of local polarization vectors in \Cref{fig:sketch_stepwise_switching} (not to scale).
            For simplicity we assume constant energies per area \(E_i\) in the following.

            \Cref{fig:sketch_stepwise_switching}~(a) shows a possible transition path from a state with two domains and two flat T180 DWs (state \RomanNumeralCaps{1}) to a single domain state (state \RomanNumeralCaps{5}) including important intermediate minima. 
            We start with two flat T180 DW of area \qtyproduct{48 x 48}{\uc} each (state \RomanNumeralCaps{1}).
            Their energy is \(2\times 48^2 \times E_{\text{DW}}\).
            If a nucleus, like the one in \Cref{fig:sketch_stepwise_switching}~(b) forms on the DW, the energy for the additional DW area is small along $\pm x$ and $\pm y$ because this configuration resembles a charge neutral T180 DW with \(E_{\text{DW}}\).
            In contrast, the oppositely polarized regions meeting along  the two $\pm z$-surfaces of a cluster are in charged head-to-head or tail-to-tail configurations resulting in a much larger DW energy per unit area ($E_c$).
            The creation of these costly charged interfaces makes the major contribution to the maximum between state \RomanNumeralCaps{1} and state \RomanNumeralCaps{2}.
            Because the flipping of the first dipole carries the highest energy of all elemental switching steps it is expected to be rate limiting for the onset of switching.
            
            A single switched unit cell on the DW can in principle grow (i) along the polarization axis ($\pm z$-direction), (ii) in the DW plane orthogonal to the polarization axis ($\pm x$-direction), and (iii) further into the domain ($\pm y$-direction), see \Cref{fig:sketch_stepwise_switching}~(b).
            The growth mode (i) along \(z\) is lowest in energy, as the size of the charged walls is not modified, and the energy for each additional switched dipole increases by only \(2E_{\text{DW}}\).
            For completely switched columns this results in the small energy penalty of \(2 \times 48 \times E_{\text{DW}}\), see \Cref{fig:sketch_stepwise_switching}~(a).
            Because the completion of a switched column annihilates the charged interfaces, it greatly reduces energy and is likely to occur frequently.
            This leads to the second minimum at state \RomanNumeralCaps{2}.

            In growth mode (ii) the switching of single dipoles along $x$ induces an energy penalty of \(2E_c\) and in case (iii) the energy increases by \(2E_c+2E_{\text{DW}}\) which is highest in energy.
            
            After the switching of a full column, the energy for further switched columns on the DW, i.e., a growth of the nucleus along $\pm x$ by full \(z\)-columns, does not depend on the number of switched columns.
            Therefore, the next minimum to the right of state \RomanNumeralCaps{2} in \Cref{fig:sketch_stepwise_switching}~(a) is on the same energy level and corresponds to a neighboring column switched.
            As this process happens back and forth at finite temperature the DW could fluctuate by switching column-wise at no net energy cost.
            Consequently, many meta-stable states with full $z$-columns switched may exist in the trajectory.
            In case all \(z\)-columns in a plane are switched the DW is flat again and energy comes back to the original level (state \RomanNumeralCaps{3}).
            
            Only when two flat DW are so close that the switching of an additional column bridges the small reversed domain, the collapse of the reversed domain may start.
            From state \RomanNumeralCaps{3} to state \RomanNumeralCaps{4} the switching column bridges the gap between the DWs.
            DW area in that scenario is constant, therefore, it does not come with a net energy penalty.
            
            Each consecutive \(z\)-column that switches next to the first bridging column enlarges the bridging segment and reduces DW area, thus, lowering energy by \(2 \times 48 \times E_{\text{DW}}\).
            The lowest energy is finally achieved when all DWs have disappeared (state \RomanNumeralCaps{5}).

            \paragraph{Calculation of energy landscape}
            
            For the construction of the schematic energy landscape we regarded only next neighbor interactions.
            Because there are also energy contributions of higher order we now detail the energy landscape using static simulations for the switching of the first two columns.
            Yet, we restrict ourselves to treating each unit cells in our simulation as rigid without relaxation of atoms and dipole moments.
            Thus, we obtain upper bounds of the relevant energies.

            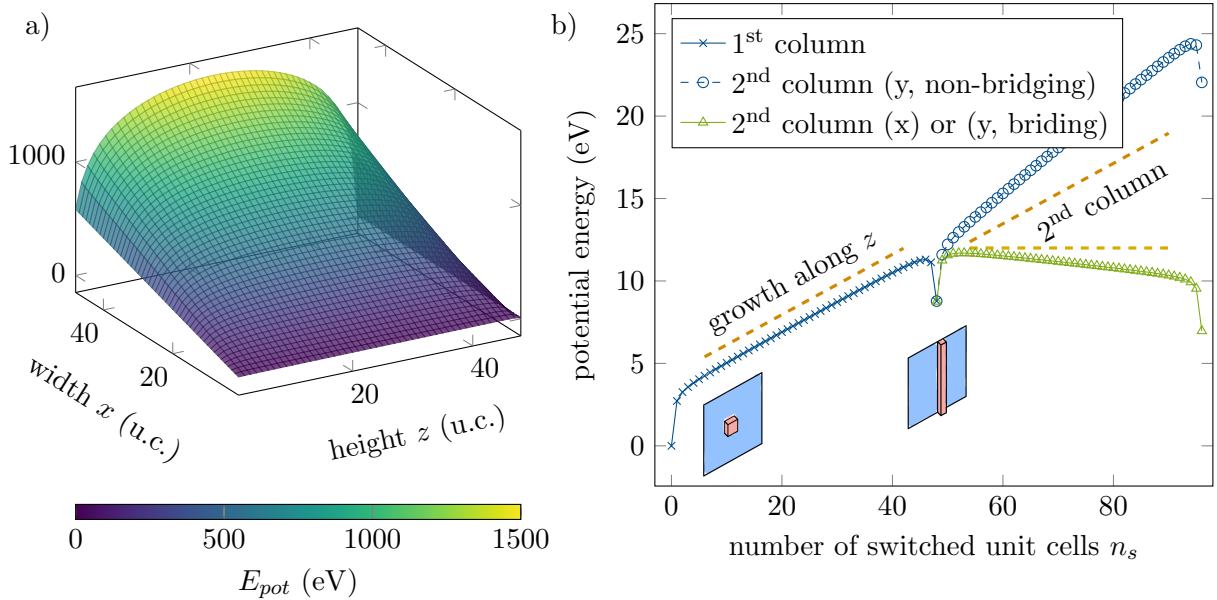
\begin{figure}[htb] 
                \centering
                \pgfplotsset{colormap/blackwhite}
                \begin{tikzpicture}
                    \coordinate (feramlandscape) at (0,0);
                    \begin{axis} [
                        width=7.5cm,
                        anchor= south west,
                        at=(feramlandscape),
                        xlabel=height $z$ in \unit{\uc},
                        ylabel=width $x$ in \unit{\uc},
                        xlabel style={sloped},
                        ylabel style={sloped},
                        colormap name=viridis, 
                        colorbar horizontal,
                        colorbar/width=2.0mm,
                        colorbar style={xlabel=$E_{pot}$ in \si{\electronvolt}},
                        /pgfplots/view={-30}{30}, 
                        point meta min=0,
                        point meta max=1500,
                        ]
                        \addplot3 [
                            surf,
                            mesh/ordering=y varies,
                            mesh/cols=48,
                            mesh/interior colormap name= viridis, 
                            colormap name = blackwhite,
                            opacity=0.75,
                            ] table [x=X, y=Y, z=Z] {feram_energy_landscape_table_scaled.dat};
                    \end{axis}
                    \node[] (labela) at (-0.5cm,4.9cm) {a)};
                    %
                    \coordinate (csenergies) at ($(feramlandscape.east)+(7.7cm, 2cm)$);
                    \begin{axis} [
                        width=9.0cm,
                        height=8.0cm, 
                        at=(csenergies),
                        anchor=west,
                        xmin=-3,
                        xmax=98,
                        xlabel=number of switched unit cells $n_s$,
                        ylabel=potential energy in \unit{\electronvolt},
                        legend pos = north west,
                        legend style={cells={anchor=west}},
                        ]
                        \addplot [tud1c, mark=x, mark size=2pt] table [x=size, y=poteng] {data/large_domain_single_column_static_pe.dat};
                        \addlegendentry{\small 1\textsuperscript{st} column};
                        \addplot [tud1c, mark=o, mark size=2pt, mark options={solid}, dashed] table [x=size, y=poteng] {data/large_domain_double_column_static_pe.dat};
                        \addlegendentry{\small 2\textsuperscript{nd} column (y, non-bridging)};
                        \addplot [tud4c, mark=triangle, mark size=2pt] table [x=size, y=poteng] {data/small_domain_double_column_static_pe.dat};
                        \addlegendentry{\small 2\textsuperscript{nd} column (x) or (y, briding)};
                        \node (plain) [anchor=south] at (11,-2.5) {\includegraphics[height=1.4cm]{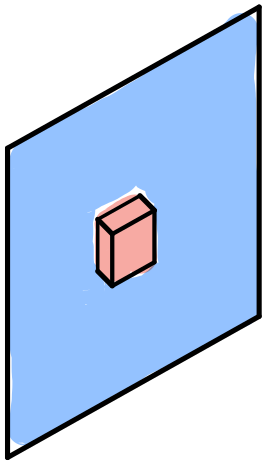}};
                        \node (plain) [anchor=north] at (48,8) {\includegraphics[height=1.4cm]{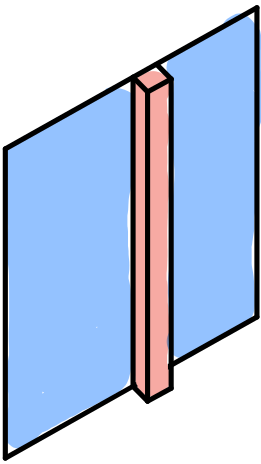}};
                        \addplot [tud7c, no marks, dashed, very thick, domain=6:42] {0.1829*x+4.3} node [pos = 0.5, sloped, above, black] {\small growth along \(z\)};
                        \addplot [tud7c, no marks, dashed, very thick, domain=54:90] {0.1829*x+2.5} node [pos = 0.6, sloped, below, black] {\small 2\textsuperscript{nd} column};
                        \addplot [tud6c, no marks, dashed, very thick, domain=54:90] {12};
                    \end{axis}
                    \node[] (labelb) at ($(labela)+(7cm, 0cm)$) {b)};
                \end{tikzpicture}
                \caption{
                    Summary of the energy landscape for switching of dipoles obtained in static simulations, i.e., the polarization $P_z$ per unit cell is fixed to its bulk value.
                    (a) Energy landscape for all possible sizes of rectangular 2D clusters/nuclei on a flat \qtyproduct{48 x 48}{\uc} DW calculated using the \heff -model. The flat wall without nucleus corresponds to the lowest point. A nucleus which spans the whole simulation cell along $x$ and half the simulation cell along $z$ corresponds to the maximum.
                    (b) Details of the energy landscape from the atomistic model for switching of one column (\(n_s = 0-48\), blue crosses) and the consecutive switching of a second column (\(n_s = 48-96\), blue circles/green triangles) on the DW.
                    The growth of the second column can either extend the cluster along \(x\) or \(y\) and can lead to the formation of a bridging segment.
                    Formation of a bridging segment as well as extension along \(x\) reduce the overall energy, while simply extending the cluster along \(y\) increases overall energy.
                }
                \label{fig:energy_landscape}
            \end{figure}

            \Cref{fig:energy_landscape}~(a) illustrates the dependency of energy on the size along $x$ and $z$ for rectangular 2-dimensional clusters on the wall as obtained in static simulations using the \heff -model.
            More detailed results for the growth of two columnar cluster along $x$, $y$, and $z$ are obtained with the atomistic model in (b).
            
            In both cases, we indeed find a large change of energy by the formation or annihilation of charged interfaces for a single flipped dipole or the completion of a full column, respectively.
            This is in line with our simple first order interaction estimate in \Cref{fig:sketch_stepwise_switching}.
            The jump from \(n_s = 1\) to 2 in \Cref{fig:energy_landscape}~(b) corresponds to double \(E_c \approx \qty{1.27}{\joule\per\square\meter} \approx \qty{1.27}{\electronvolt\per\square\uc}\), i.e., the energy of the charged interface.
            After the first unit cell on the DW has switched, the energy penalty for growth of the column along \(z\) ($n_s = 3 - 45$) is three times \(E_{\text{DW}} \approx \qty{91.5}{\milli\joule\per\square\meter} \approx \qty{91.5}{\milli\electronvolt\per\square\uc}\) per unit cell, as extracted from \Cref{fig:energy_landscape}~(b).

            For the atomistic model, these values are approximately an order of magnitude larger than the results from \Cref{subsec:model_validation} where the atomic positions were relaxed.
            We confirmed in a separate calculation that the reduction of polarization magnitude close to the interface, put simply, the \textit{smearing} of the DW, is the reason for this overestimation.
            As a result, the slopes in \Cref{fig:energy_landscape} (a) and (b) are much too steep, cf. discussion on the Miller-Weinreich model in literature \cite{Miller1960, Shin2007, Liu2016b, Liu2017}.
            Note that the curve for the first switched column (blue crosses in \Cref{fig:energy_landscape}~(b)) basically details the transition from state \RomanNumeralCaps{1} to state \RomanNumeralCaps{2} in \Cref{fig:sketch_stepwise_switching}~(a).

            Taking into account only next neighbor unit cells in \Cref{fig:sketch_stepwise_switching} we expected constant energy for the switching of the second column along \(x\), i.e., a horizontal line for the green triangles in \Cref{fig:energy_landscape}~(b) for $n_z = 48-96$.
            The corresponding curve, however, slightly decreases in energy.
            The reason for this is the favorable interaction of second-nearest neighboring unit cells with the cluster of switched dipoles.

            However, if the second column were to grow deeper into the domain (along \(y\)) we would observe a steeper slope than during the growth of the first column (blue empty circles, $n_z = 50 - 94$) \Cref{fig:energy_landscape}~(b).
            The reason is an unfavorable energy contribution from second-nearest neighbor unit cells.

            Only when the growth along \(y\) forms a bridging segment (transition from state \RomanNumeralCaps{3} to state \RomanNumeralCaps{4} in \Cref{fig:sketch_stepwise_switching}) the energy follows a path of slightly decreasing energy.
            In fact, this situation is identical to a growth of the nucleus along the \(x\)-directions (green triangles in \Cref{fig:sketch_stepwise_switching}~(b)) because the number of anti-parallel unit cells in the nearest and second-nearest neighbor shell is identical.

            These details on the energy landscape also lead to the conclusion that once a single dipole has been reversed, it either relaxes back on short time scales, or grows predominantly along the \(z\)-direction. This is exactly our observation from \Cref{fig:polarization_switching_series}.
            Since the lowest energy cost is associated to growth along $z$, needles of switched dipoles (instead of single dipoles) form almost instantaneously and needle-like clusters of switched polarization occur, see \Cref{fig:polarization_switching_series} and Supplemental Material 5 at \_ .
    
            Note that \Cref{fig:sketch_stepwise_switching} only gives one possible path for domain collapse. 
            Due to chaotic thermal fluctuations each of the process can occur in either direction.
            Fluctuation of the DW on different time scales and of different magnitude are the result.
            The collapse of a thin domain can, therefore, follow a variety of different pathways.
            Additionally, which pathway is taken may change with temperature.
            We suspect that this is, in fact, the origin of the non-Arrhenius behavior demonstrated above.

\newpage
\clearpage

\section{Discussion \& Summary}

    Using two independent computational models, we have observed that very small ferroelectric domains in barium titanate can collapse spontaneously.
    This effect becomes more pronounced the closer the temperature comes to the phase transition and the closer the DWs are initially.
    Our simulations showed that thermal fluctuations are the reason why DWs that do not interact initially come into contact leading to the annihilation of very small domains.
    Different intermediate steps of the domain collapse were identified but the analytic description of the collapse probability is left to further work.
    We expect that this aspect is critical for the discussion of the lower size limits to electronic components based on ferroelectrics.
    
    Further simulations are required to firmly establish a limit in terms of domain size and temperature beyond which domain collapse becomes unlikely.
    For applications relying on very high densities of DWs it could, however, be beneficial to introduce locations of DW pinning to increase possible DW density even further.
    In this context research into the pinning by point defects and doping can be found in literature but also using dislocations as pinning defects appears to be a feasible approach.

    Due to the model character of \ch{BaTiO3} we expect that our results are transferable to other ferroelectric perovskites.
    Yet, it remains to be shown that the observations on T180 DWs also hold true for other DW types such as \qty{90}{\degree} DW.
    Some of our preliminary calculations suggest that this is indeed the case but that time and length scales are changed due to the significantly wider T90 DWs.

\section*{Acknowledgments}
    A. Klomp is funded by Deutsche Forschungsgemeinschaft (DFG) through SPP 1599.
    This work was supported by the Hessian State Ministry for Higher Education, Research and the Arts under the LOEWE collaborative project "FLAME". 
    Calculations for this research were conducted on the Lichtenberg high performance computer of the TU Darmstadt.
    R. Khachaturyan, T. Wallis and A. Gr\"unebohm acknowledge financial support by DFG via the Emmy Noether group GR4792/2.
    R. Khachaturyan and A. Gr\"unebohm acknowledge fruitful discussion with M. Stricker.
    
\bibliography{zotero_export}

\clearpage

\appendix

\section{Fluctuation inside domain and at DW}
\label{app:1}
    
    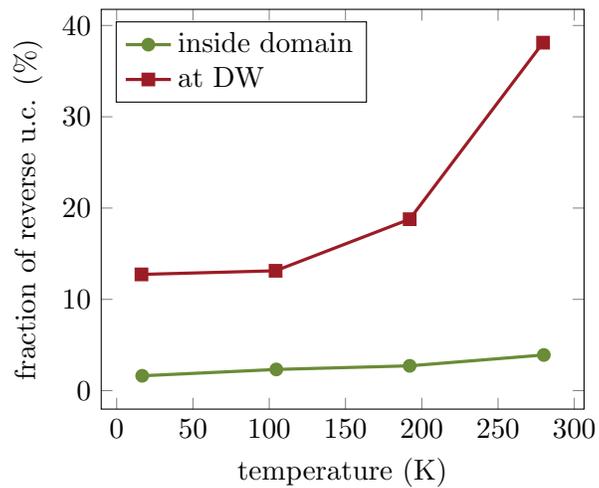
\begin{figure}[htb] 
        \centering
        \begin{tikzpicture}[/pgfplots/tick scale binop=\times]
            \begin{axis} [
                width=8cm,
                anchor = west,
                xlabel=temperature in \unit{\kelvin},
                ylabel=fraction of reverse u.c. in \unit{\percent},
                legend pos=north west,
                legend style={cells={anchor=west}},
                ]
                \addplot [tud4d, very thick, mark=*] coordinates {
                    (16.72354948805461, 1.6288951841359847)
                    (104.64163822525597, 2.322946175637391)
                    (192.08191126279863, 2.719546742209637)
                    (280.00000000000006, 3.9093484419263476)
                };
                \addlegendentry{inside domain};
                \addplot [tud9d, very thick, mark=square*] coordinates {
                    (16.245733788395903, 12.73371104815864)
                    (104.16382252559728, 13.130311614730878)
                    (192.08191126279863, 18.781869688385267)
                    (279.52218430034134, 38.11614730878187)
                };
                \addlegendentry{at DW};
            \end{axis}
        \end{tikzpicture}
        \caption{
            \label{fig:num_reversed_cells}
            Number of reversed polarization vectors inside a domain and at the DW.
            More unit cells with reversed polarization can be found at the interface and the fraction increases with temperature.
            Data from \heff -calculation.
            Atomistic results show an identical trend.
        }
    \end{figure}

\newpage
\clearpage

\section{Local atomistic relaxations at DW}
\label{app:2}

    The coarse-grained \heff -model cannot reproduce changes in interatomic distances at the DW. 
    Thus, we test if such local relaxations are even important for T180 DWs using the atomistic core-shell model.
    To this end, we study the radial distribution functions (RDF) of the \ch{Ti}-\ch{O} distance, the \ch{Ba}-\ch{O} distance and the angular distribution function of \ch{O}-\ch{Ti}-\ch{O}.
    The example of the \ch{Ti}-\ch{O} RDF is shown in \Cref{fig:rdf}, which is in agreement to literature \cite{Tinte2000, Qi2016}.
    For the RDF at the DW only \qty{2}{\uc} on each side of the interface are considered.
    Relaxations at the DW seem to be negligible or absent due to the abrupt nature of the DW.
    Consequently, the atomic relaxations play a minor at T180 DWs and the \heff~model is expected be a good approximation at finite temperature.

    \begin{figure}[htb] 
        \centering
        \begin{tikzpicture}[/pgfplots/tick scale binop=\times]
            \begin{axis} [
                name=pdfgraph,
                width=8cm,
                anchor = west,
                xmin=1.65,
                xmax=2.45,
                ymin=-0.05,
                ymax=1.25,
                xlabel=distance in \unit{\angstrom},
                ylabel=normalized PDF,
                legend pos=north east,
                legend style={cells={anchor=west}},
                ]
                \addplot [tud1c, very thick, no marks, dashed] table [x=PairSeparationDistance, y=cubic400K] {data/normalized_TiO_pdf.dat};
                \addlegendentry{cubic}; 
                \addplot [tud4c, very thick, no marks] table [x=PairSeparationDistance, y=tetragonal299Kbulk] {data/normalized_TiO_pdf.dat};
                \addlegendentry{tetragonal}; 
                \addplot [tud9c, very thick, no marks] table [x=PairSeparationDistance, y=tetragonal299Kdw] {data/normalized_TiO_pdf.dat};
                \addlegendentry{T180 DW}; 
            \end{axis}
        \end{tikzpicture}
        \caption{
        (b) Normalized Ti-O pair distribution function of cubic (blue) mono-domain tetragonal (green), 2 f.u. layers left and right of bi-domain tetragonal phase (red) and mono-domain orthorhombic (yellow).
        }
        \label{fig:rdf}
    \end{figure}

\newpage
\clearpage

\section{DW width and energy as a function of temperature}
\label{app:3}

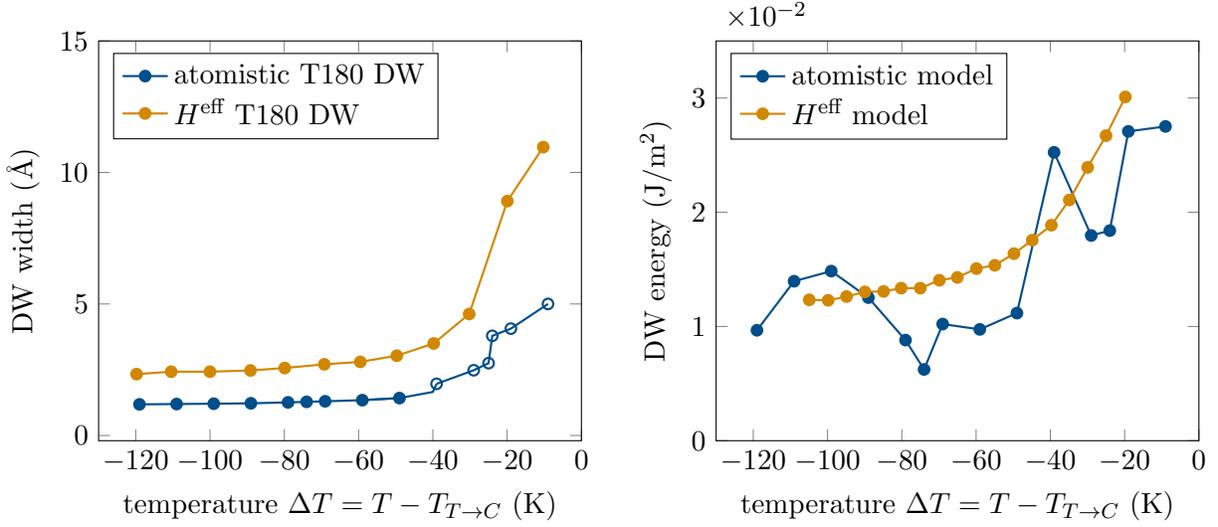
\begin{figure}[htb] 
        \centering
        \begin{tikzpicture}[/pgfplots/tick scale binop=\times]
            \begin{axis} [
                name=dwwidthgraph,
                width=8cm,
                xlabel={temperature \(\Delta T = T - T_{T \rightarrow C}\) in \unit{kelvin}},
                ylabel=DW width in \unit{\angstrom}, 
                xmin=-130,
                xmax=0,
                ymax=15,
                legend pos= north west,
                legend style={cells={anchor=west}},
                ]
                \addplot [x filter/.code={\pgfmathparse{\pgfmathresult-349}}, tud1c,mark=*, thick] table [x=Temp, y=dwwidthAng, restrict x to domain=-119:-49] {data/dw_width_T180_with_temp_v4.dat};
                \addlegendentry{atomistic T180 DW};
                \addplot [x filter/.code={\pgfmathparse{\pgfmathresult-349}}, tud1c, no marks, thick, forget plot] table [x=Temp, y=dwwidthAng, restrict x to domain=-59:-34] {data/dw_width_T180_with_temp_v4.dat};
                \addplot [x filter/.code={\pgfmathparse{\pgfmathresult-349}}, tud1c ,mark=o, thick, forget plot] table [x=Temp, y=dwwidthAng, restrict x to domain=-39:-9] {data/dw_width_T180_with_temp_v4.dat};
                \addplot [x filter/.code={\pgfmathparse{\pgfmathresult-300}}, y filter/.code={\pgfmathparse{\pgfmathresult*10}}, tud7c, thick, mark=*] table [x=temp, y=width] {data/dwwidth_180deg_Heff_extracted.dat}; 
                \addlegendentry{\heff~T180 DW};
            \end{axis}
            \begin{axis} [
                name=dwenergygraph,
                width=8cm,
                at={($(dwwidthgraph)+(5cm,0cm)$)},
                anchor = west,
                xmin=-130,
                xmax=0,
                ymin=0,
                ymax=0.035,
                xlabel={temperature \(\Delta T = T - T_{T \rightarrow C}\) in \unit{kelvin}},
                ylabel=DW energy per area in \unit{\joule\per\square\meter},
                legend pos=north west,
                legend style={cells={anchor=west}},
                ]
                \addplot [x filter/.code={\pgfmathparse{\pgfmathresult-349}}, tud1c, mark=*, thick,] table [x=temperature, y=gamma, restrict x to domain=-119:-9] {data/dw_energy_180deg_head2tail.dat};
                \addlegendentry{atomistic model};
                \addplot [x filter/.code={\pgfmathparse{\pgfmathresult-300}},tud7c, mark=*, thick, ] table [x=temp, y=gamma,] {data/dw_energy_180deg_Heff_extracted.txt}; 
                \addlegendentry{\heff~model};
            \end{axis}
        \end{tikzpicture}
        \caption{
        Left: Width of T180 DW over temperature. Data for effective Hamiltonian model is extracted from Ref.~\cite{Grunebohm2020}.
        Right: Temperature dependence of T180 DW energy density for atomistic and \heff -model. In the higher temperature regime the fluctuation of dipole moments lets the DW appear rough.
        }
        \label{fig:dw_width_energy_over_T}
    \end{figure}

    In \Cref{fig:dw_width_energy_over_T}~(a) we display the width of the DW obtained by the atomistic and the \heff -model.
    The course of the curves is in good agreement.
    Well within the ferroelectric phase (40~K below \( T_{T \rightarrow C}\)), we find a T180 wall thicknesses of \(d_{\rm DW} \approx \qty{0.4}{\uc} \approx \qty{1.6}{\angstrom}\) and  \(\qty{0.6}{\uc} \approx \qty{2.5}{\angstrom}\) for our atomistic and coarse-grained models, respectively, using a tanh-fit.
    In agreement, atomically sharp walls have also been predicted by DFT calculations (\qty{2.8}{\angstrom} at \qty{0}{\kelvin} \cite{Grunebohm2012}).
    Furthermore, our results deviate by less than \qty{15}{\percent} from the DFT results at \qty{0}{\kelvin} by \textcite{Padilla1996} and a width of \qty{3.2}{\angstrom} has been predicted by Landau theory at room temperature \cite{Marton2010}.
    
    \Cref{fig:dw_width_energy_over_T}~(b) compares the temperature dependency of the DW energy per area for the \heff -model from \textcite{Grunebohm2020} and the atomistic simulation.
    These energies are computed from a comparison of energies for a two-domain and a single-domain simulation cell.
    The evolution of the DW properties with time are strongly affected by stochastic processes and, thus, large deviations between different configurations at one temperature might occur.
    In full agreement with this, the atomistic results based on a single simulation run each, show a large noise level.
    Nevertheless, one can see a clear trend for an increase of the energy with temperature.
    The energy in the coarse-grained simulations have been obtained by heating up one domain wall and, thus, show less noise. 
    Also, due to the reduced number of degrees of freedom in the latter model this model shows reduced thermal noise.

\newpage
\clearpage
    
\section{DW energy from fluctuations}
\label{app:4}

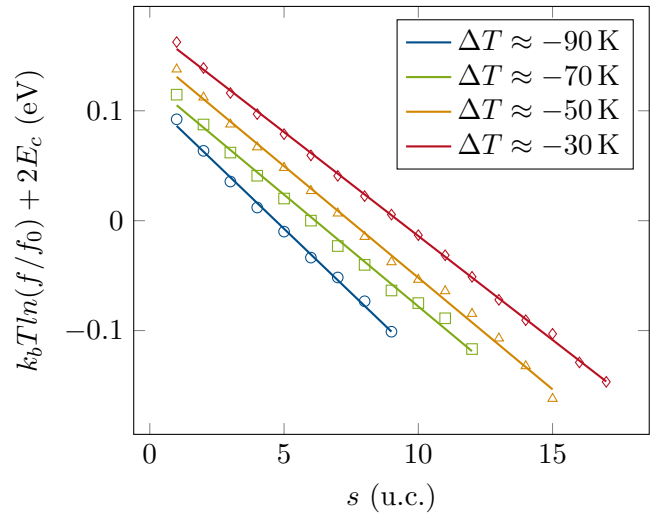
\begin{figure}[htbp] 
        \centering
        \begin{tikzpicture}[/pgfplots/tick scale binop=\times]
            \begin{axis} [
                anchor = west,
                xlabel= \(s\) in \unit{\uc},
                ylabel= \(k_b T ln{\left(f/f_0\right)} + 2 E_c\) in \unit{\electronvolt},
                legend pos=north east,
                legend style={cells={anchor=west}},
                ]
                \addplot [tud1c, only marks, mark=o, forget plot] table [x=cluster_size, y=log_freq_data] {data/cluster_size_log_fit_260K.dat};
                \addplot [tud1c, no marks, thick,] table [x=cluster_size, y=log_freq_fit] {data/cluster_size_log_fit_260K.dat};
                \addlegendentry{\( \Delta T \approx \qty{-90}{\kelvin} \)};
                \addplot [tud4c, only marks, mark=square, forget plot] table [x=cluster_size, y=log_freq_data] {data/cluster_size_log_fit_280K.dat};
                \addplot [tud4c, no marks, thick,] table [x=cluster_size, y=log_freq_fit] {data/cluster_size_log_fit_280K.dat};
                \addlegendentry{\( \Delta T \approx \qty{-70}{\kelvin} \)};
                \addplot [tud7c, only marks, mark=triangle, forget plot] table [x=cluster_size, y=log_freq_data] {data/cluster_size_log_fit_300K.dat};
                \addplot [tud7c, no marks, thick,] table [x=cluster_size, y=log_freq_fit] {data/cluster_size_log_fit_300K.dat};
                \addlegendentry{\( \Delta T \approx \qty{-50}{\kelvin} \)};
                \addplot [tud9c, only marks, mark=diamond, forget plot] table [x=cluster_size, y=log_freq_data] {data/cluster_size_log_fit_320K.dat};
                \addplot [tud9c, no marks, thick,] table [x=cluster_size, y=log_freq_fit] {data/cluster_size_log_fit_320K.dat};
                \addlegendentry{\( \Delta T \approx \qty{-30}{\kelvin} \)};
            \end{axis}
        \end{tikzpicture}
        \caption{
        Logarithm of frequency of reversed clusters in a single domain structure over cluster size.
        The slope is proportional to the DW energy.
        }
        \label{fig:cluster_size_analysis}
    \end{figure}

    We developed \Cref{eq:frac_reversed_dipoles_energy} to extract the temperature dependence of the DW energy per unit area from the number and size of reversed clusters in a single-domain state.
    In \Cref{fig:cluster_size_analysis}, we display the number of reversed needle-like clusters \(f\) in normalized and logarithmic representation over the size \(s\) of the cluster.
    Clusters are groups of unit cells with reversed polarization that are connected along the polarization axis.
    From the slope of the linear regressions in \Cref{fig:cluster_size_analysis} we estimate \(E_{\text{DW}}\).

\newpage
\clearpage

\section{Nucleation and propagation of domain collapse}
\label{app:5}

    \begin{figure} [htbp] 
        \centering
        \begin{tikzpicture}
            \draw[draw=black, fill=tud1c, opacity=0.15] (0.0, 2.0) rectangle ++(5.2,-16);
            \draw[draw=black, fill=tud1c, opacity=0.4] (0.0, -14) rectangle ++(5.2,-0.8);
            \node[inner sep=3pt, rounded corners=0.1cm, fill={black!10}, opacity=0.9, text opacity=1,] (scenarioA) at (2.6,-14.4) {scenario A};
            \draw[draw=black, fill=tud4c, opacity=0.15] (5.6, 2.0) rectangle ++(5.2,-16);
            \draw[draw=black, fill=tud4c, opacity=0.4] (5.6, -14) rectangle ++(5.2,-0.8);
            \node[inner sep=3pt, rounded corners=0.1cm, fill={black!10}, opacity=0.9, text opacity=1,] (scenarioB) at (8.2,-14.4) {scenario B};
            \draw[draw=black, fill=tud7c, opacity=0.15] (11.2, 2.0) rectangle ++(5.2,-16);
            \draw[draw=black, fill=tud7c, opacity=0.4] (11.2, -14) rectangle ++(5.2,-0.8);
            \node[inner sep=3pt, rounded corners=0.1cm, fill={black!10}, opacity=0.9, text opacity=1,] (scenarioC) at (13.8,-14.4) {scenario C};
            \node [anchor=west] (senarioAframe90000) at (0,0) {\includegraphics[width=5cm]{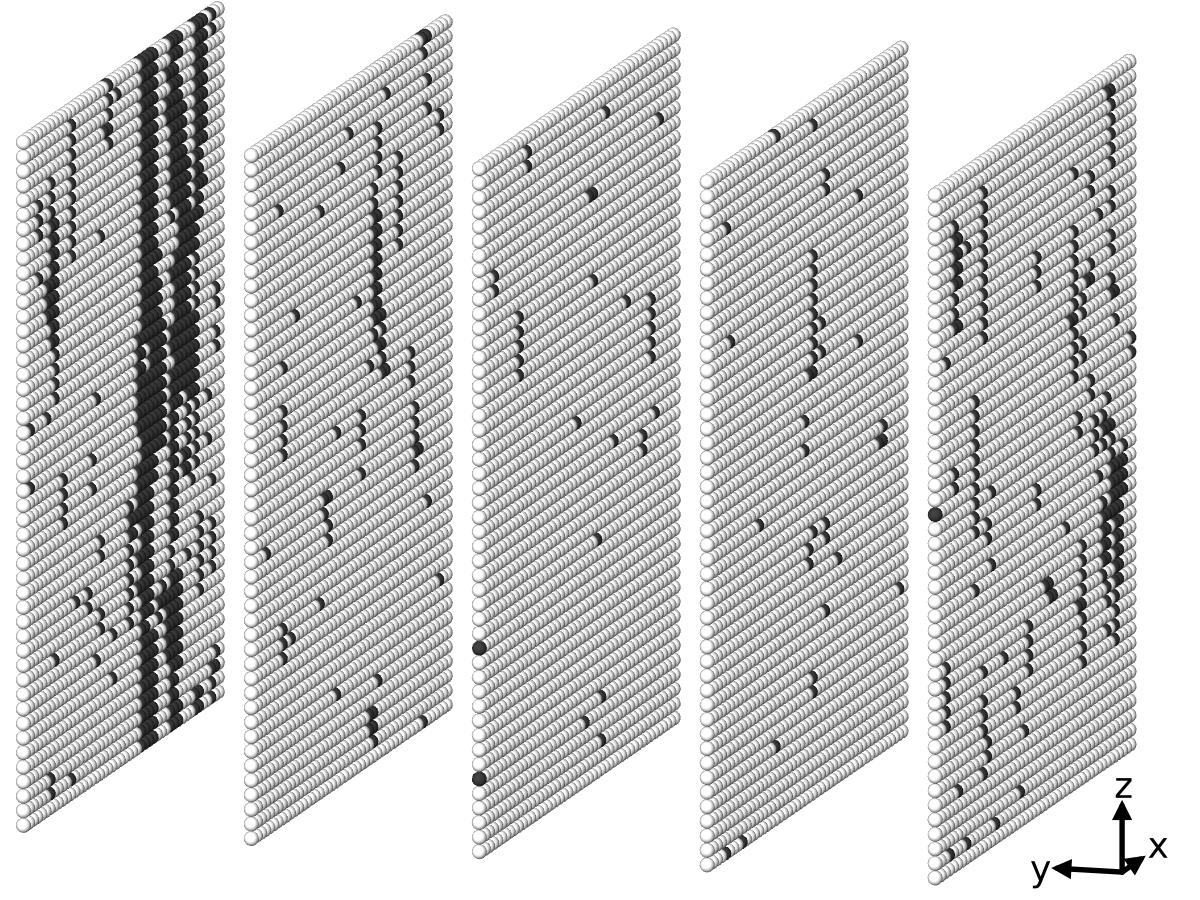}};
            \node[inner sep=3pt, rounded corners=0.1cm, fill=tud1c, opacity=0.7, text opacity=1,] (labelA1) at ($(senarioAframe90000.south west)+(0.7cm,0.5cm)$) {\color{white}\qty{90}{\pico\second}};
            \node [anchor=west] (senarioAframe92000) at (0,-4) {\includegraphics[width=5cm]{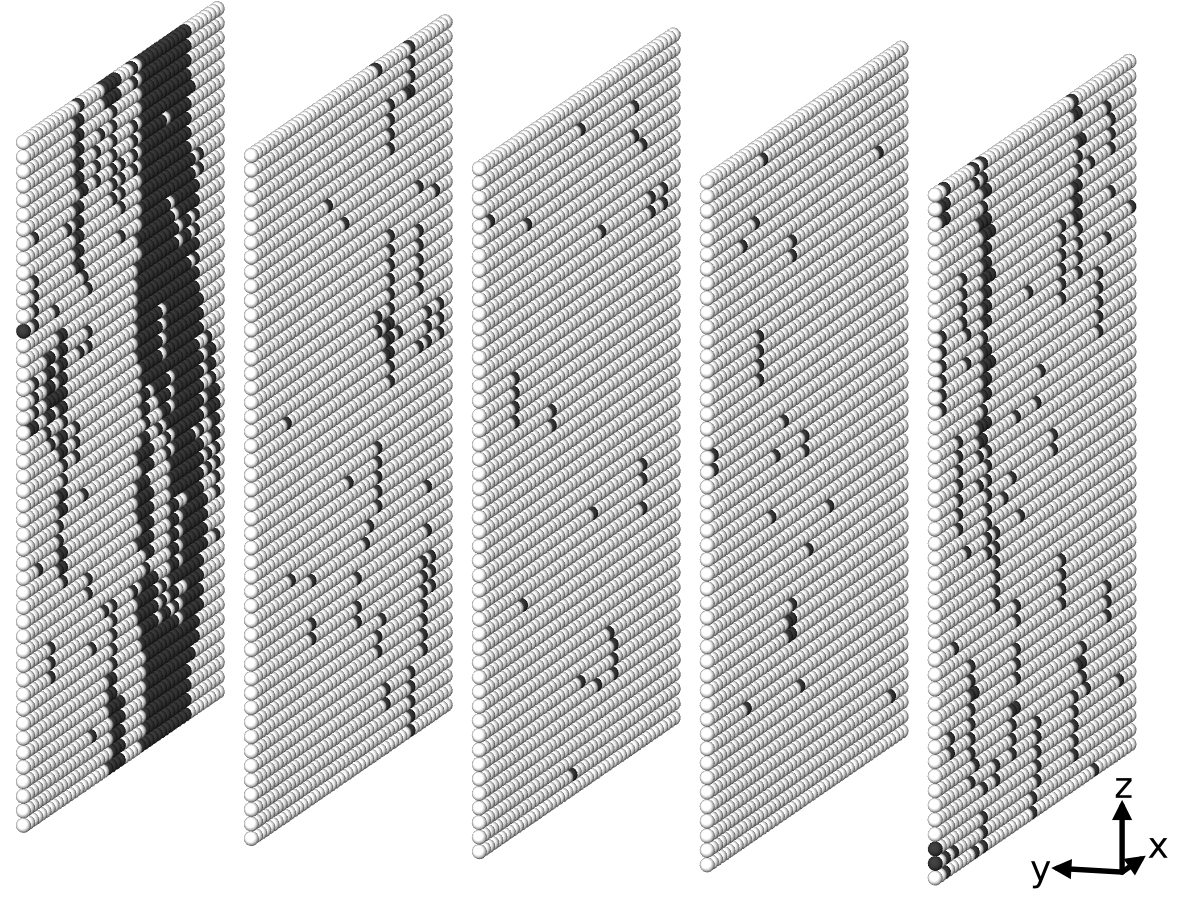}};
            \node[inner sep=3pt, rounded corners=0.1cm, fill=tud1c, opacity=0.7, text opacity=1,] (labelA2) at ($(senarioAframe92000.south west)+(0.7cm,0.5cm)$) {\color{white}\qty{92}{\pico\second}};
            \node [anchor=west] (senarioAframe94000) at (0,-8) {\includegraphics[width=5cm]{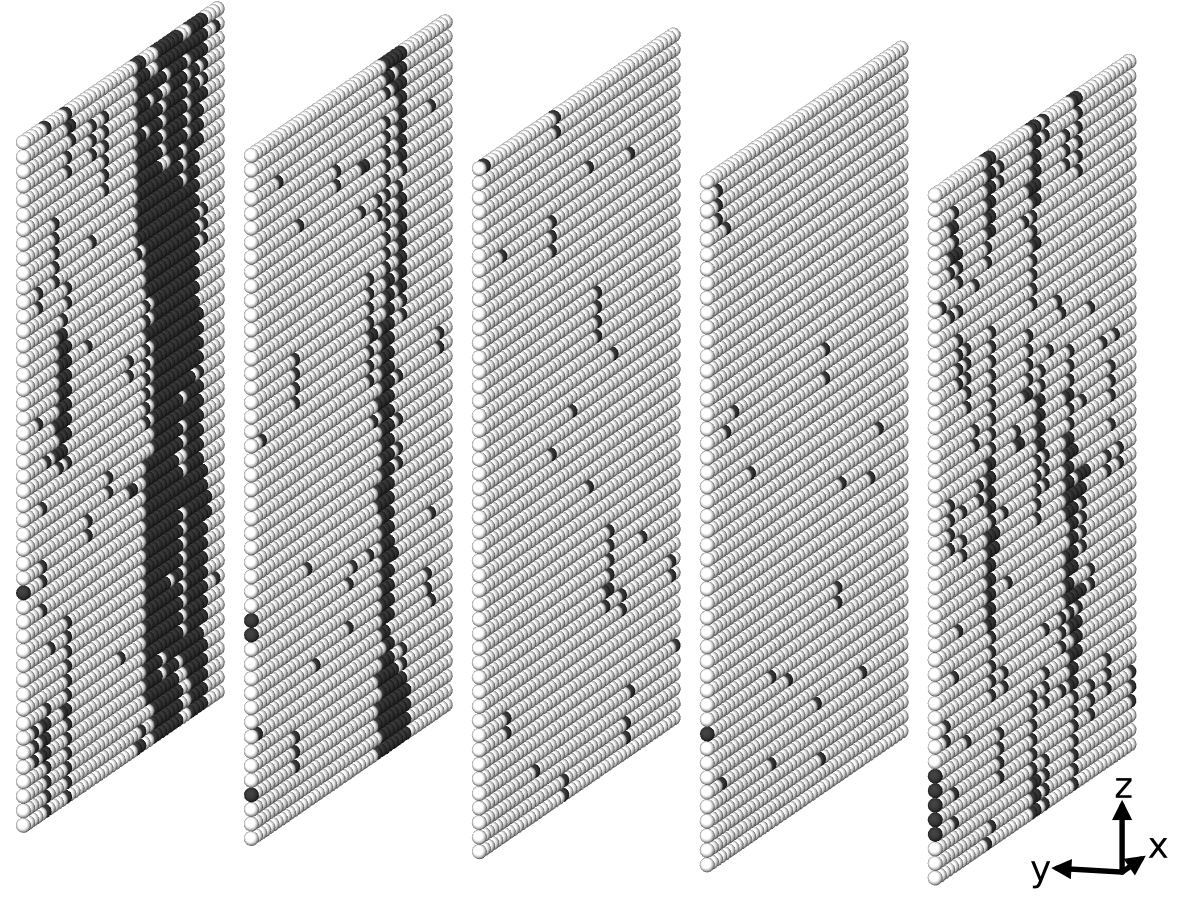}};
            \node[inner sep=3pt, rounded corners=0.1cm, fill=tud1c, opacity=0.7, text opacity=1,] (labelA3) at ($(senarioAframe94000.south west)+(0.7cm,0.5cm)$) {\color{white}\qty{94}{\pico\second}};
            \node [anchor=west] (senarioAframe96000) at (0,-12) {\includegraphics[width=5cm]{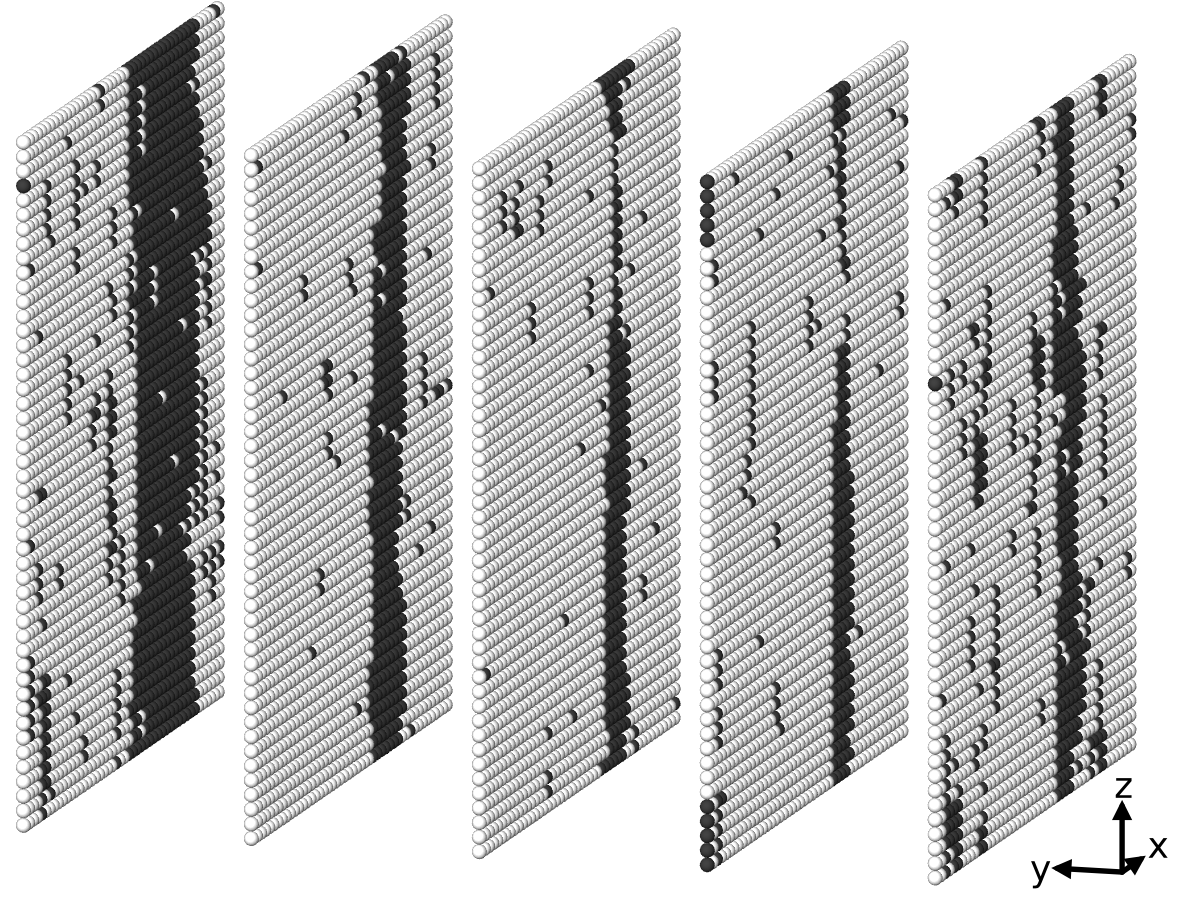}};
            \node[inner sep=3pt, rounded corners=0.1cm, fill=tud1c, opacity=0.7, text opacity=1,] (labelA4) at ($(senarioAframe96000.south west)+(0.7cm,0.5cm)$) {\color{white}\qty{96}{\pico\second}};
            \node [anchor=west] (senarioBframe118000) at (5.6,0) {\includegraphics[width=5cm]{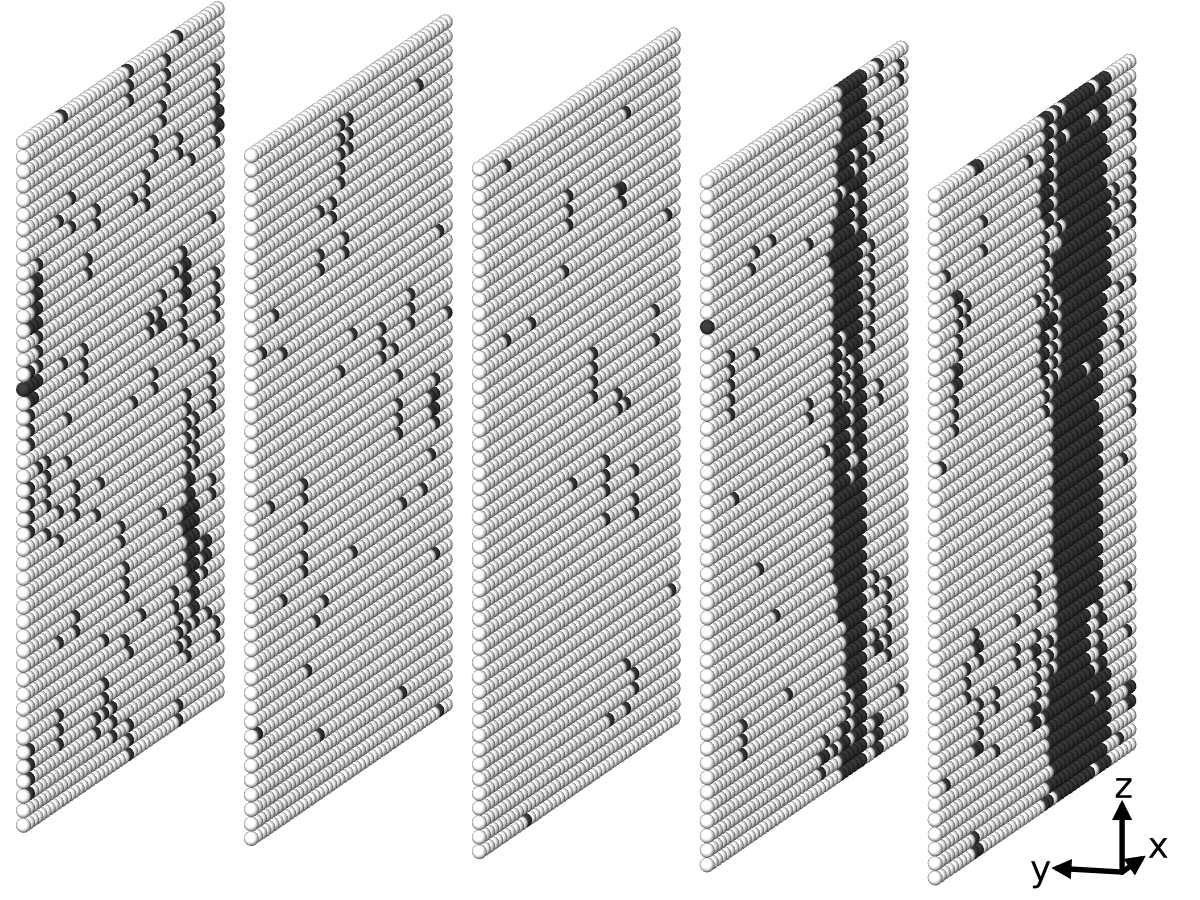}};
            \node[inner sep=3pt, rounded corners=0.1cm, fill=tud4c, opacity=0.7, text opacity=1,] (labelB1) at ($(senarioBframe118000.south west)+(0.7cm,0.5cm)$) {\color{white}\qty{118}{\pico\second}};
            \node [anchor=west] (senarioBframe120000) at (5.6,-4) {\includegraphics[width=5cm]{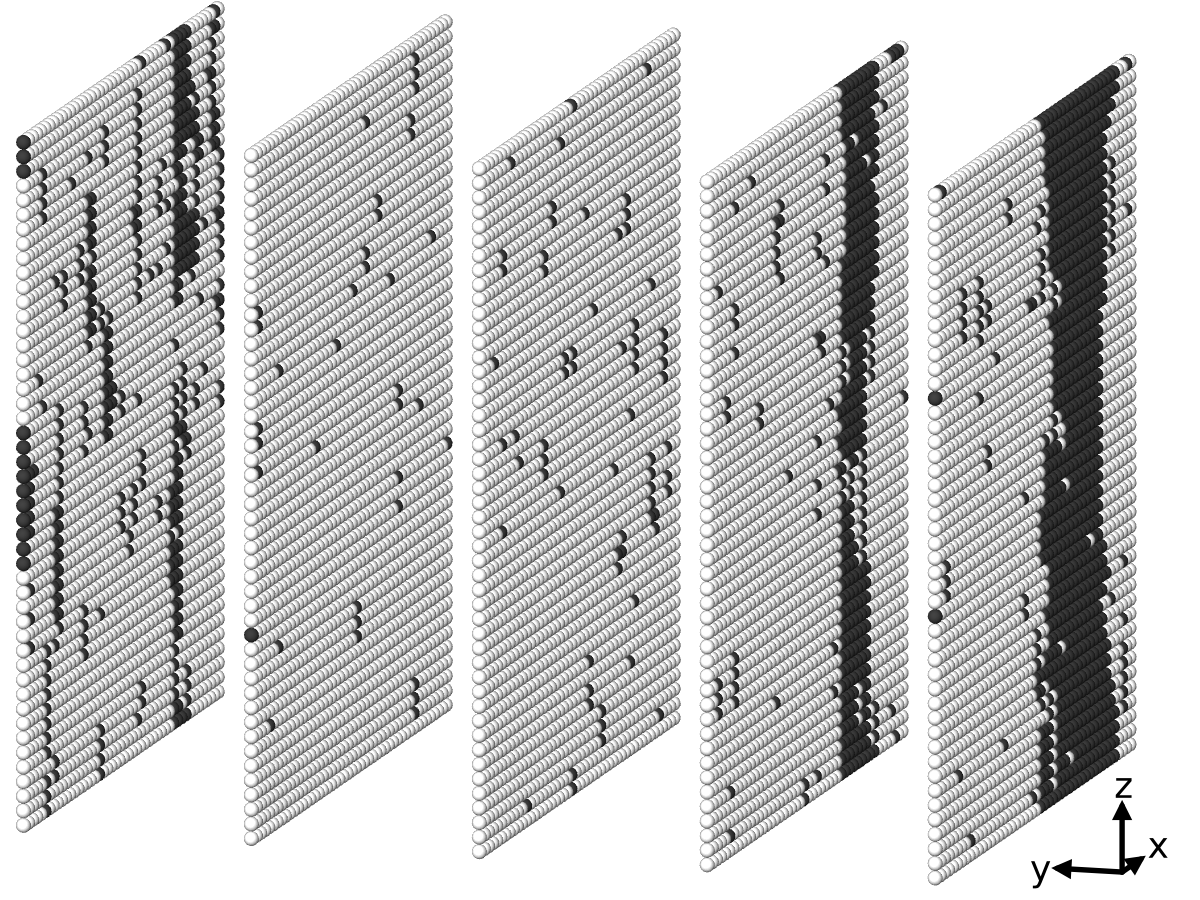}};
            \node[inner sep=3pt, rounded corners=0.1cm, fill=tud4c, opacity=0.7, text opacity=1,] (labelB2) at ($(senarioBframe120000.south west)+(0.7cm,0.5cm)$) {\color{white}\qty{120}{\pico\second}};
            \node [anchor=west] (senarioBframe122000) at (5.6,-8) {\includegraphics[width=5cm]{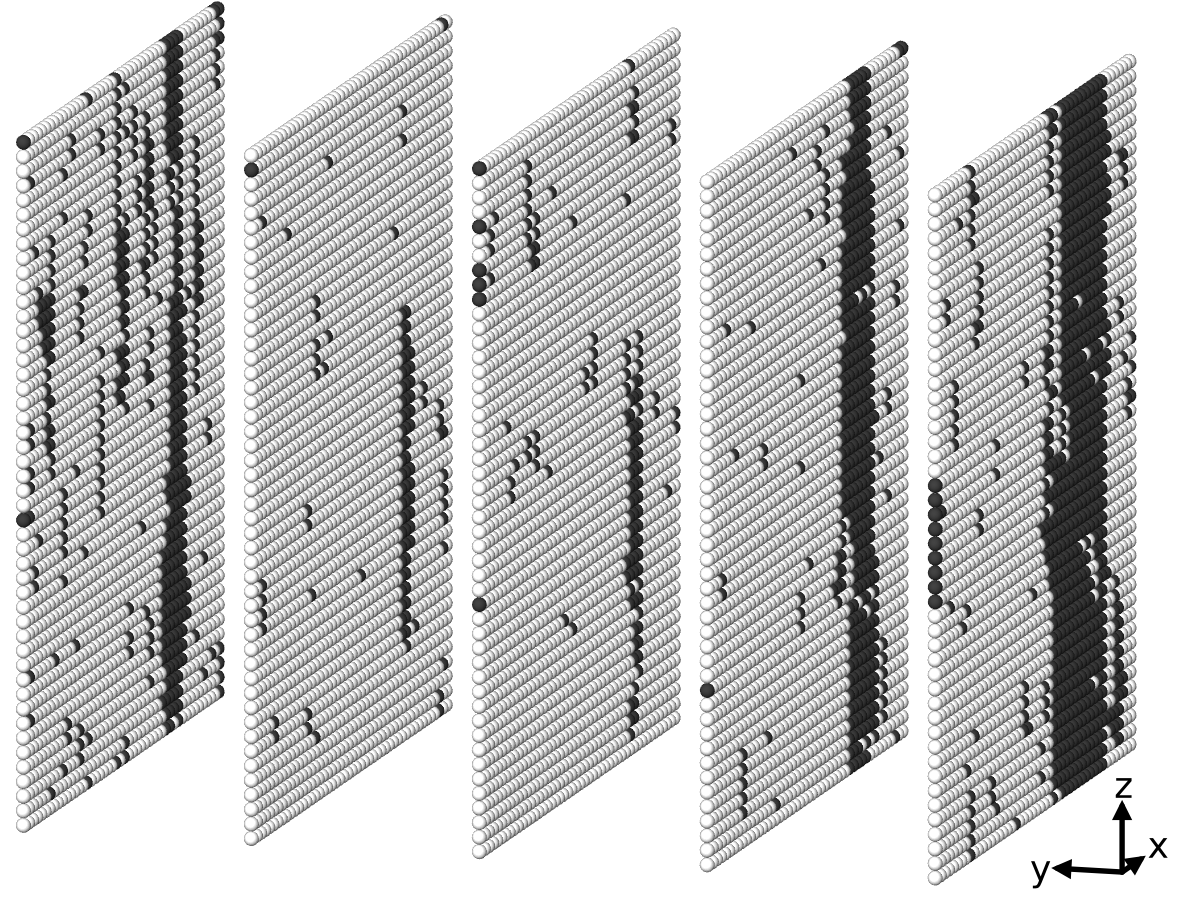}};
            \node[inner sep=3pt, rounded corners=0.1cm, fill=tud4c, opacity=0.7, text opacity=1,] (labelB3) at ($(senarioBframe122000.south west)+(0.7cm,0.5cm)$) {\color{white}\qty{122}{\pico\second}};
            \node [anchor=west] (senarioBframe124000) at (5.6,-12) {\includegraphics[width=5cm]{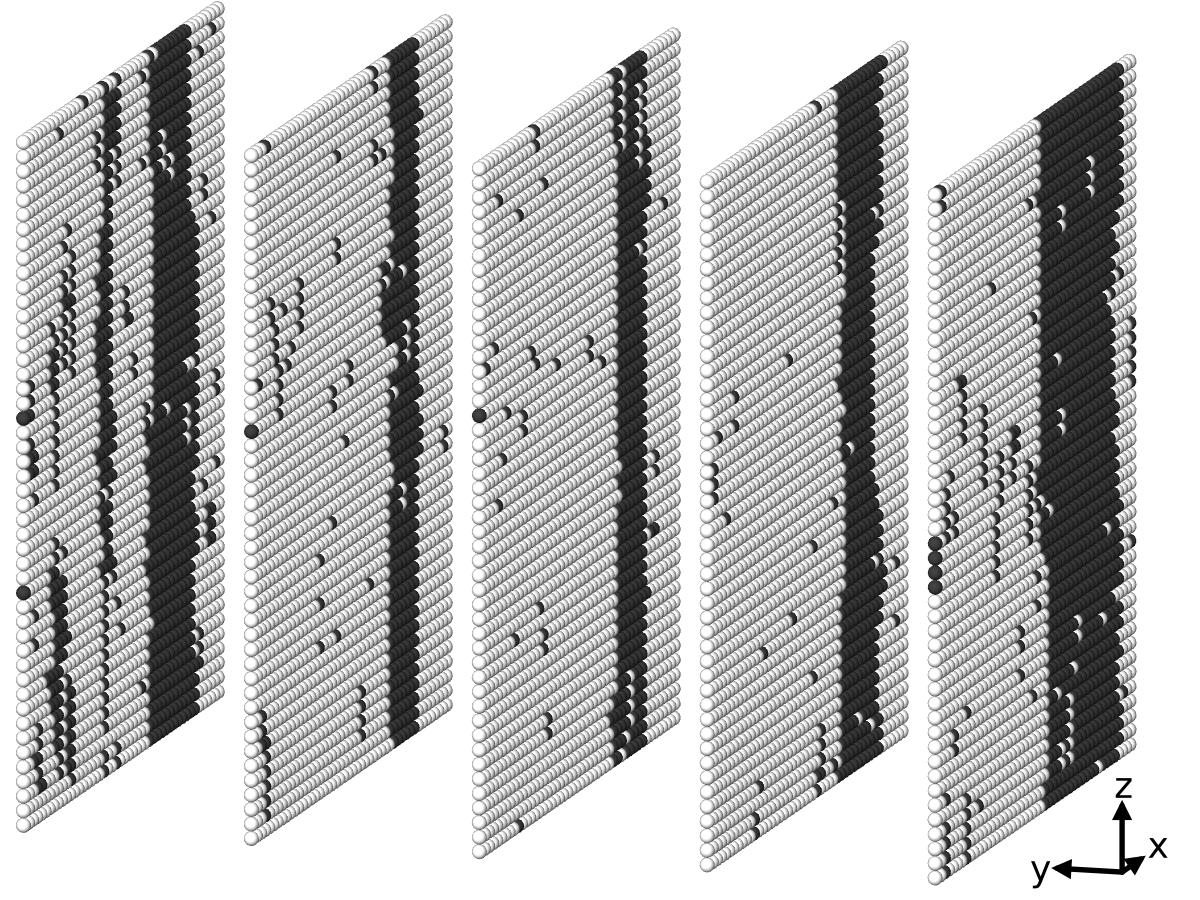}};
            \node[inner sep=3pt, rounded corners=0.1cm, fill=tud4c, opacity=0.7, text opacity=1,] (labelB4) at ($(senarioBframe124000.south west)+(0.7cm,0.5cm)$) {\color{white}\qty{124}{\pico\second}};
            \node [anchor=west] (senarioCframe31000) at (11.2,0) {\includegraphics[width=5cm]{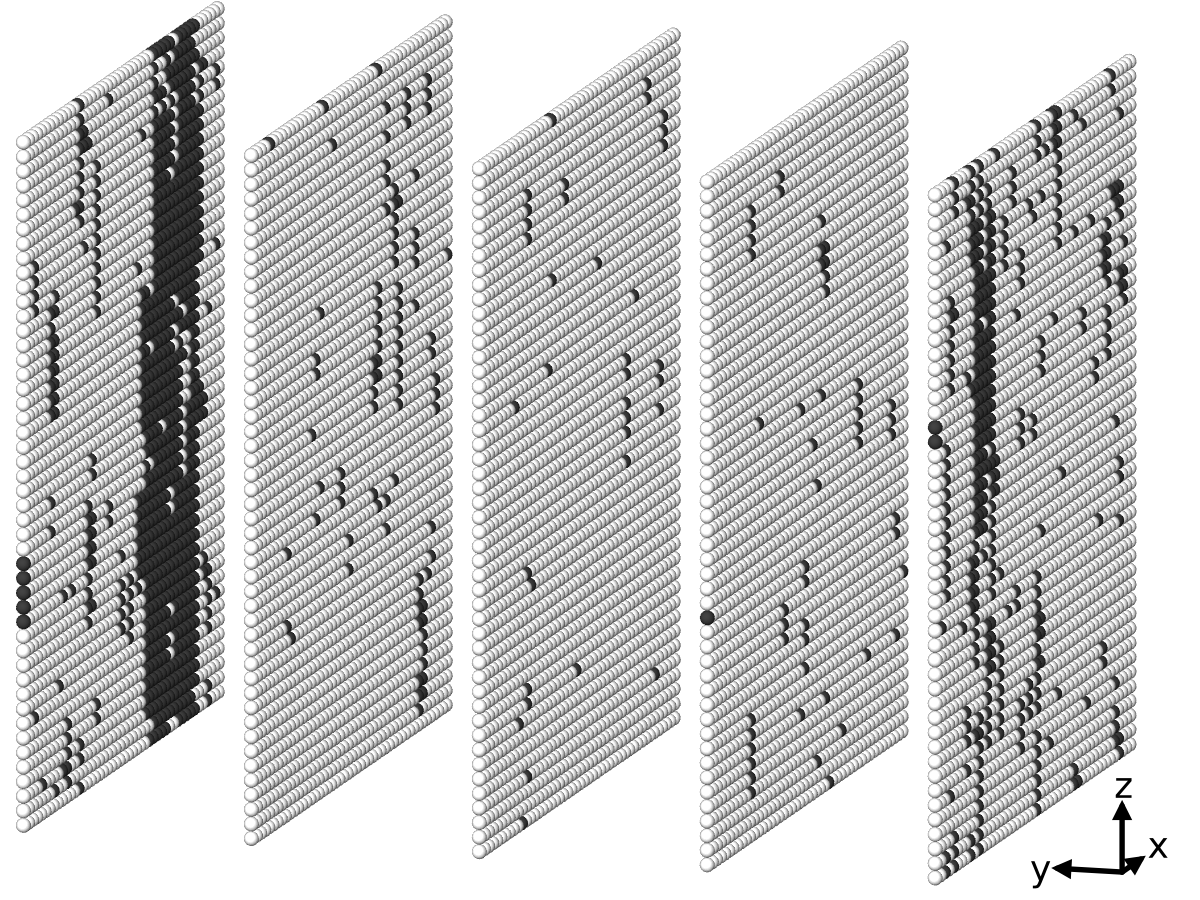}};
            \node[inner sep=3pt, rounded corners=0.1cm, fill=tud7c, opacity=0.7, text opacity=1,] (labelC1) at ($(senarioCframe31000.south west)+(0.7cm,0.5cm)$) {\color{white}\qty{31}{\pico\second}};
            \node [anchor=west] (senarioCframe71000) at (11.2,-4) {\includegraphics[width=5cm]{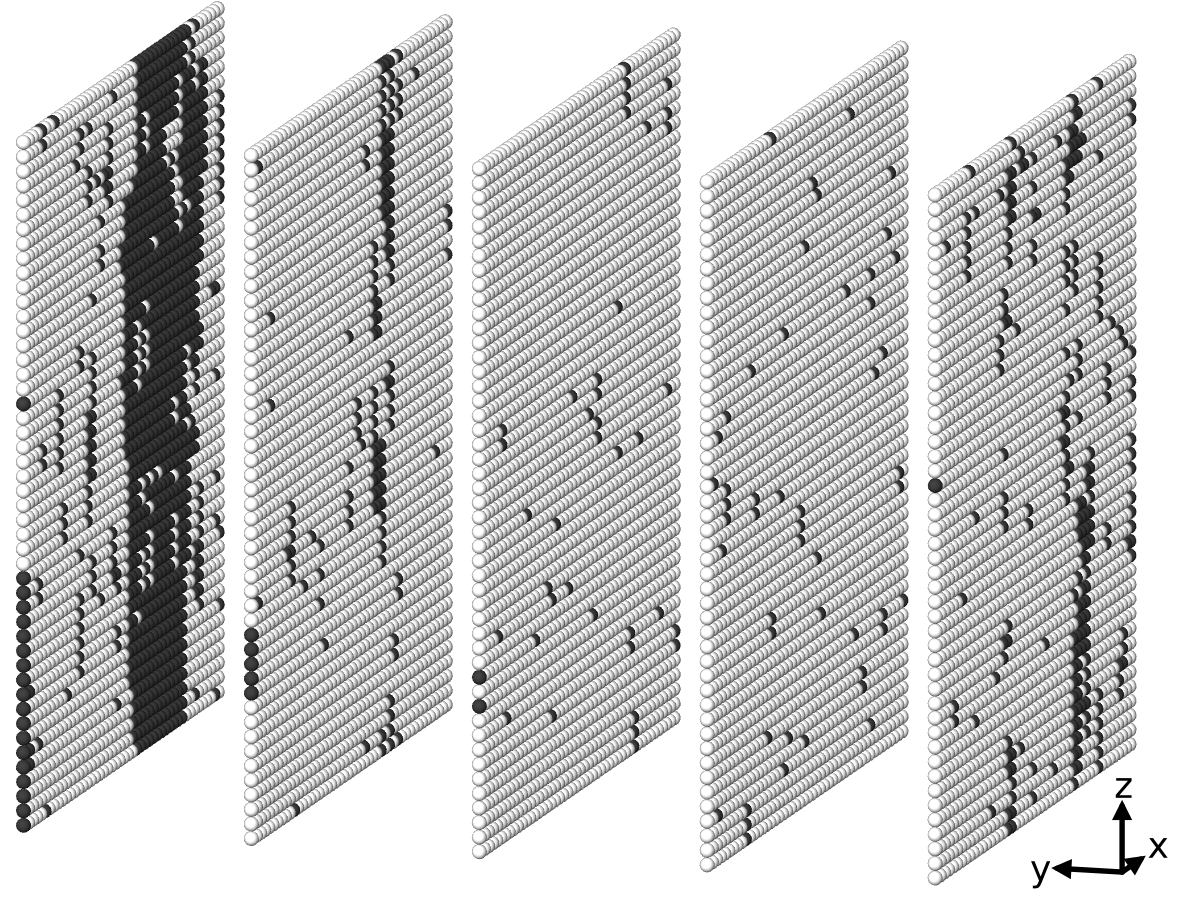}};
            \node[inner sep=3pt, rounded corners=0.1cm, fill=tud7c, opacity=0.7, text opacity=1,] (labelC2) at ($(senarioCframe71000.south west)+(0.7cm,0.5cm)$) {\color{white}\qty{71}{\pico\second}};
            \node [anchor=west] (senarioCframe121000) at (11.2,-8) {\includegraphics[width=5cm]{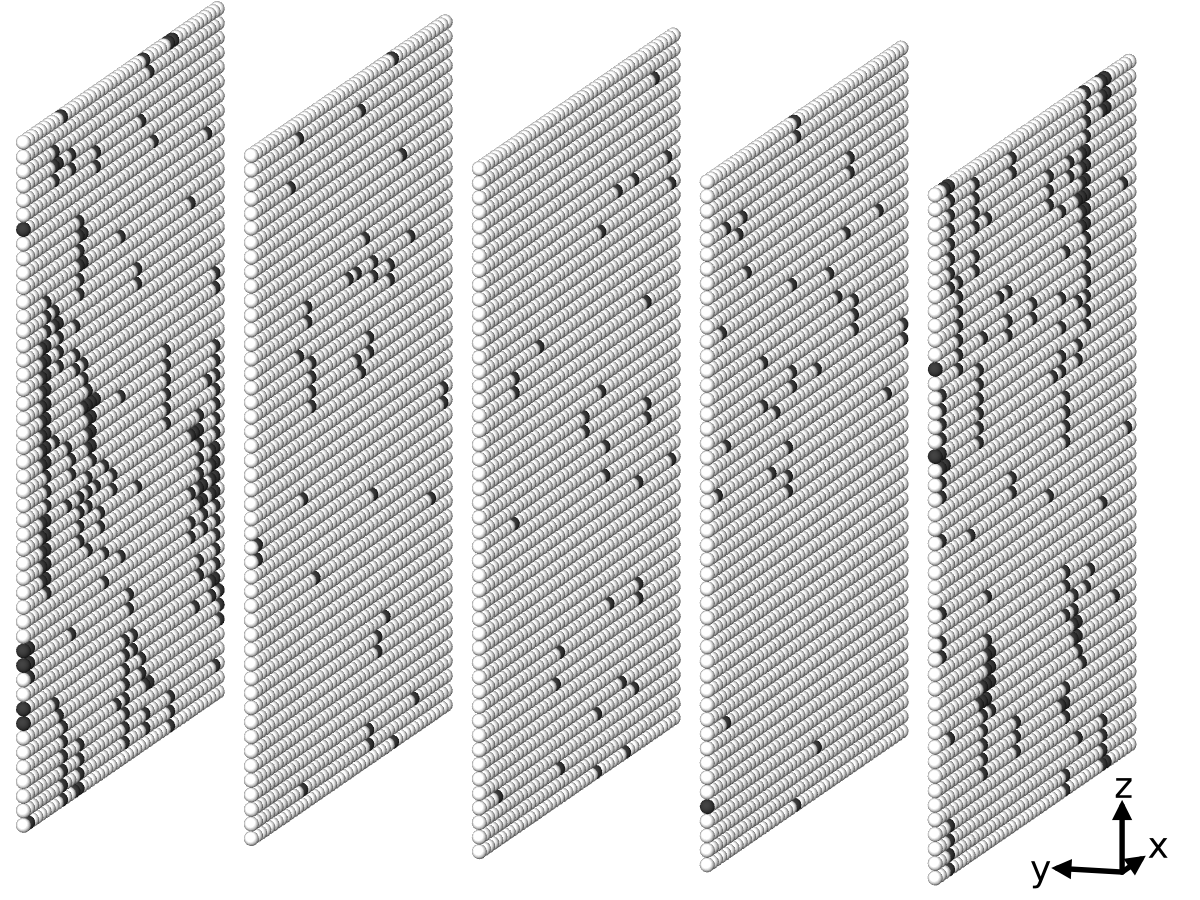}};
            \node[inner sep=3pt, rounded corners=0.1cm, fill=tud7c, opacity=0.7, text opacity=1,] (labelC3) at ($(senarioCframe121000.south west)+(0.7cm,0.5cm)$) {\color{white}\qty{121}{\pico\second}};
            \node [anchor=west] (senarioCframe171000) at (11.2,-12) {\includegraphics[width=5cm]{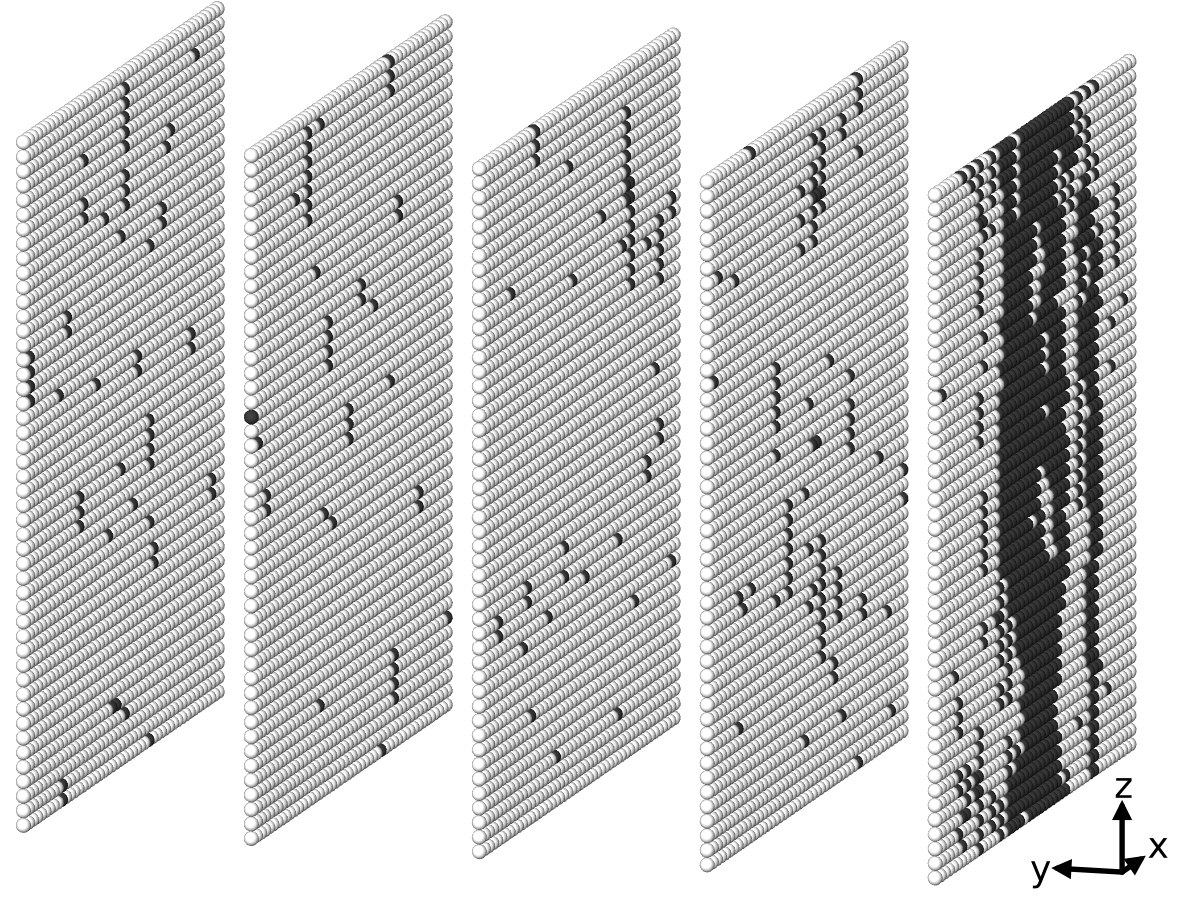}};
            \node[inner sep=3pt, rounded corners=0.1cm, fill=tud7c, opacity=0.7, text opacity=1,] (labelC4) at ($(senarioCframe171000.south west)+(0.7cm,0.5cm)$) {\color{white}\qty{171}{\pico\second}};
        \end{tikzpicture}
        \caption{
        Layer-wise polarization for different scenarios evolving from top to bottom, each with an initial domain thickness of \qty{5}{\uc}.
        Light and dark color distinguishes positive and negative \(P_z\).
        Initially all layers were light.
        (A) Clusters of reversed dipoles are visible on both surfaces. One cluster quickly penetrates a large portion of the reversed domain.
        (B) The bridging element does not encompass the whole column when it first appears.
        (C) Large clusters form on either of the DW but collapse does not occur. 
        }
        \label{fig:nucleation_propagation_feram}
    \end{figure}
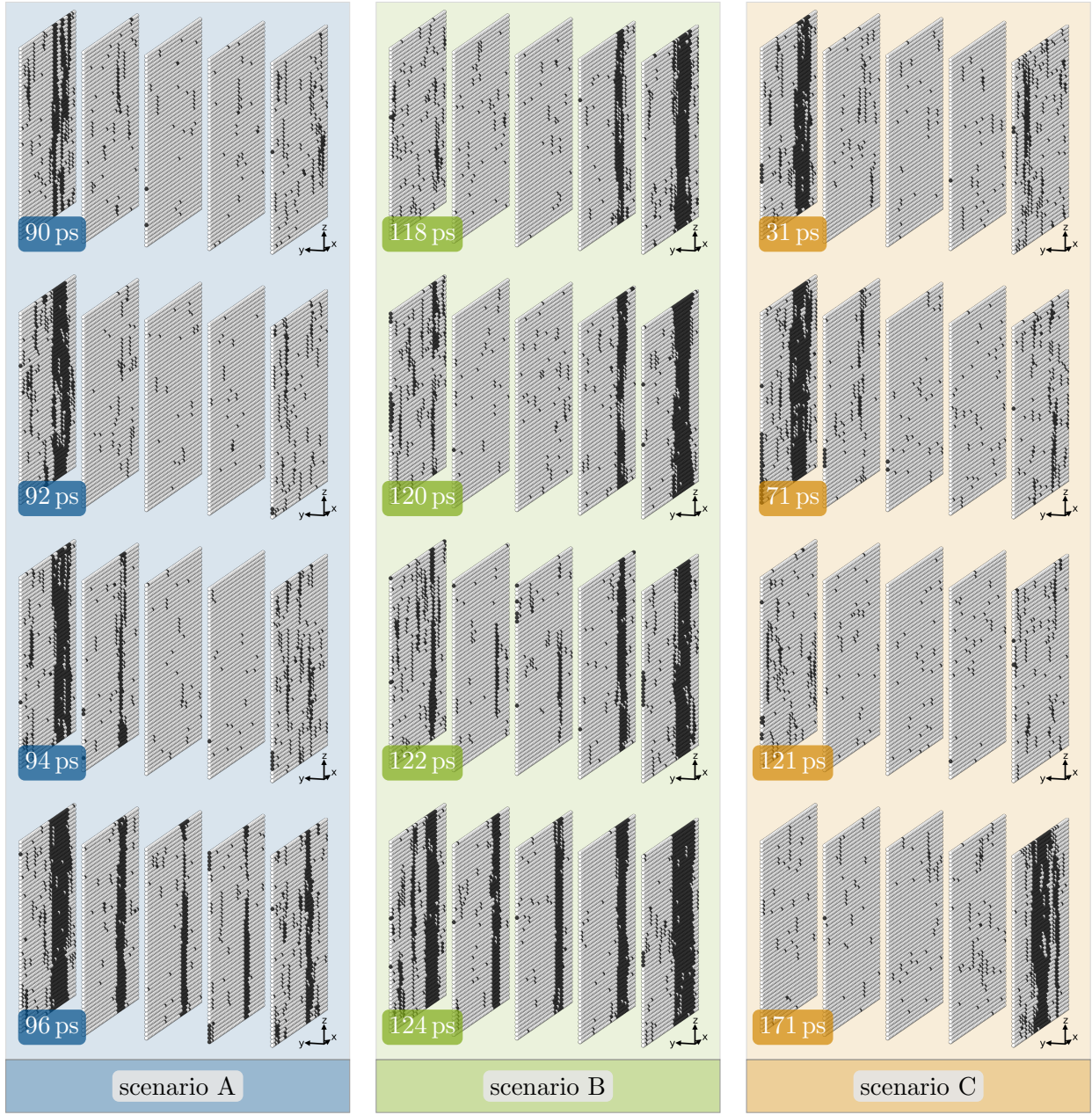

\newpage
\clearpage

\section{Energy change during domain collapse}
\label{app:6}

    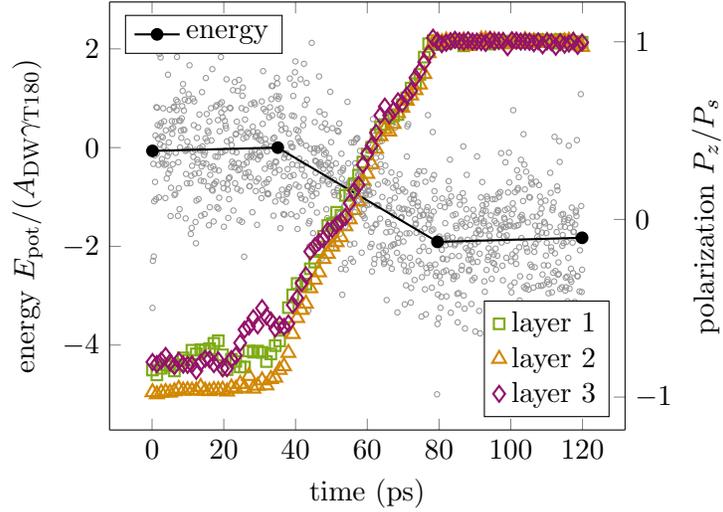
\begin{figure}[htb] 
        \centering
        \begin{tikzpicture}[/pgfplots/tick scale binop=\times]
            \begin{axis}[
                name=energyatomistic,
                anchor = west,
                axis y line*=left,
                set layers,
                mark layer=axis background,
                xlabel={time in \unit{\pico\second}},
                ylabel={energy \(E_{\text{pot}}/(A_{\text{DW}}\gamma_{\text{T180}})\)},
                legend pos=north west,
                legend style={cells={anchor=west}},
                ]
                \addplot [black!40, mark=o, only marks, forget plot, mark size=1] table [x=time, y=potEng] {data/energy_kinetics_CS_315K_sample07.dat};
                \addplot [black, mark=*, thick,] table [x=time, y=potEng] {data/energy_kinetics_CS_315K_sample07_pwlf.dat};
                \addlegendentry{energy};
            \end{axis}
            \begin{axis}[
                name=layersatomistic,
                anchor = west,
                hide x axis,
                axis y line*=right,
                ylabel=polarization $P_z / P_s$,
                ytick={-1,0,1},
                legend pos=south east,
                legend style={cells={anchor=west}},
                ]
                \addplot [tud4c, mark=square, only marks, thick, mark size=2] table [x=time, y=avgP] {data/polarization_kinetics_L1_CS_315K_sample07.dat};
                \addlegendentry{layer 1};
                \addplot [tud7c, mark=triangle, only marks, thick, mark size=3] table [x=time, y=avgP] {data/polarization_kinetics_L2_CS_315K_sample07.dat};
                \addlegendentry{layer 2};
                \addplot [tud10c, mark=diamond, only marks, thick, mark size=3] table [x=time, y=avgP] {data/polarization_kinetics_L3_CS_315K_sample07.dat};
                \addlegendentry{layer 3};
            \end{axis}
        \end{tikzpicture}
        \caption{
        Kinetics of layer-wise polarization of every layer of a reversed domain (domain size of 3 unit cells at \qty{-34}{\kelvin}) together with the change in potential energy. Potential energy is normalized by the energy of one domain wall and polarization is normalized by the spontaneous polarization.
        This data origins from the atomistic model; the \heff -approach shows qualitatively identical results.
        }
        \label{fig:polarization_energy_kinetics_CS}
    \end{figure}

\end{document}